\documentclass[namedreferences]{solarphysics}
\usepackage[optionalrh]{spr-sola-addons} 
\usepackage{graphicx}        
\usepackage{color}           
\usepackage{url}             
\usepackage{pdflscape}

\begin{document}

\begin{article}

\begin{opening}

\title{Time Evolution Altitude of an Observed Coronal Wave}

\author{C.~\surname{Delann\'ee}$^{1}$, G.~\surname{Artzner}$^{2}$, B. ~\surname{Schmieder}$^{1}$, S. ~\surname{Parenti}$^{3}$}
\runningauthor{Delann\'ee et al.}
\runningtitle{Altitude of a coronal wave}

   \institute{$^{1}$ LESIA, Observatoire de Paris, site de Meudon, 5 place Jules Jansen, F-92195 Meudon cedex
                     email: \url{cecile.delannee@obspm.fr}}
   \institute{$^{2}$ Institut d'astrophysique spatiale, b\^at 121, Université paris sud, 91405 Orsay UMR 8617 email: \url{guy.artzner@ias.u-psud.fr}}
   \institute{$^{3}$ Royal Observatory of Belgium, 3 av. Circulaire, 1180 Bruxelles, Belgium
                     email: \url{s.parenti@oma.be}}

\begin{abstract}
The nature of coronal wave fronts is deeply debated. They are observed in several wavelength bandpasses in spectra, and are frequently interpreted as magnetosonic waves propagating in the lower solar atmosphere. However, they can be attributed to the line of sight projection of the edges of coronal mass ejections. Therefore, the altitude estimation of these features is crucial to discriminate in favor of one of these two interpretations. We take advantage of a set of observations obtained from two different points of view by EUVI/SECCHI/STEREO on December, 7th 2007 to derive the time evolution of the altitude of a coronal wave front. We develop a new technique to compute the altitude.

We find that the observed brightness has an increasing altitude during 5 minutes, then the altitude decreases slightly back to the low corona. We interpret the evolution of the altitude as following: the increase of altitude of the wave front is linked to the rise of a bubble like structure whether it is a magnetosonic wave front or a CME in the first phase. During the second phase, the observed brightness is mixed with the brightening of the underlying magnetic structures as the emission of the plasma of the wave front fades due to the plasma dilution with the altitude.

\end{abstract}
\end{opening}

\section{Introduction}
     \label{S-Introduction}
Coronal wave fronts are bright arcs propagating on the solar disc observed in a wide range of plasma temperature. They were first discovered by Moreton (1964) in H$\alpha$ images. In this reference, a wave agent is called to understand how flares can perturb a distant filament. The velocity of this wave was derived to be about 1000 km s$^{-1}$. Later, many of these distant perturbations, associated with flares, were observed. Nowadays, as the H$\alpha$ observations have better spatial and temporal resolution than before, the first instants of these waves are clearly observed. They are bright arcs followed by dark arcs in the blue wing, dark arcs followed by bright arcs in the red wing (Smith and Harvey 1971) or bright arcs in the core of the H$\alpha$ line (see e.g. Asa\"i et al. 2012). Most of the theoritical models of this phenomenon are based on magnetosonic wave or shock modes propagating on the solar surface (see e.g. Shen and Liu 2012). Uchida (1968, 1970) proposed that a fast-mode MHD weak shock wave can intersect the chromosphere to produce the wave observed in H$\alpha$.

With the SoHO satellite and its Extreme ultra-violet Imaging Telescope (EIT, Delaboudini\`ere et al. 1996), the coronal wave came to be also observable in EUV wavelength bandpass, as bright arcs (Thompson et al. 1998 and references in Patsourakos and Vourlidas, 2012). Later on, several sets of observations showed that a single coronal wave can emit in EUV and in H$\alpha$ (Thompson et al. 2000 and references to the velocity problem given below), suggesting a possible multithermal structure. However, the velocity of these new EUV waves seemed to be slower (200-500 km s$^{-1}$, Long 2012) than what was derived from the H$\alpha$ observations only (Biesecker et al. 2002, Eto et al. 2002).
The difference in velocities were resolved by simultaneous observations showing a full overlay of the H$\alpha$, EUV and soft X-ray coronal wave fronts (Vr\v{s}nak et al. 2006, Warmuth et al. 2005, Narukage et al. 2002, Kumar et al. 2013 and references therein that show the difficulty to understand the observations themselves). Adding detailed studies of the shapes, dynamics, and the relation with observations of flares and coronal mass ejections, some authors found that the coronal waves are deeply connected to coronal mass ejections (Biesecker et al. 2002), rather to flares (Chen 2006), and that they are deeply connected to the surrounding magnetic field (Delann\'ee et al. 2007). In that sense, Delann\'ee et al. (2008) proposed that the coronal waves are in fact not waves at all but the edges of the coronal mass ejections observed projected onto the solar disc, as confirmed by the MHD simulation that they performed.

The altitude of a coronal wave could be a discriminating parameter between these two models (namely, coronal magnetosonic wave or shocks, and the line of sight projection of the edges of the
coronal mass ejection). If the altitude of the wave front is quite stable, therefore, the wave front could be a structure lying inside a layer of the stratified solar atmosphere. If the wave front continuously increases to quite large altitude, therefore, it could be the signature of the ejection of material. Patsourakos and Vourlidas (2008) analyzed a coronal wave observed with EUVI/SECCHI/STEREO on December, 7th 2007. Their study led to the conclusion that the wave fronts are really wave structures propagating at 90$\pm$7 Mm above the solar surface, leading the authors to exclude the prediction of an altitude of 300-500 Mm for the wave front made in Delann\'ee et al. (2008). 

In this study, we use the same data set enriched by multiwavelength and dynamical study to derive the evolution of the altitude of the observed coronal wave versus time.

\section{Observations}
\label{observation}

We analyze the data obtained on December, 7th 2007 between 04:30 UT and 05:15 UT, as they present the studied wave front. The wave, starting at 04:33 UT, is accompanied by a B class flare beginning at 04:30 UT and a weak CME at 05:54 UT as shown in the CDAW CME list (\url{http://cdaw.gsfc.nasa.gov/CME_list/}). The observations are described in Patsourakos and Vourlidas (2008). We remind here just their main characteristics.

The two spacecraft, Ahead (A) and Behind (B), of the Solar Terrestrial Relations Observatory (STEREO) mission, obtained clear observations of a coronal wave that day. They were separated by an angle of $\approx 45^\circ$, giving the opportunity to have two points of view of the same structure to derive the altitude of the coronal wave, using three methods that will be described later on. We analyze the Extreme Ultra-Violet Imager (EUVI/SECCHI/STEREO, W\"usler et al. 2004) observations, an instrument on board STEREO, through several filters centered at: (a) 171 \AA, (b) 195 \AA, (c) 284 \AA~and (d) 304 \AA, giving an idea of the temperature of the observed structure as they are centered, respectively, on the emission line of (a) Fe{\sc ix} and Fe{\sc x} emitted by a plasma at the typical temperature of 1 MK, of (b) Fe{\sc xii} emitted at 1.5 MK, of (c) Fe{\sc v} emitted at 2 MK and of (d) He{\sc ii} emitted at 0.05 MK. However, we point out that the filters do not provide real spectroscopic signatures of the observations as each bandpasse includes several emission lines. The analyzed observations are processed with the standard secchi$\_$prep routine present in the SolarSoft. 

Unfortunately no H$\alpha$ data are available for that day, so that we cannot investigate more the geometry of the wave. Nevertheless, we can use the soft X-Ray Telescope (XRT/Hinode, Golub et al. 2007, Kano et al. 2008) observations. These observations are obtained using several filters. We use four images obtained using  Al\_poly, C\_poly, Ti\_poly, Al\_poly/Ti\_poly, with an exposure time respectively of  23.1, 23.1, 8.2, 16.4 seconds. Using these filters, we estimate the temperature of the observed wave front (see Section \ref{sec:temperature}). The Hinode XRT data is processed using the standard XRT software available on SolarSoft. The saturated and bloomed pixels are discarded from the analysis.
The temperature sensitivity of these bands covers the range $6.1\leq \log T\leq7.5$ (Narukage et al. 2011). 

XRT images have large and small field of views. The small XRT field of view is 526$\times$526 arcsec$^{2}$. As the Hinode point of view is different than the STEREO A and B ones, the active region emitting the wave front is located near the center of the observed solar disc in Hinode images. 
To compare the XRT observations with the EUVI A and B observations, we use observations obtained with the Extreme ultra-violet Imaging Telescope on board the Solar and Heliospheric Observatory (EIT/SoHO, Delaboudini\`ere et al. 1996). The EIT filters are centered on the same wavelengths than the EUVI A and B filters. The EIT field of view is the whole sun as the large XRT field of view.

\begin{figure}
\includegraphics[width=2.95cm,height=2.95cm]{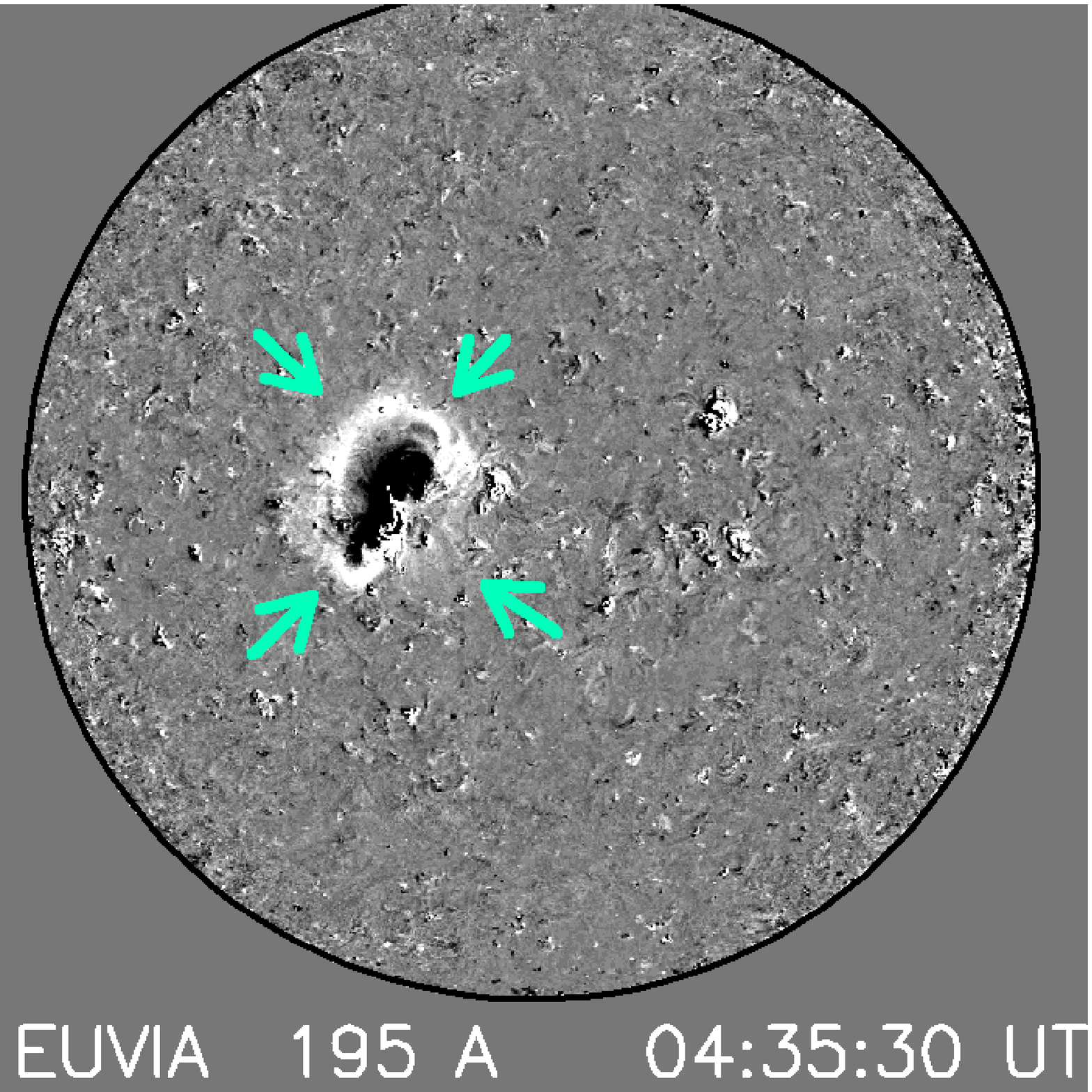}
\includegraphics[width=2.95cm,height=2.95cm]{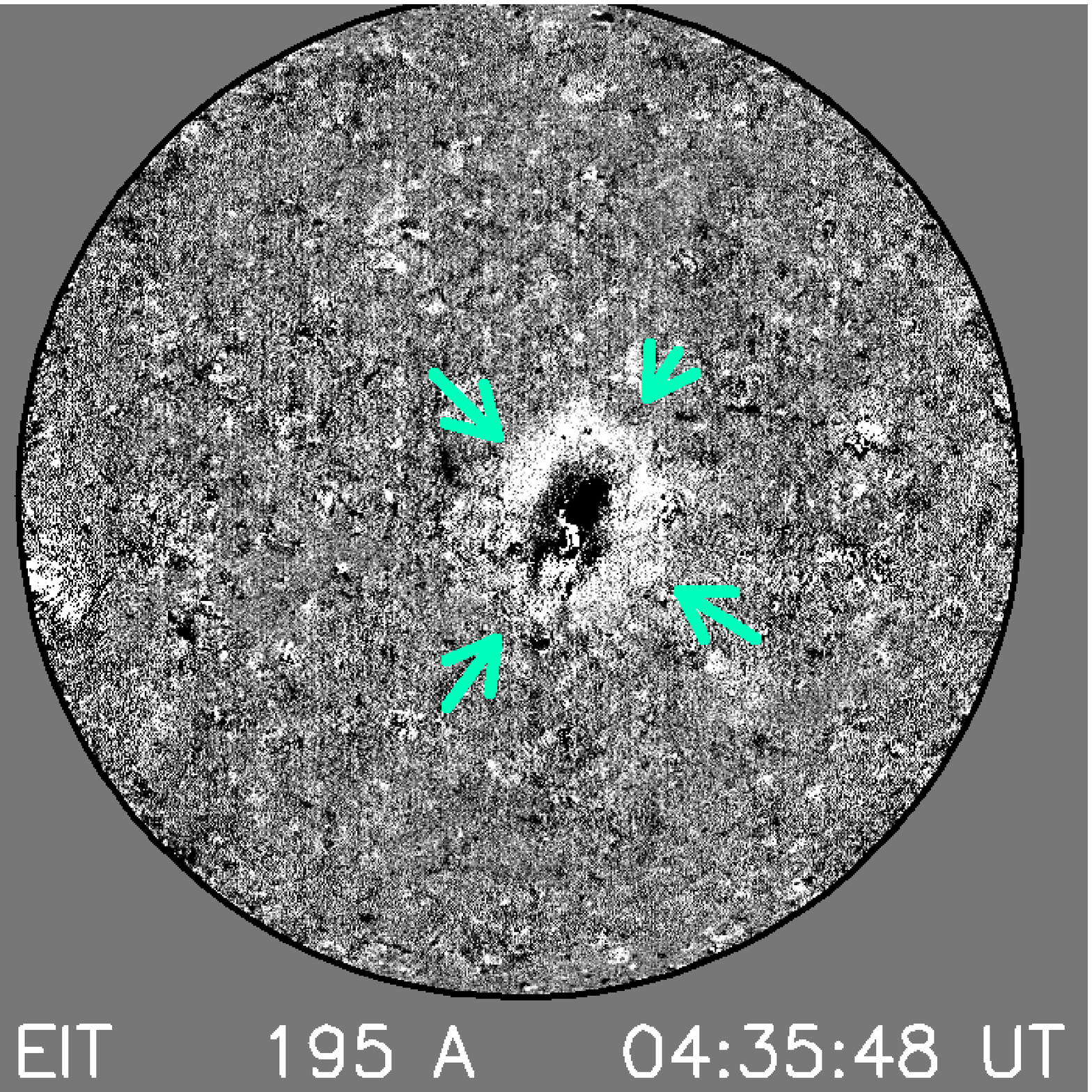}
\includegraphics[width=2.95cm,height=2.95cm]{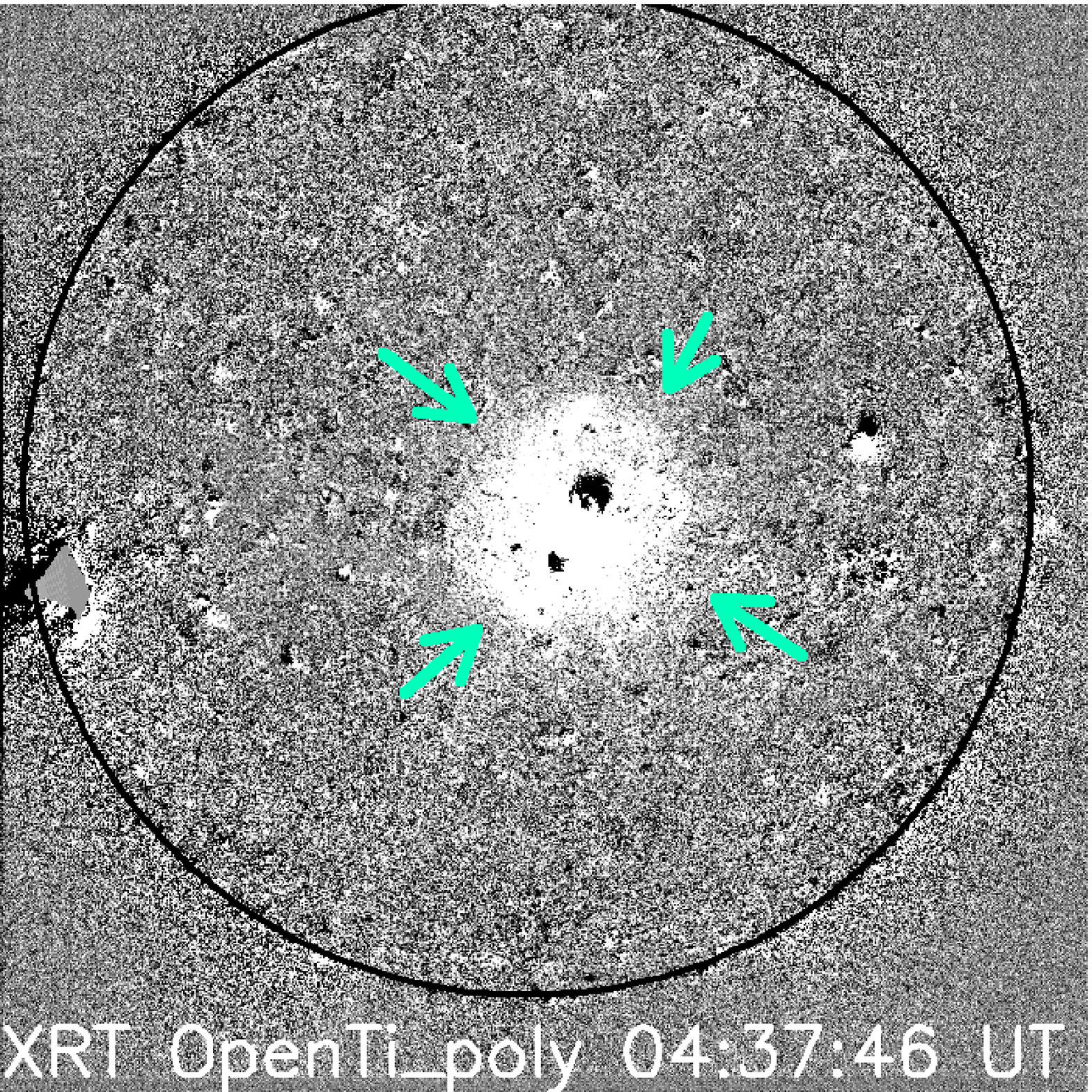}
\includegraphics[width=2.95cm,height=2.95cm]{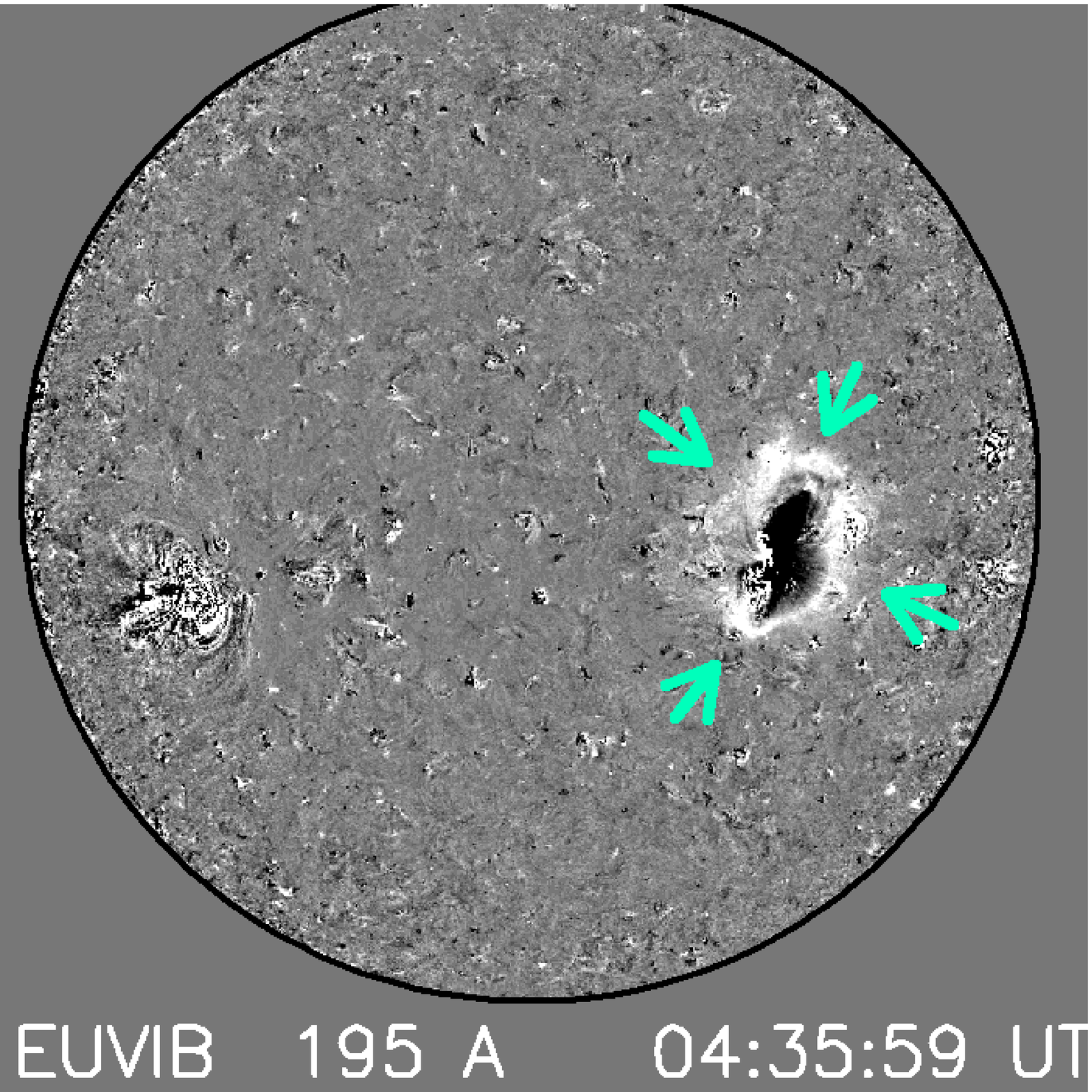}
\caption{ Images obtained through filter centered at  195 \AA~in EUVI/SECCHI/STEREO A (first image), EIT (second image) and EUVI/SECCHI/STEREO B (last image). The XRT (third) image is obtained using the Ti\_poly filter. The time delay between the images is about 2 min. The images are difference of image with a pre-event image corrected from the solar rotation. The intensity scaling for each image is chosen to reveal the wave front. The images display the full sun. The flaring active region is about the sun center, slightly apart in the EUVI A and EUVI B fields of view depending on the position of the spacecraft. The wave front is the white ellipse, marked by the arrows, around the flaring active region.}
   \label{XRTEIT}
   \end{figure}

Figure \ref{XRTEIT} shows four observations of the whole sun using EUVI A 195 \AA, EIT 195 \AA, XRT Ti\_poly and EUVI B 195 \AA, from left to right order. The presented images are difference of an image obtained at the studied date with a pre-event image corrected from the solar rotation. The scaling of intensity is adjusted to enhanced the wave front. The time delay for this data set is about 2 min. Using the velocities given in Section \ref{morphology}, the wave front propagates through about 18 Mm which corresponds to 10 EIT pixels and 0.3 mm in the panels shown in Figure \ref{XRTEIT}, during these 2 min. The wave front appears in the 4 observations. X-rays wave front location is coherent enough with the EUV wave front to assume that the same wave is observed using the X ray and EUV instruments. We note here that the morphology of the wave front, as viewed from Hinode, i.e. from the Earth, is rather symmetrical along the central meridian that passes by the flare site. This symmetry corresponds to the line of sight integration consideration given in Section \ref{morphology}.

\section{Observability of the wave front}
\subsection{Morphologies from the two STEREO spacecraft points of view}
\label{morphology}

\begin{figure}
\centerline{\includegraphics[width=5cm,height=5cm]{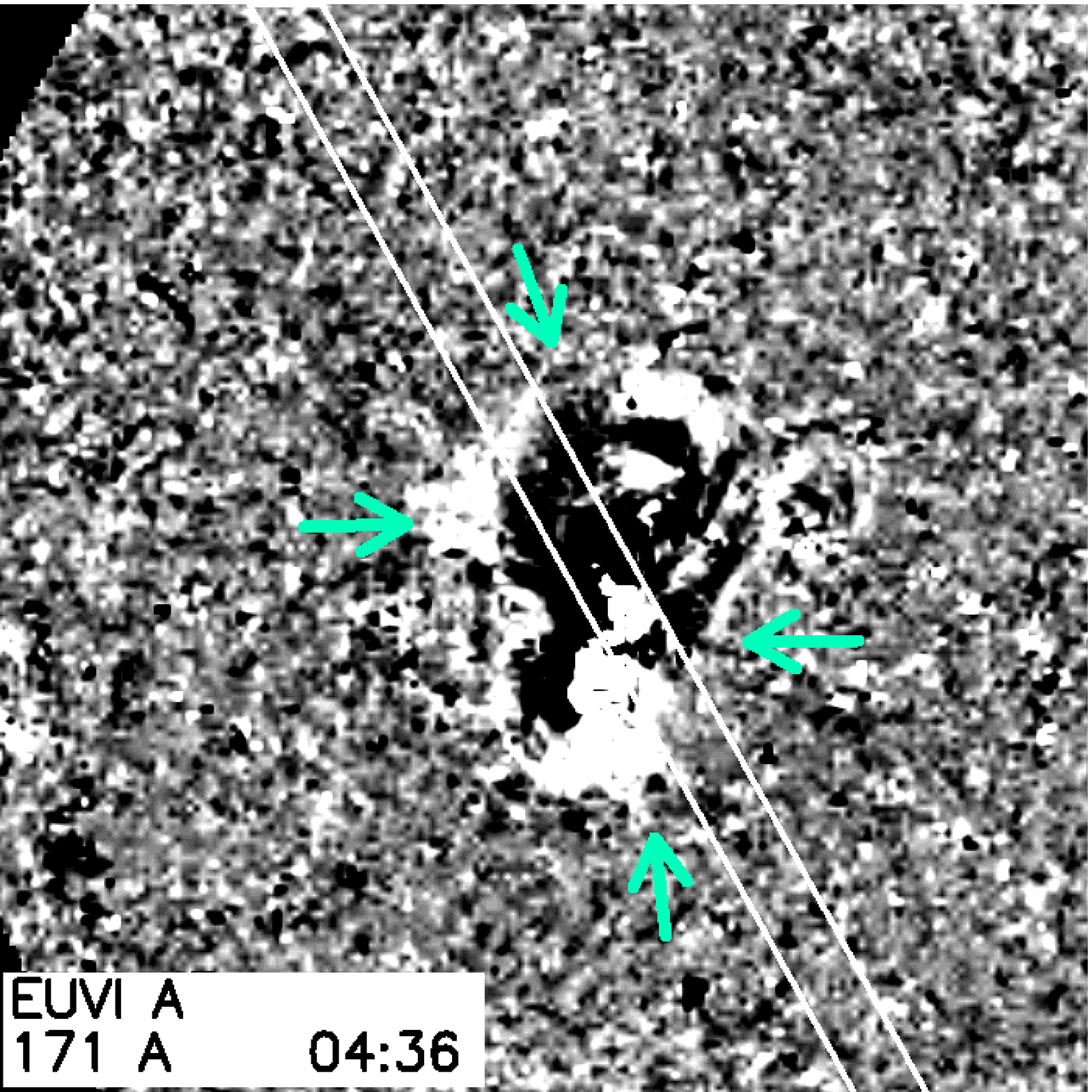}
\includegraphics[width=5cm,height=5cm]{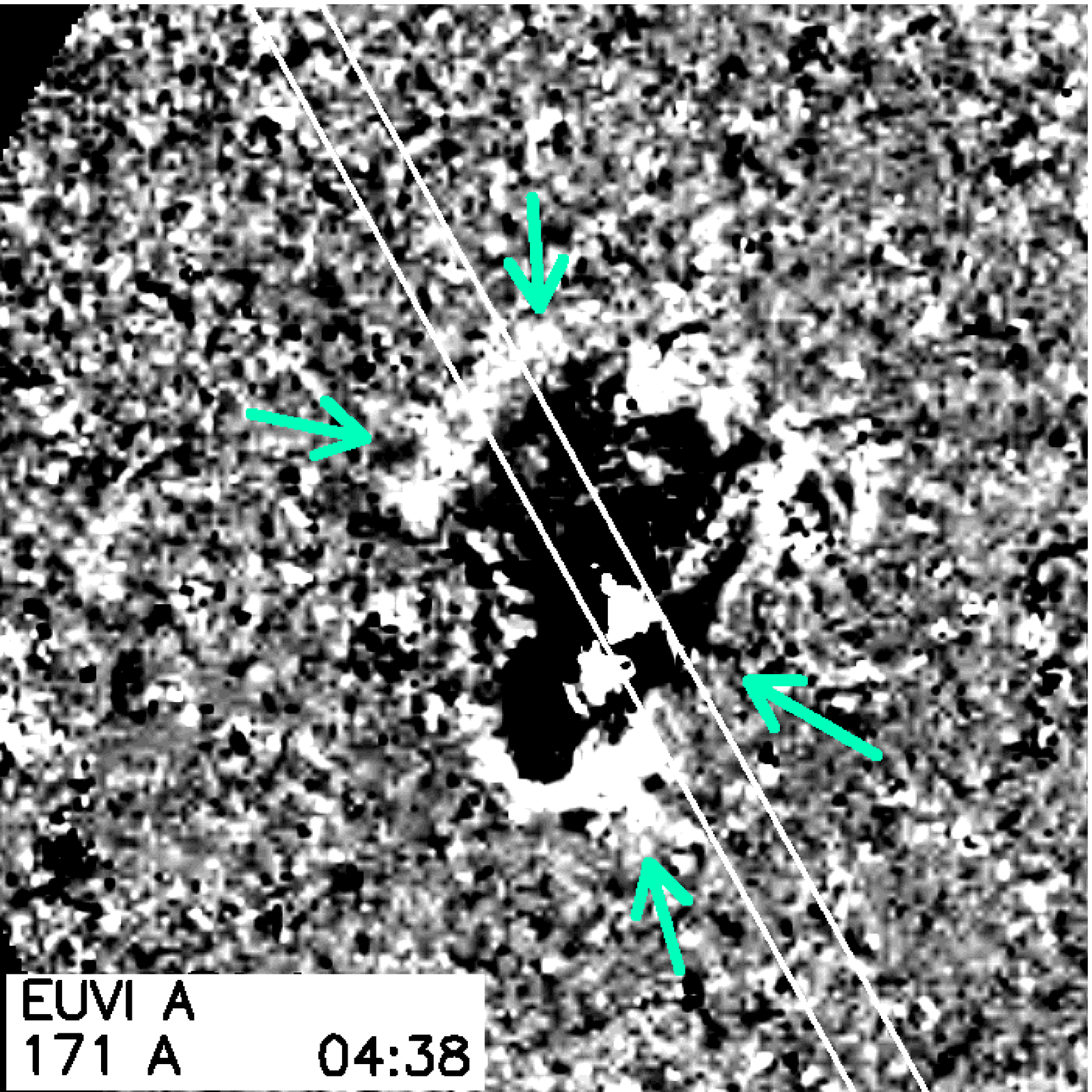}}
\centerline{\includegraphics[width=5cm,height=5cm]{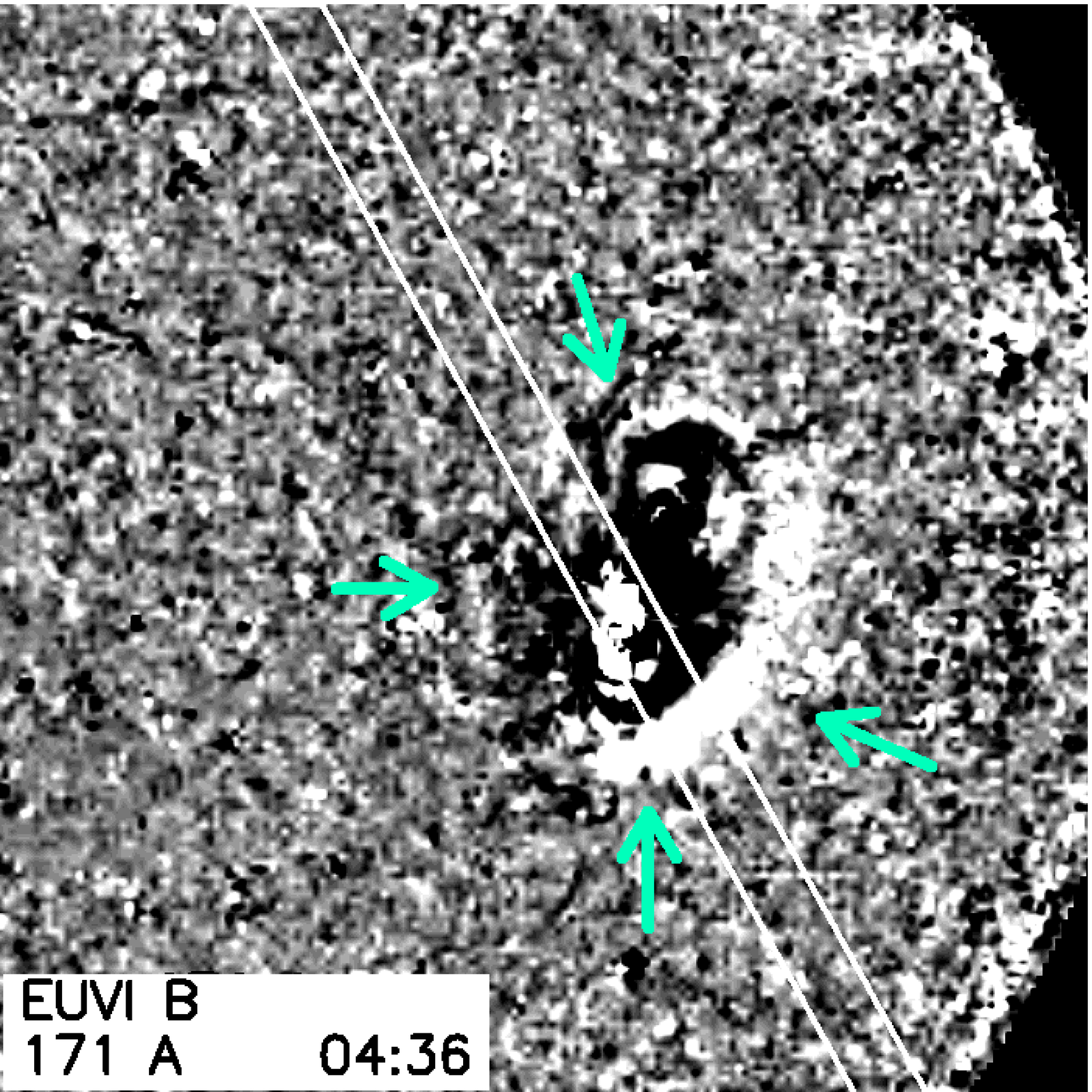}
\includegraphics[width=5cm,height=5cm]{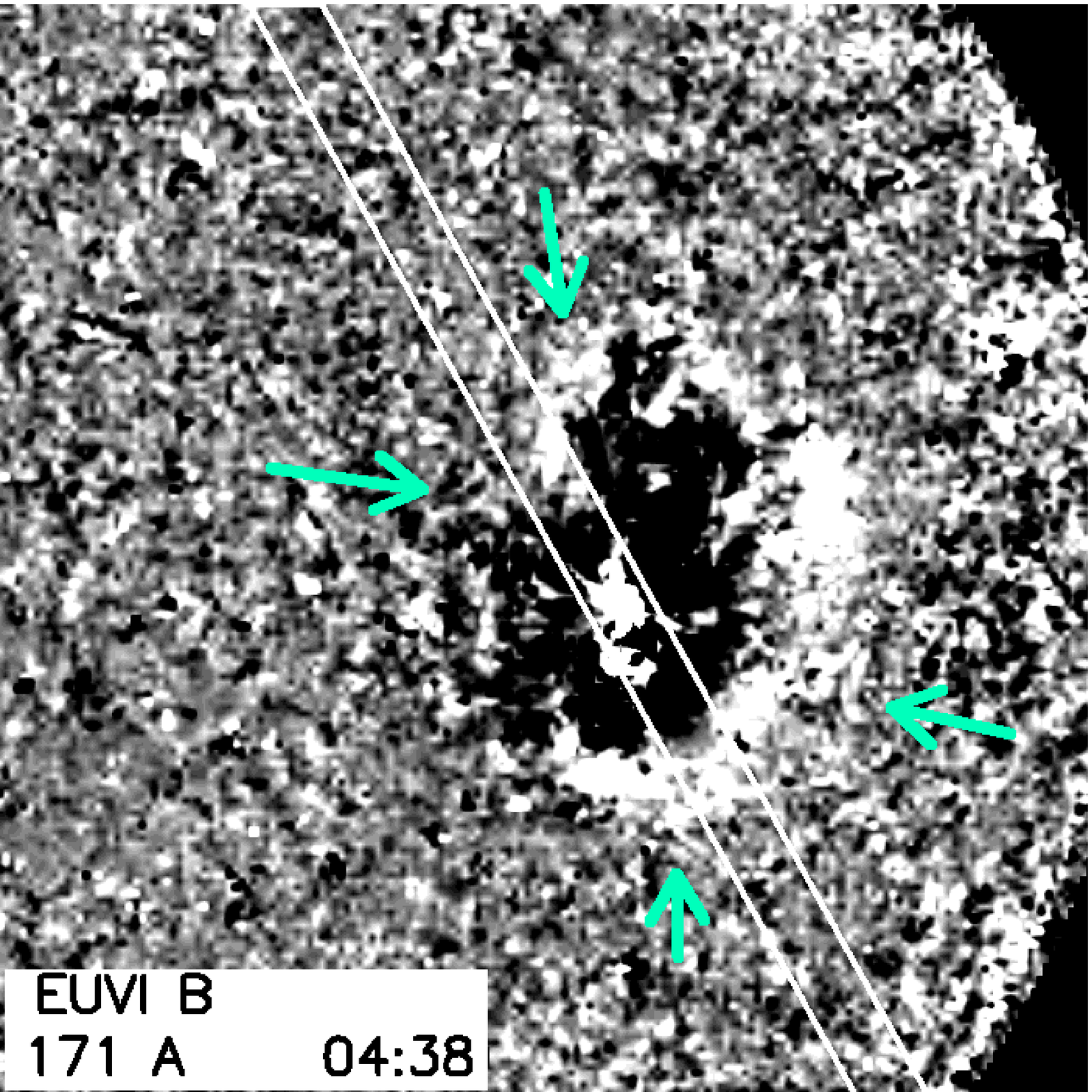}}
\centerline{\includegraphics[width=5cm,height=5cm]{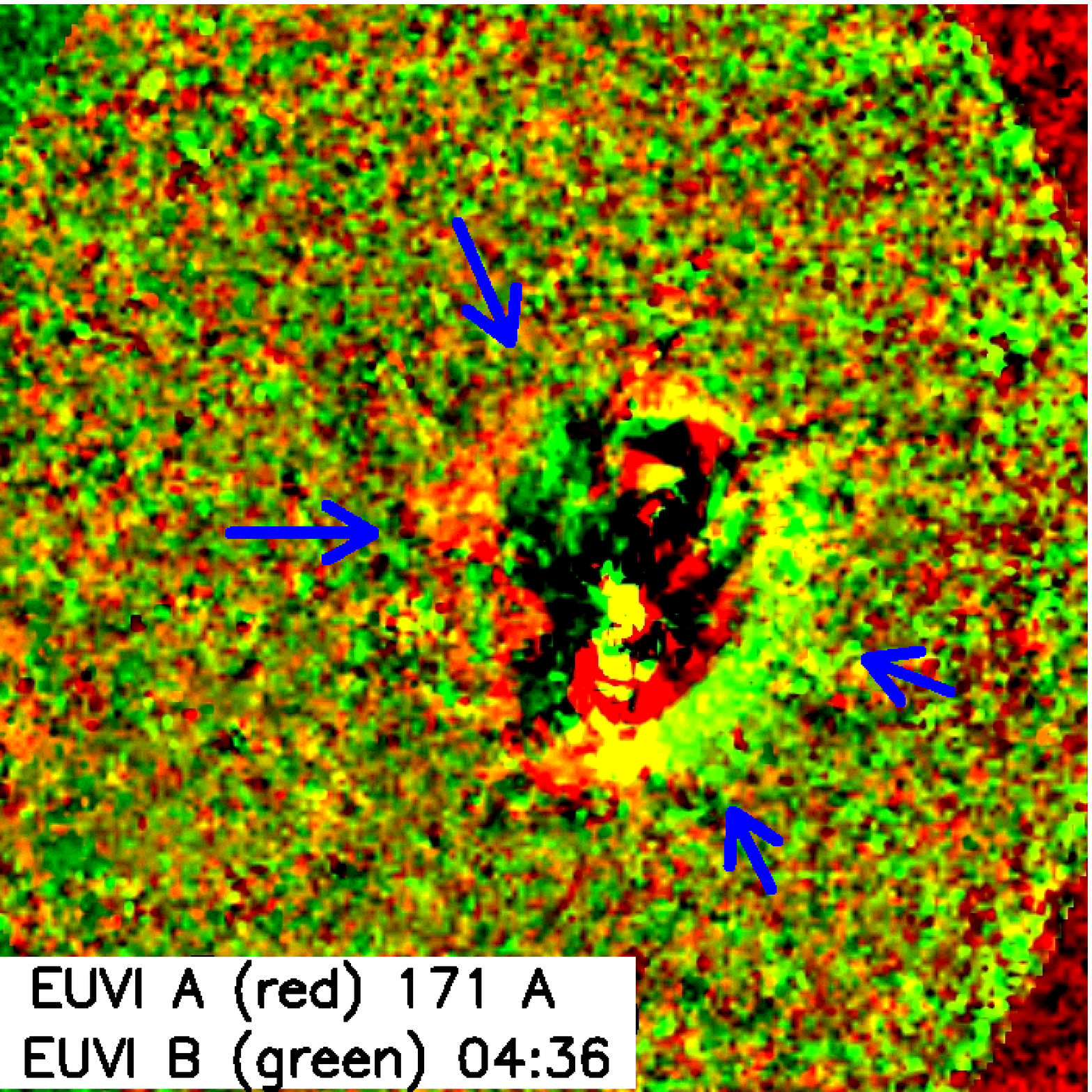}
\includegraphics[width=5cm,height=5cm]{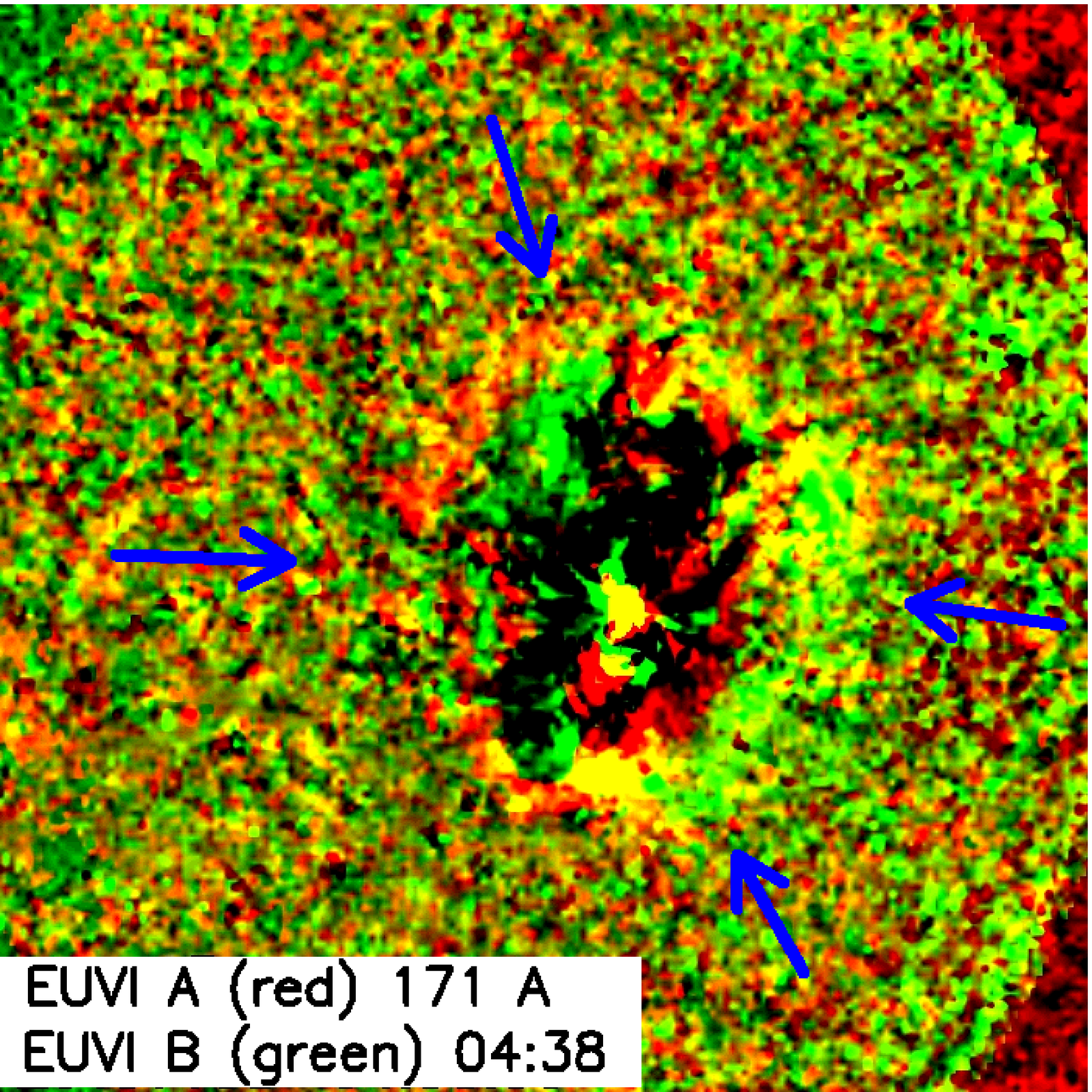}}
\centerline{\includegraphics[width=6.5cm,height=5cm]{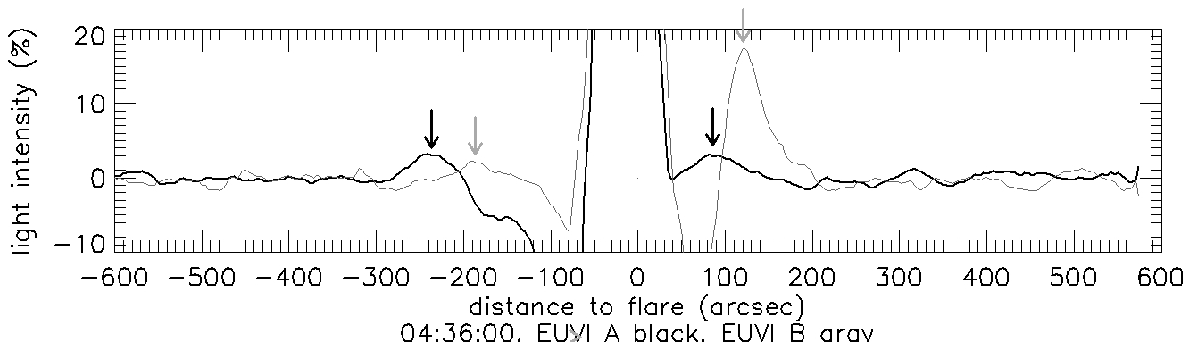}
\includegraphics[width=6.5cm,height=5cm]{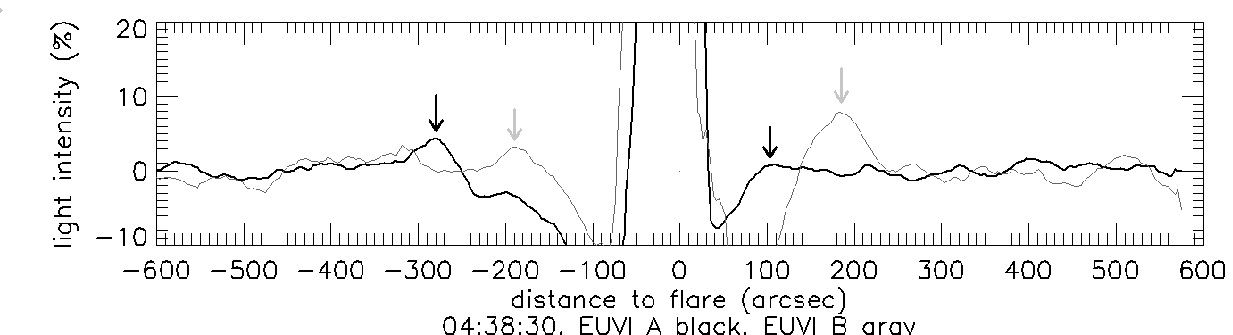}}
\caption{Derotated based difference images of observation obtained with EUVI/SECCHI/STEREO using the filter centered at 171 \AA. First (second) line presents the observations obtained with STEREO A (B, respectively). Third line presents the overlay of the two first images using the red (green) channel to display the STEREO A (B, respectively) observation. Colored arrows indicate the outer edge of the wave front. Fourth line: percentage of intensity variation, from the image at 04:31:00 UT, averaged over sliding squares comprised inside the parallelogram shown on each images of the two first lines. The black (gray) line stands for the EUVI A (B, resp.) data. The displayed range of values cuts of the flare emission. The arrows highlight the coronal wave front propagation over distances from the flare site. First (second) column presents the observations at 04:36:00 UT (04:38:30 UT resp.).}
   \label{superposition}
   \end{figure}

When the wave front is observed from two different points of view, it shows up different morphologies that are emphasized in Figure \ref{superposition}.
All the images in Figure \ref{superposition} are processed as follows: first, the solar rotation is corrected to be coherent with a pre-event observation obtained at 04:31:00 UT used as a reference image, second the brightness of each image is subtracted with the reference image, third a hyperbolic tangent is applied to the difference intensity in order to smooth its spatial variations, fourth the displayed values are comprised between $\pm 0.1$ in order to search for very low intensity variations. The observations obtained with STEREO B are scaled to the ones obtained with STEREO A. We display only a portion of the Sun (1113$\times$1113 arcsec$^2$) centered on the erupting active region, overlaying the observations from the two spacecraft of the active region. The first and second columns are the time sequence obtained with the process described above. The first line shows the observations obtained by STEREO A, the second by STEREO B and the third line displays the STEREO A observations in the red channel and the STEREO B ones, in the green channel. Therefore, the yellow pixels of the image are due to overlay of green and red pixels, mainly in the flare site.

From the images obtained during the first 8 minutes of propagation of the coronal wave front (from 04:33:59 UT to 04:41:29 UT), we clearly see that the two eastern and western sides of the coronal wave front do not show up the same intensities when viewed by the two different spacecraft: STEREO A observations have a bright western front and a faint eastern front while the STEREO B observation have a bright eastern front and a faint western front (see Figure \ref{superposition}). 

\begin{table}
\begin{tabular}{|c|c|c|c|c|c|c|c|c|c|c|c|c|}
\cline{1-13}
time&\multicolumn{6}{c|}{eastern}&\multicolumn{6}{c|}{western}\\
(UT) &\multicolumn{6}{c|}{edge}&\multicolumn{6}{c|}{edge}\\
\cline{2-13}
&\multicolumn{3}{c|}{STEREO}&\multicolumn{3}{c|}{STEREO}&\multicolumn{3}{c|}{STEREO}&\multicolumn{3}{c|}{STEREO}\\
&\multicolumn{3}{c|}{Ahead}&\multicolumn{3}{c|}{Behind}&\multicolumn{3}{c|}{Ahead}&\multicolumn{3}{c|}{Behind}\\
\cline{2-13}
&d&wd&I&d&wd&I&d&wd&I&d&wd&I\\
04:&('')&('')&(\%)&('')&('')&(\%)&('')&('')&(\%)&('')&('')&(\%)\\
\cline{1-13}
33 &-200&10&-1&-175&?&-1&68&?&1&120&30&9\\
36 &-230&40&5&-195&40&0&130&?&1&178&50&19\\
38 &-290&40&5.5&-210&40&2&160&?&1&200&40&7\\
41 &-350&40&9&-310&?&3&160&?&1&220&?&2\\
43 &-390&60&7&-320&20&4&?&?&?&?&?&?\\
46 &-420&?&1&-350&40&4&?&?&?&?&?&?\\
\cline{1-13}
\end{tabular}
\label{table}
\caption{Estimated parameters of the wave front from the mean intensity variation in 171 \AA~presented in Figure \ref{superposition}: its distance to the flare (d), its wideness (wd) and its percentage intensity increase (I). Question marks are written where these parameters could not be estimated due to the lack of evidence of the wave front signal.}
\end{table}

In order to derive some quantities from the observations, we average the light intensity over squares comprised in the slice drawn on the images in the Figure \ref{superposition}. We shift pixels by pixels the square of integration along the slice to obtain a curve of a mean intensity, along the slice, around the flare site. Then, we compute the percentage of mean intensity variation from the pre-event image to the studied image. Varying the width of the squares of integration, we obtain different results: the wider the width is, the flatter the plot is; the smaller the width is, and the noisier the plot is. After trying several widths for the squares of integration, we find that 10$\times$10 pixels give a plot having the best ratio signal to noise of the wave front. We choose the slice to cut the wave front and the flare site as this latter seems to be the origin of the wave. As the wave front is asymmetric, we try to find the best inclination to have the strongest signal of the wave front in our plot. The results are presented in Figure \ref{superposition}, fourth line. These plots show clearly the propagation of the coronal wave, but the exact morphological parameters of the wave are difficult to derive. We estimate the distance (d) of the wave front from the flare, the full width at half maximum (wd) of the wave front, its brightness (I) (see Table 1) over the time. Estimates are made by hand as the peak of intensity is sometimes very difficult to localize numerically, therefore the derived numbers are quite disputable due to errors of measure. To reduce the errors of measure, we constantly compare the four rows of the Figure \ref{superposition}. We estimate the error on the values about 30\%.

We find that during the first 15 min. (04:31 - 04:46 UT), the eastern side of the coronal wave front is wider and brighter seen by STEREO A than seen by STEREO B, and its western side is wider and brighter seen by STEREO B than seen by STEREO A. The brightness increases for several minutes for both parts of the wave front then decreases, more precisely, the increase lasts 8 (10) min. (04:33 - 04:41 (04:43) UT) for the eastern edge observed by STEREO A (B, resp.), and 3 min. (04:33 - 04:36 UT) for the western edge observed by STEREO B. No brightness variation is clearly measurable for the observation of the western edge by STEREO A. The two points of view show similar speed of the same edge of the coronal wave: from the values given in Table 1 we find about 200 km/s for the eastern edge and about 140 km/s for the western edge. However, we note that the speed of each edge is not constant during its propagation.

\begin{figure}
ice cone model \hfill dome model
\centerline{\includegraphics[width=5cm]{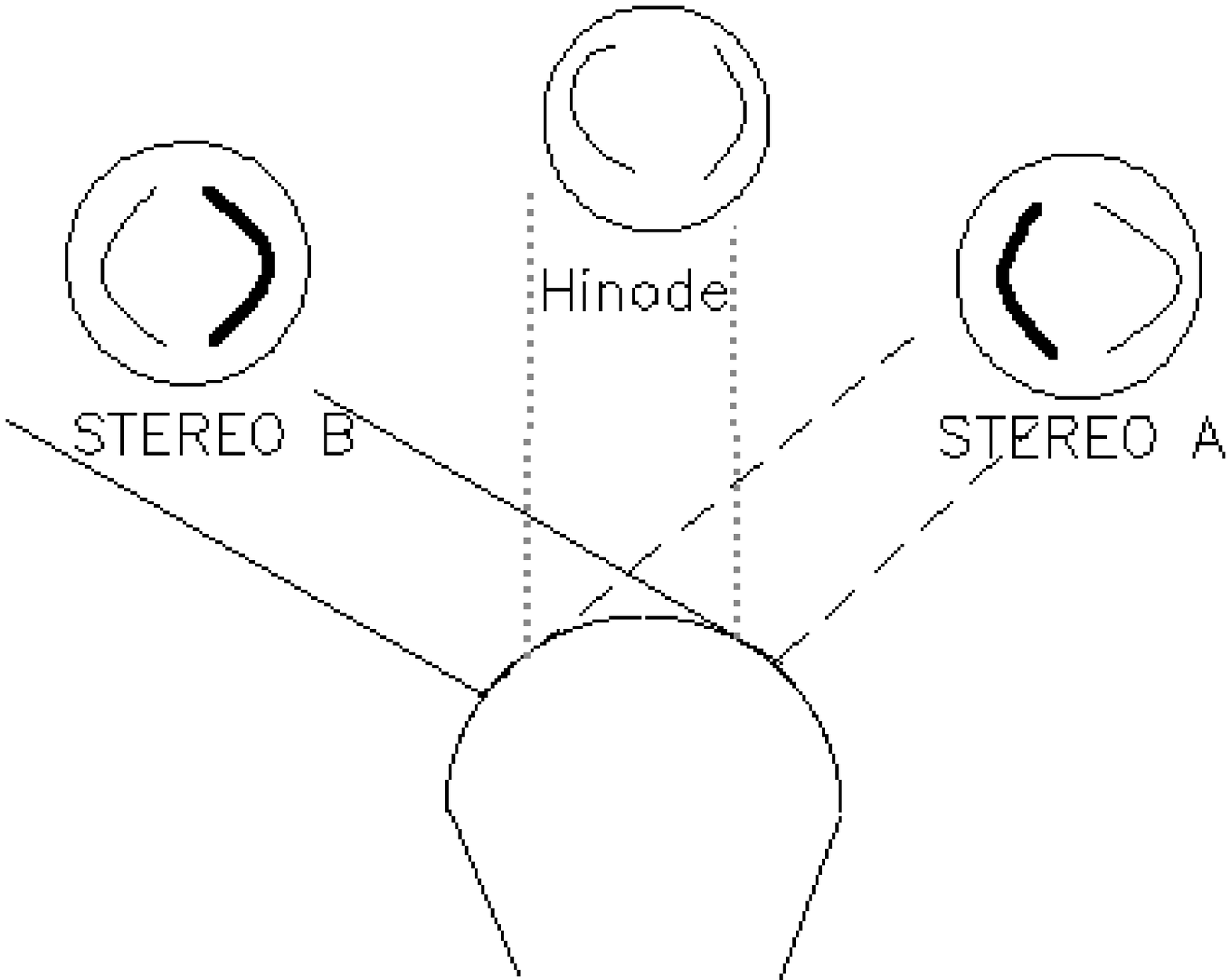}
\hfill
\includegraphics[width=5cm]{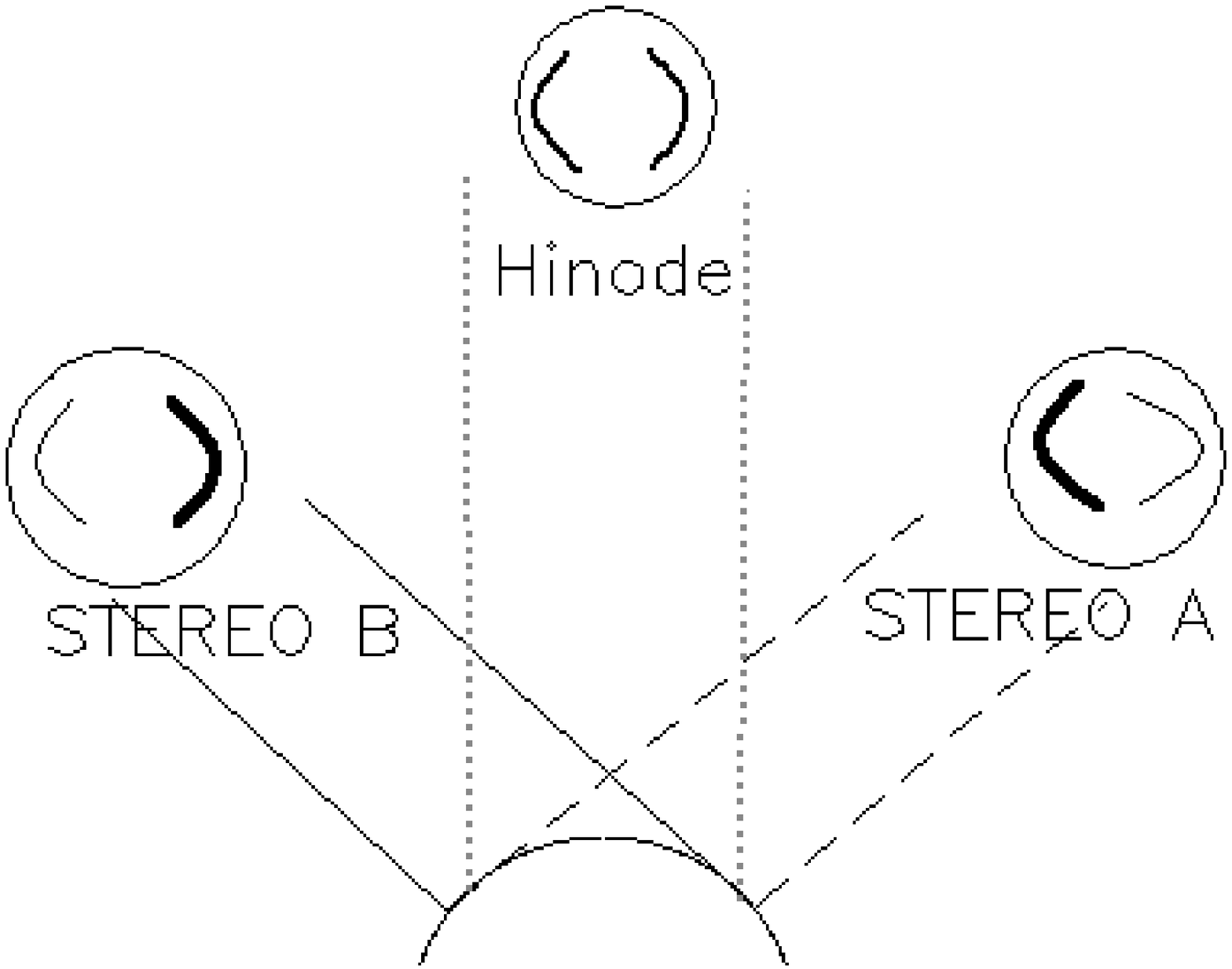}}
\caption{Sketches of the possible geometries of the wave front that fit its observed morphologies as seen from both points of view of the STEREO pair (sides of each panel) and hinode (middle of each panel): ice cone (left panel) and dome (right panel). Six light rays are drawn to define the lines of sight from the three spacecraft of the two edges of the wave front. The circles drawn above the spacecraft names represent the solar limb in their field of view. In the circles are drawn the two edges of the wave front. The same edge is large in one STEREO field of view and narrow in the other one in the STEREO field of view. The two edges have the same widness in the Hinode field of view.}
 \label{sketchdomecone}
 \end{figure}

The appearance of the wave front depends on its geometry and the point of view. As the plasma is optically thin in the used wavelength, the light emitted by each layer of plasma along the line of sight is integrated along each layer. To understand the morphology of the wave front, we try to find some geometries that fit the observed wave front imposing that they have quite smooth parameters and that they are bubble like. We find at least two morphologies interpreting the observed properties: a dome or an ice cone (see sketch in Figure \ref{sketchdomecone}). The ice cone can be a model of the morphology of a CME (Thernisien, 2011) and the dome, a model of the morphology of the magnetosonic wave (Veronig et al., 2010). These morphologies are certainly not the only ones as we do not work through a demonstration process, instead we try to find, into the usual morphologies given in the literature, the line of sight that may produce the same observed morphologies.

For the two possibilities that we propose, the point of view implies that the lines linking them to the borders of the dome or ice cone meet one border with a nearly flat angle and the other border with a larger angle. The point of view is crucial to produce the observed morphology: if we shift this point of view, the rendering of the integration along the line of sight may be completely different.
We cannot certify that the position of one spacecraft would imply that the line of sight presented in the sketch of Figure \ref{sketchdomecone} would be the most frequent one. Therefore, there should be other observations that would show different characteristics concerning the morphology as seen from the different points of view.

During the first 15 min. (04:31 - 04:46 UT), the eastern side is farther from the active region as seen in STEREO A than as seen in STEREO B, and the western side is farther as seen from B than from A, meaning that the structure is not on the solar surface but rather in altitude (see Figure \ref{superposition} and Table \ref{table}). The shift increases during the first 5 (3) min. then decreases, meaning that the altitude of the eastern (western, resp.) edge of the coronal wave increases during this time interval then tends to go back to the solar surface. The shift seems also larger for the eastern edge than for the western one meaning that the eastern edge is higher than the western one. However, due the very low brightness of the eastern (western) edge observed by the STEREO B (A, resp.) leading to large uncertainties, these results have to be verified with the other following methods.

\subsection{Using several temperature sensitive filters}
\label{sec:temperature}

\begin{figure}
\centerline{\includegraphics[width=3.5cm,height=3.5cm]{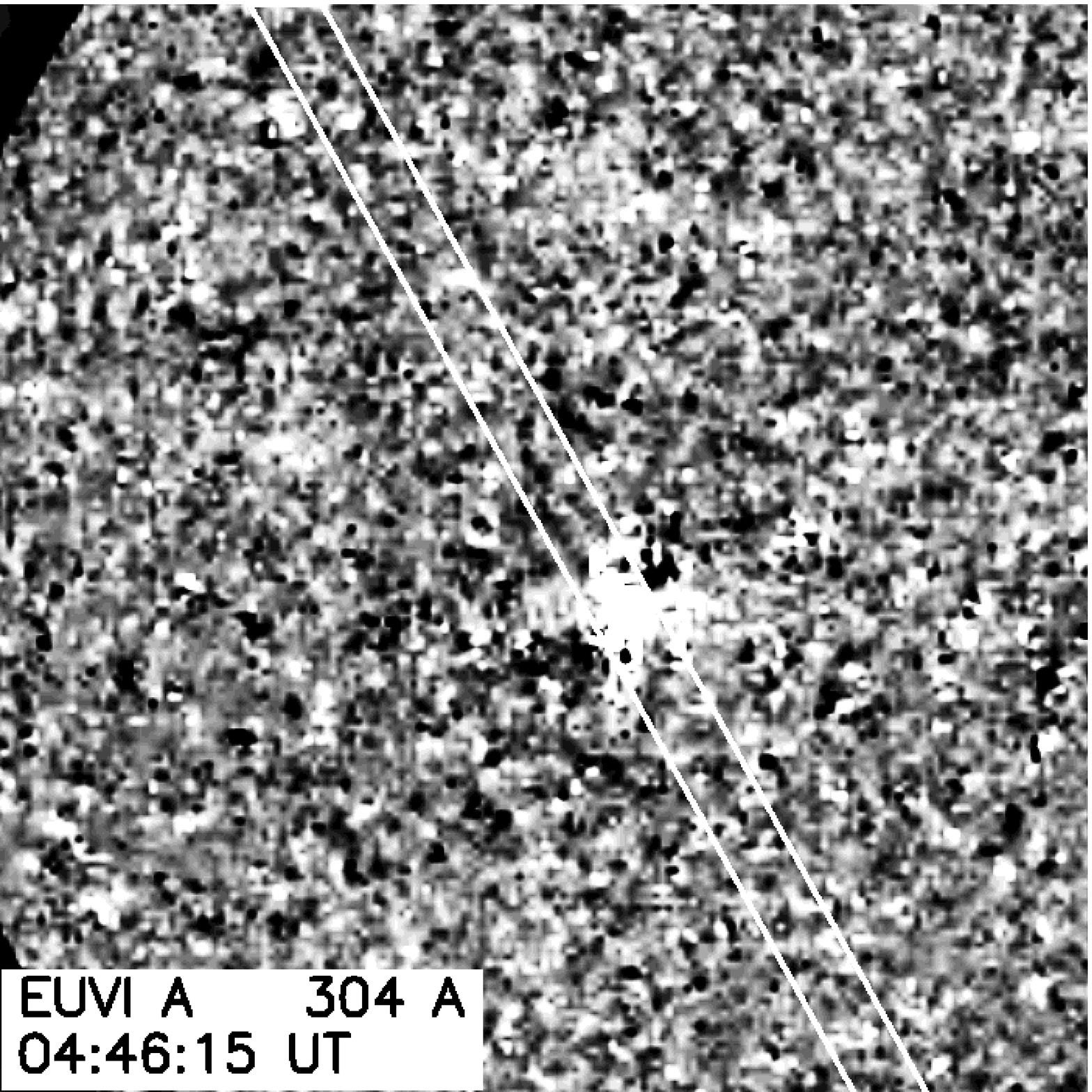}
\includegraphics[width=3.5cm,height=3.5cm]{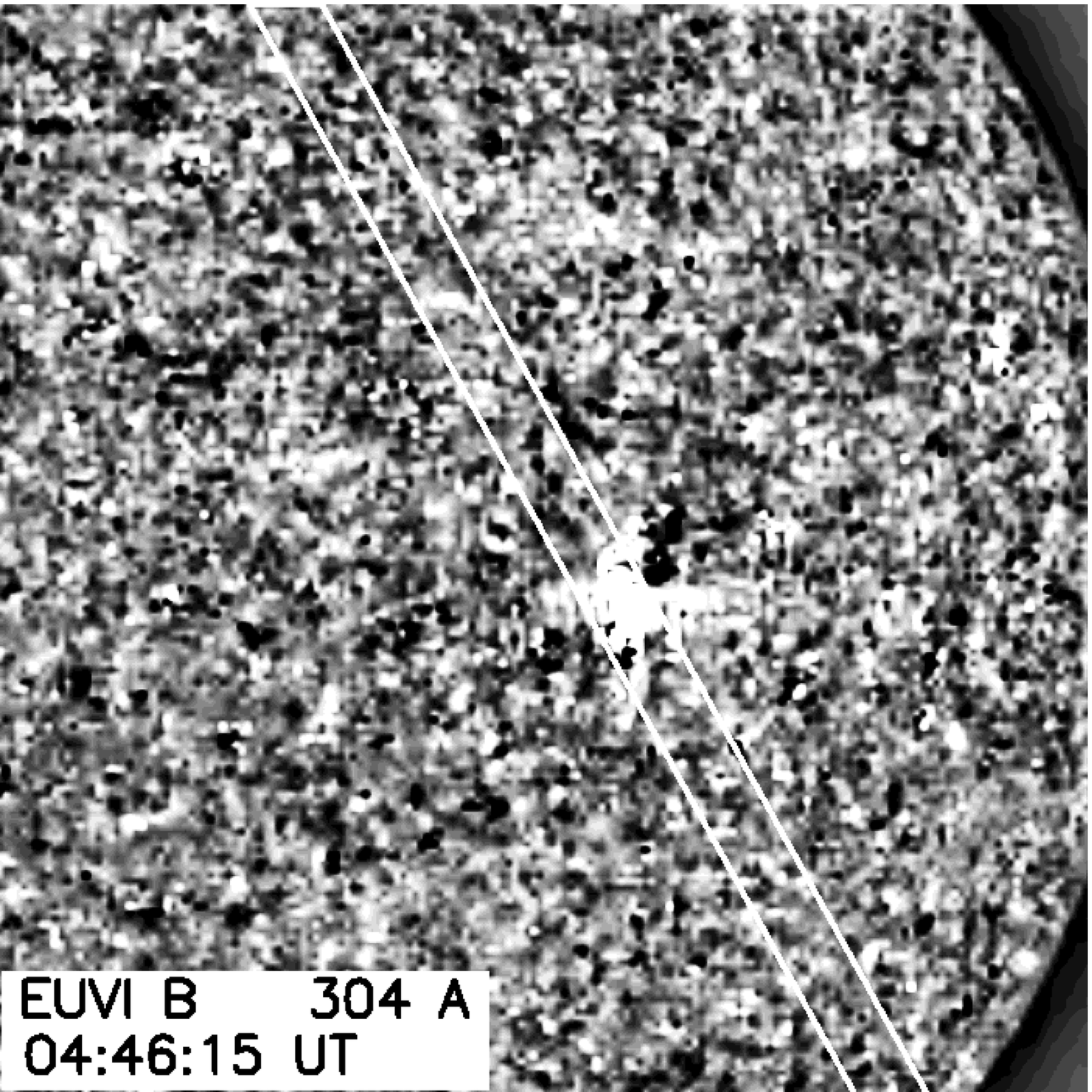}
\includegraphics[width=5cm,height=3.5cm]{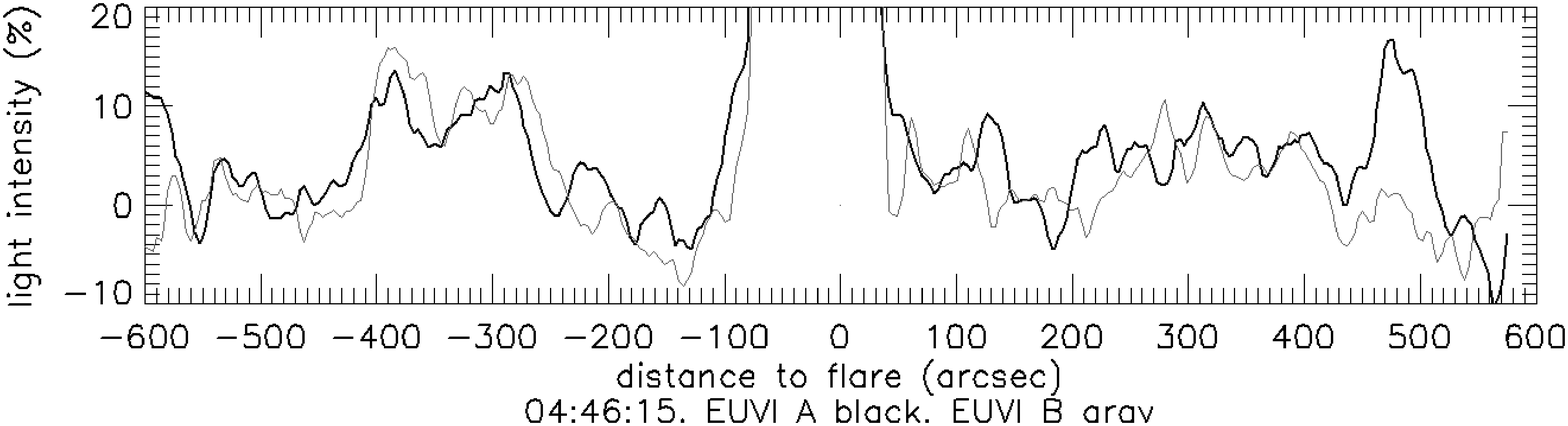}}
\centerline{\includegraphics[width=3.5cm,height=3.5cm]{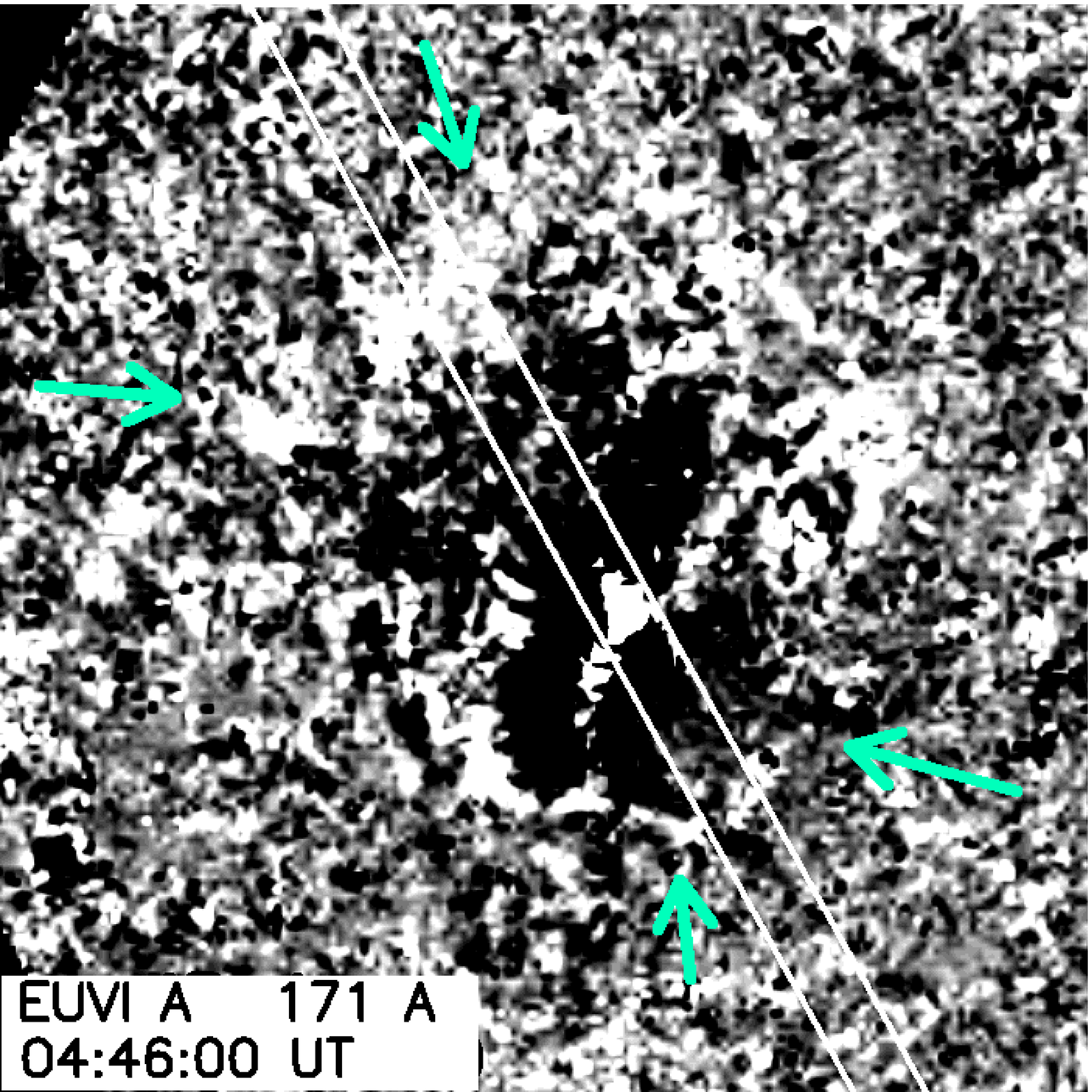}
\includegraphics[width=3.5cm,height=3.5cm]{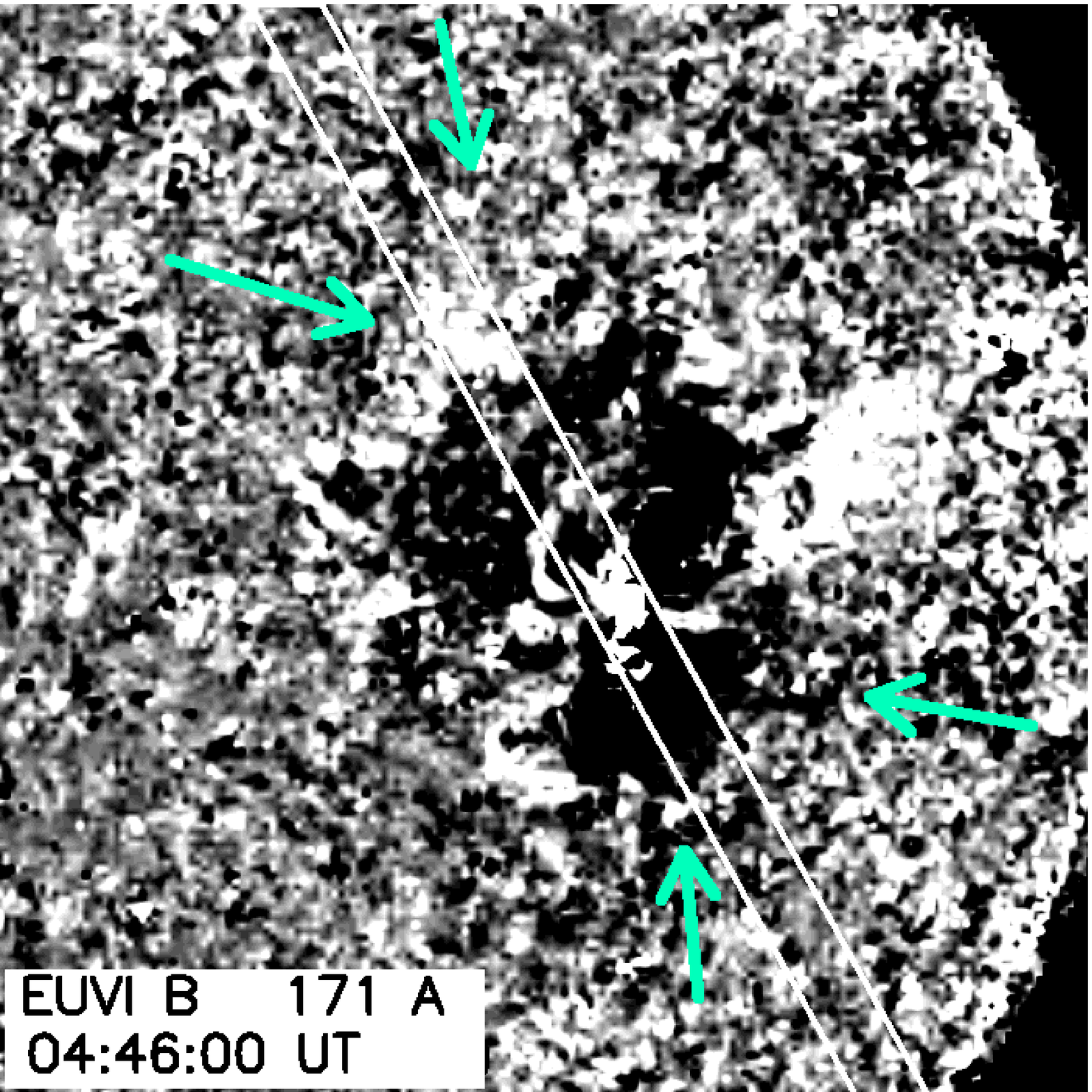}
\includegraphics[width=5cm,height=3.5cm]{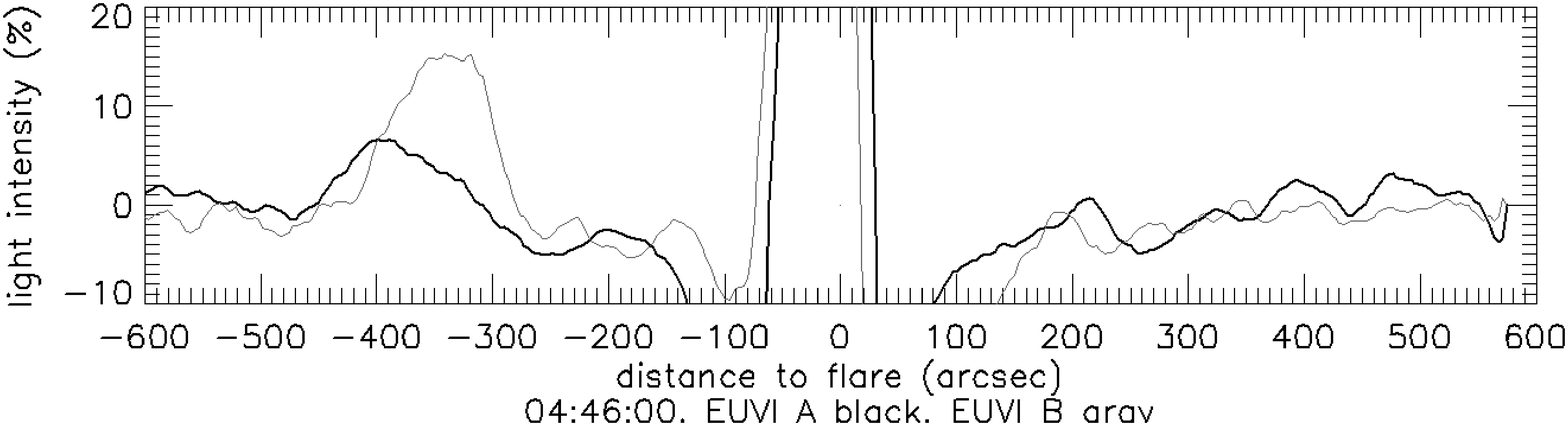}}
\centerline{\includegraphics[width=3.5cm,height=3.5cm]{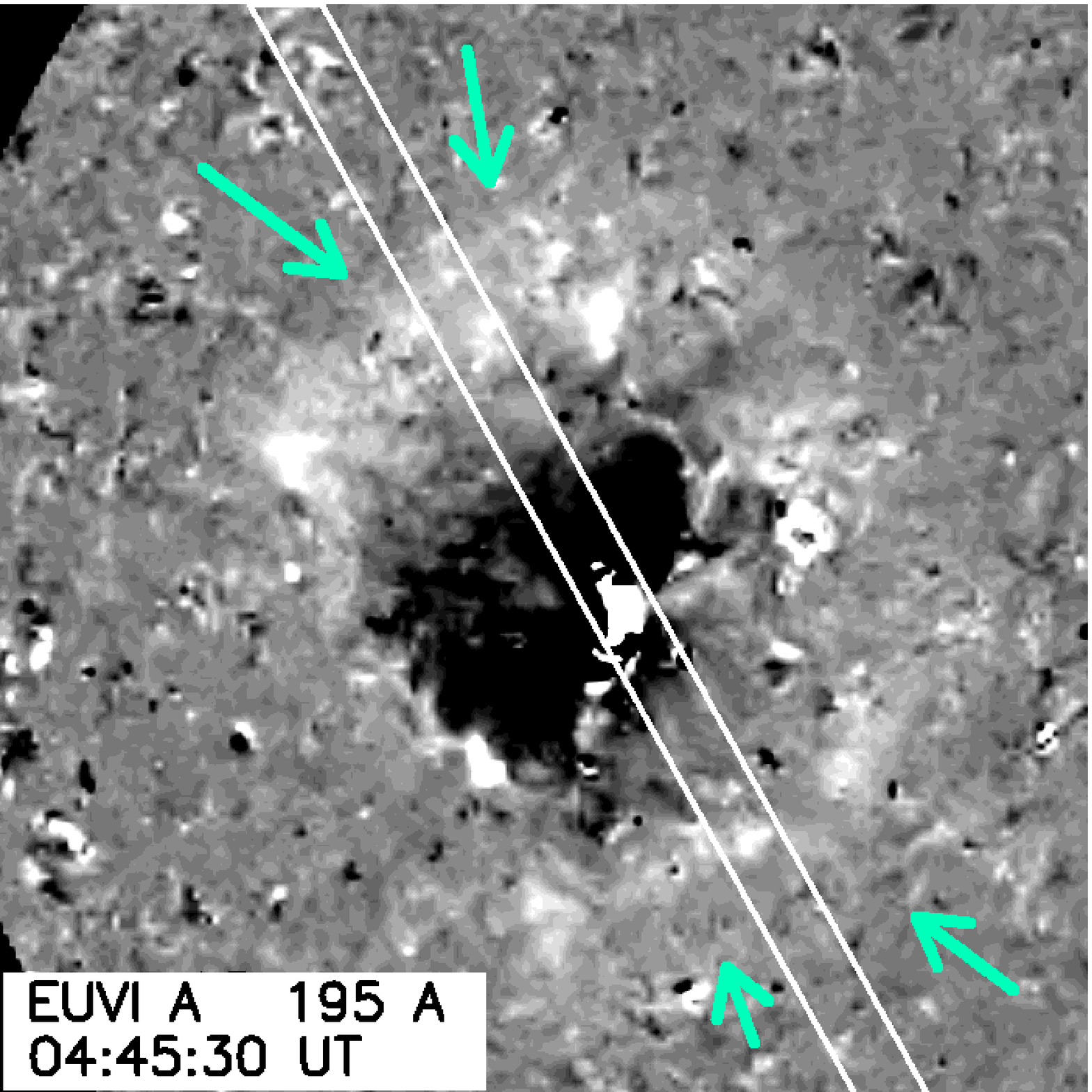}
\includegraphics[width=3.5cm,height=3.5cm]{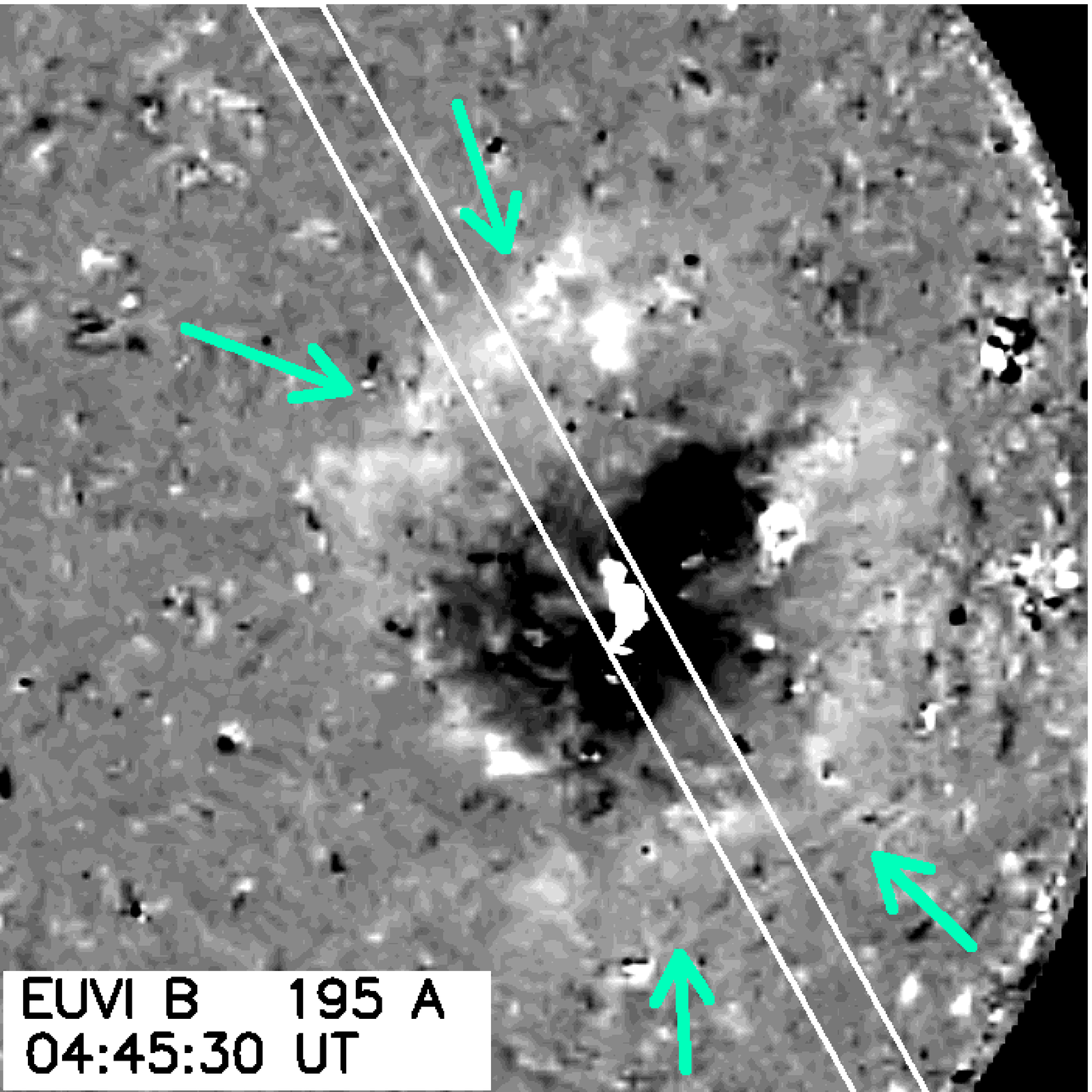}
\includegraphics[width=5cm,height=3.5cm]{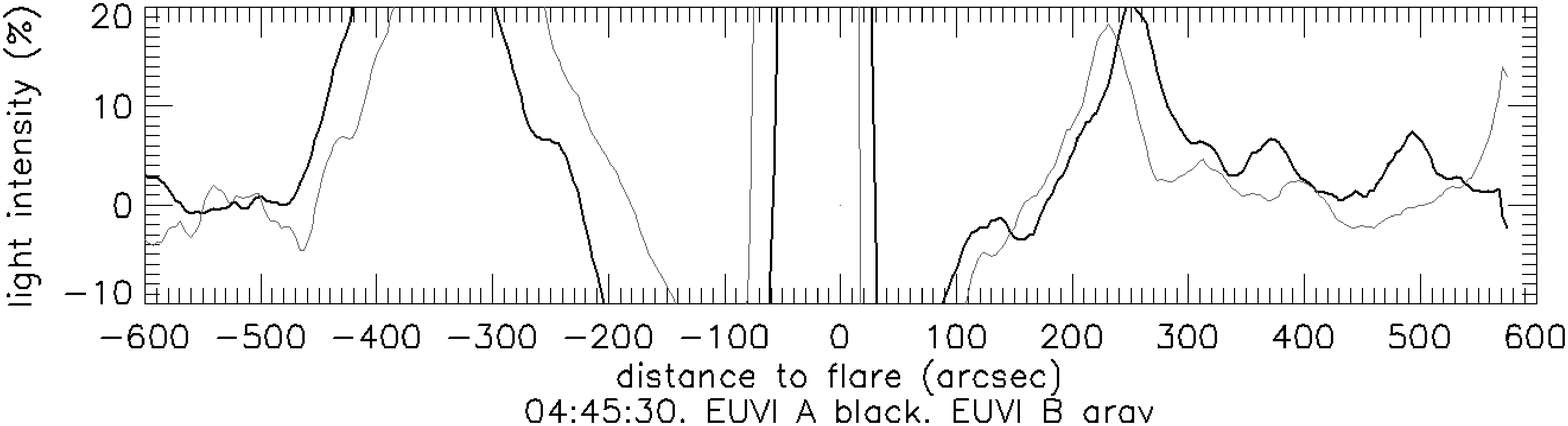}}
\centerline{\includegraphics[width=3.5cm,height=3.5cm]{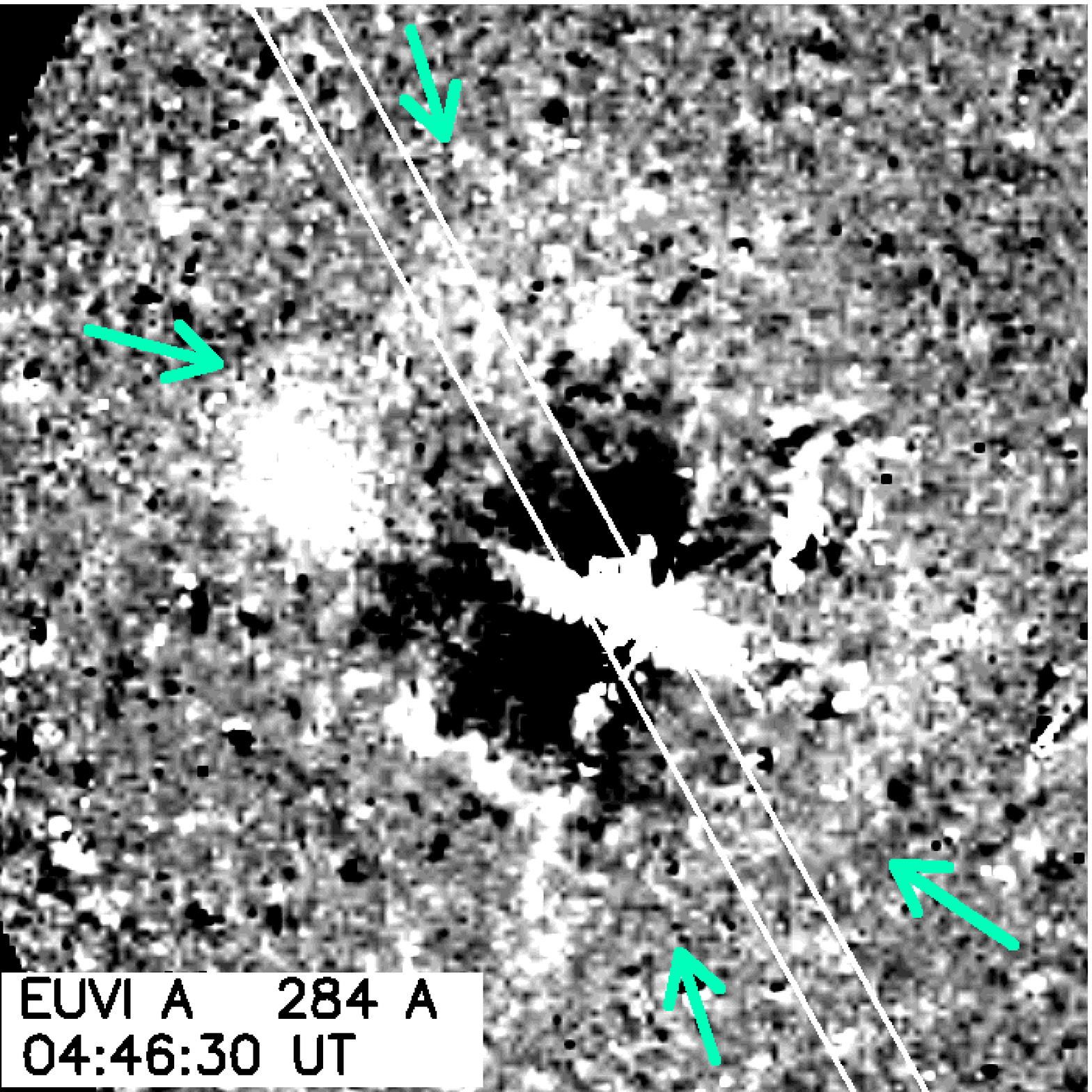}
\includegraphics[width=3.5cm,height=3.5cm]{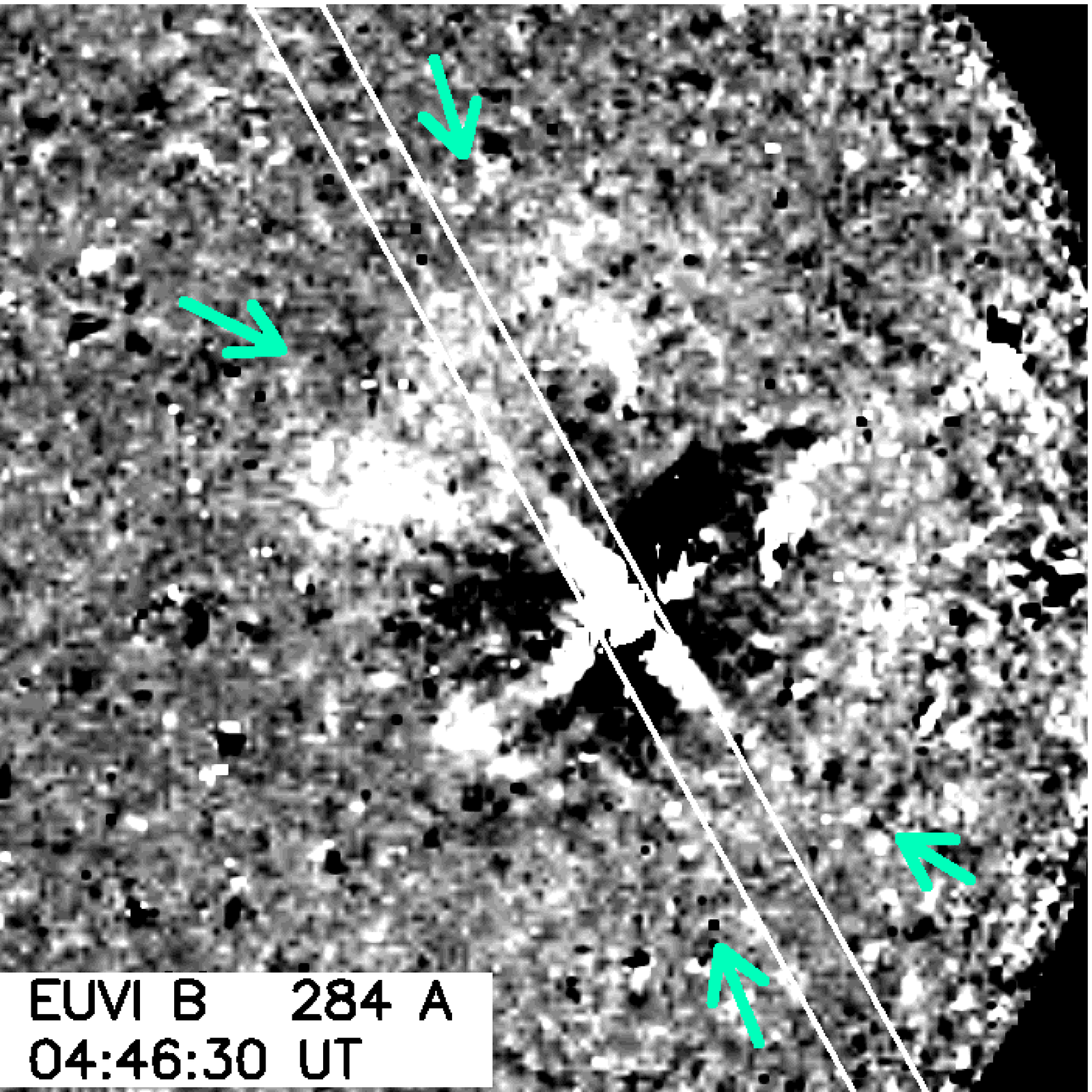}
\includegraphics[width=5cm,height=3.5cm]{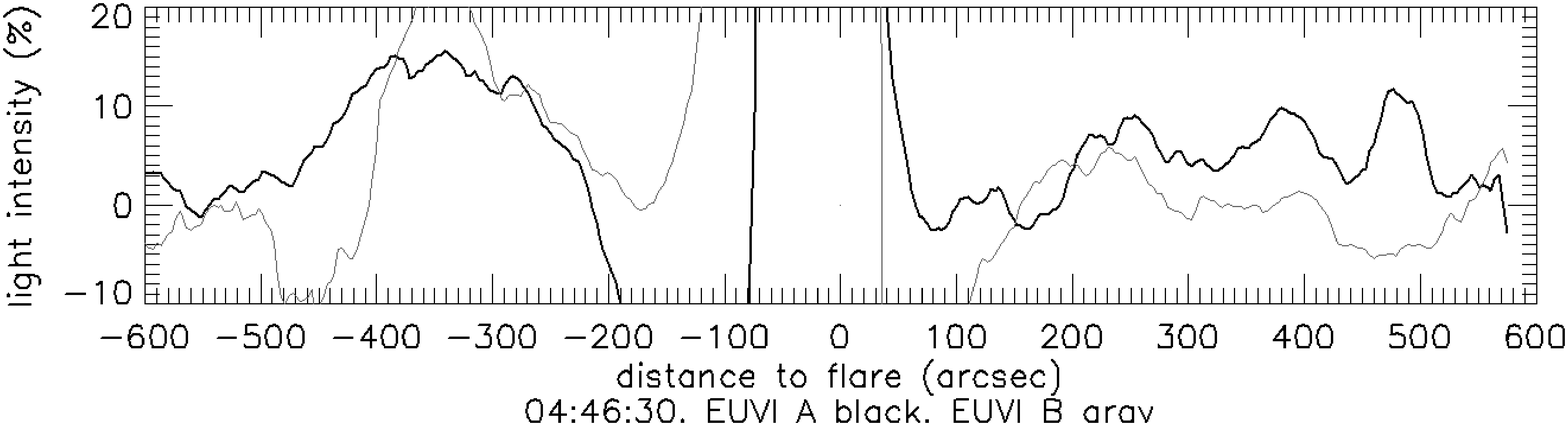}}
\caption{Images obtained through filters centered at 304 \AA~(first line), 171 \AA~(second line), 195 \AA~(third line) and 284 \AA~(fourth line), sensitive to the temperature of the emitting plasma. The first (second) column present observations of EUVI/SECCHI/STEREO A (B, resp.). The observations are obtained in a time interval of 1 min. All these images are processed as described in Section \ref{observation}. The intensity scaling is different for each filter to reveal the wave front. Arrows show the outer edge of the wave front. The plots (third column) are percentage variation of a mean intensity, integrated over square sliding along the slice drawn on the images, over the mean intensity obtained in a pre-event observation, permitting a better comparison of the emission in each wavelength. The black (gray) line of the percentage variation stands for EUVI/SECCHI/STEREO A (B resp.). The wave front is clearly observed in the three hotter EUV wavelengths.}
 \label{temperature1}
 \end{figure}

In Figure \ref{temperature1} we display difference images (the reference images are at 04:26:30 UT in 284 \AA, 04:26:15 UT in 304 \AA, 04:31:00 UT in 171 \AA, 04:25:30 in 195 \AA, all images corrected from the solar rotation to 04:31:00 UT) of the observations obtained with the two STEREO spacecraft (two first columns) and the percentage variation of the mean intensity (third column) through the four available bandpasses of EUVI at 04:46 UT. The presented observations are almost simultaneous, in 1 min. time delay. The observations are put from the lower (upper row) to the higher temperatures
(bottom row). The wave front is not exactly the same when observed through different filters: it is not observable in 304 \AA~and the brightest in 195 \AA. Table 2 shows the same parameters as in Table 1, but derived from the plots displayed in Figure \ref{temperature1}.

If the wave front is located at roughly the same location, with the same bright patches, its intensity and wideness are quite higher through 195 \AA~than through the other filters. In 171 \AA~its intensity is very low and in 284 \AA~it is in the middle range. The wave front structure seems to have a temperature ranging 1.5-2 MK. As the center of the emission shows up at the same location in the three bandpasses in the two STEREO points of view, these emissions seem to come from plasma lying at the same altitude. However, as more plasma emits in 195 \AA~than in the other bandpass, leading to a higher integration of the emission along the line of sight, the mean altitude of the plasma emitting in 195 \AA~is possibly affected.

\begin{table}
\begin{tabular}{|c|c|c|c|c|c|c|c|c|c|c|c|c|}
\cline{1-13}
wave&\multicolumn{6}{c|}{eastern}&\multicolumn{6}{c|}{western}\\
length &\multicolumn{6}{c|}{edge}&\multicolumn{6}{c|}{edge}\\
\cline{2-13}
&\multicolumn{3}{c|}{STEREO}&\multicolumn{3}{c|}{STEREO}&\multicolumn{3}{c|}{STEREO}&\multicolumn{3}{c|}{STEREO}\\
&\multicolumn{3}{c|}{Ahead}&\multicolumn{3}{c|}{Behind}&\multicolumn{3}{c|}{Ahead}&\multicolumn{3}{c|}{Behind}\\
\cline{2-13}
&d&wd&I&d&wd&I&d&wd&I&d&wd&I\\
&\tiny{('')}&\tiny{('')}&\tiny{(\%)}&\tiny{('')}&\tiny{('')}&\tiny{(\%)}&\tiny{('')}&\tiny{('')}&\tiny{(\%)}&\tiny{('')}&\tiny{('')}&\tiny{(\%)}\\
\cline{1-13}
171 \AA&-420&?&1&-350&40&4&?&?&?&?&?&?\\
(04:46:00)& & & &    &  & & & & & & & \\
\cline{1-13}
195 \AA&-420&120&15.5&-350&70&30&330&30&13&330&70&14\\
(04:45:30)& & & &    &  & & & & & & & \\
\cline{1-13}
284 \AA&-430&100&9&-340&90&10&350&?&8&370&?&10\\
(04:46:30)& & & &    &  & & & & & & & \\
\cline{1-13}
\end{tabular}
\label{tab:temperature}
\caption{Same estimated parameters of the wave front as in Table 1 in three different wavelengths in observations shown in Figure 3. Question marks are written when these parameters could not be estimated due to the lack of evidence of the wave front signal.}
\end{table}

\begin{figure}
\centerline{\includegraphics[height=4.cm]{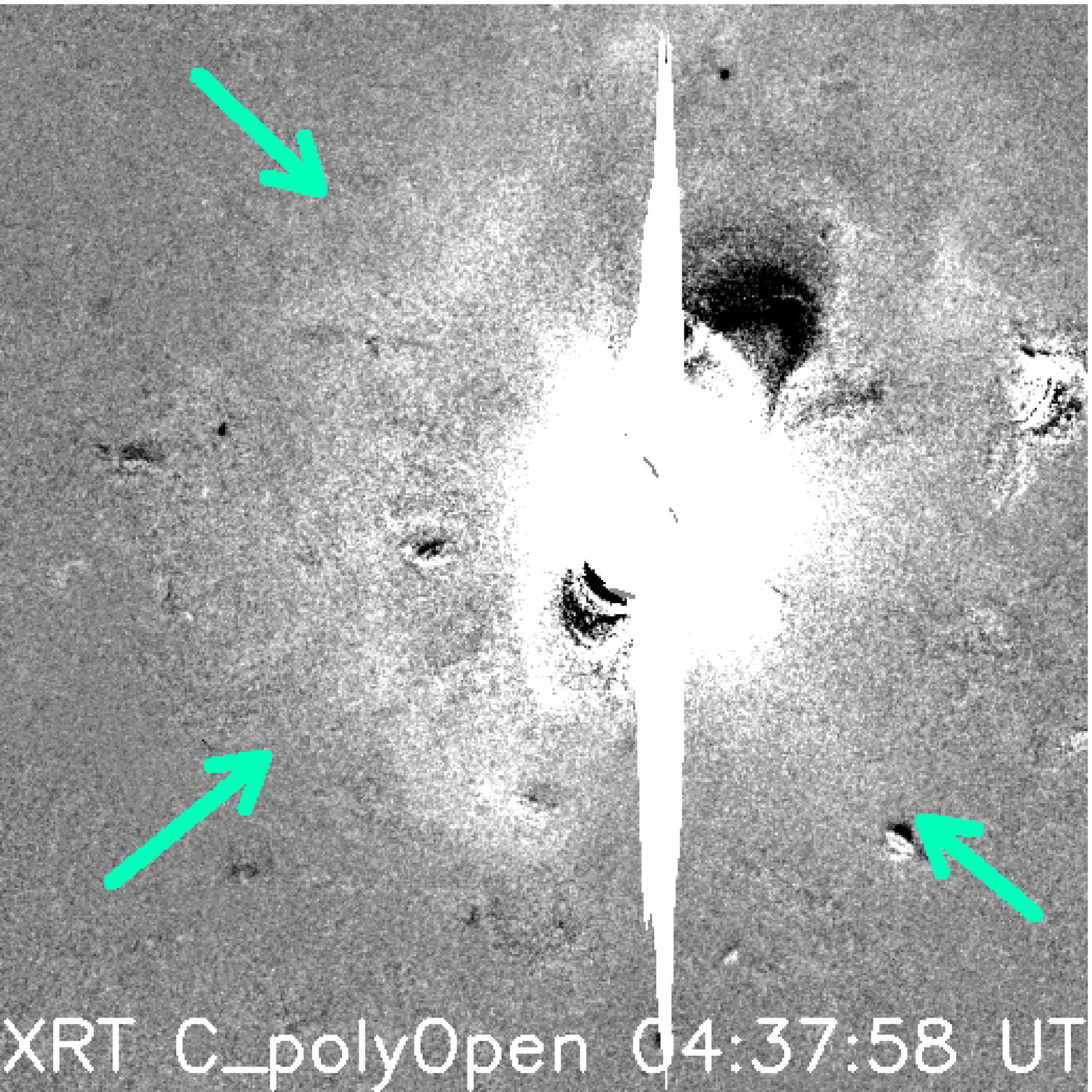}
\includegraphics[height=4.cm]{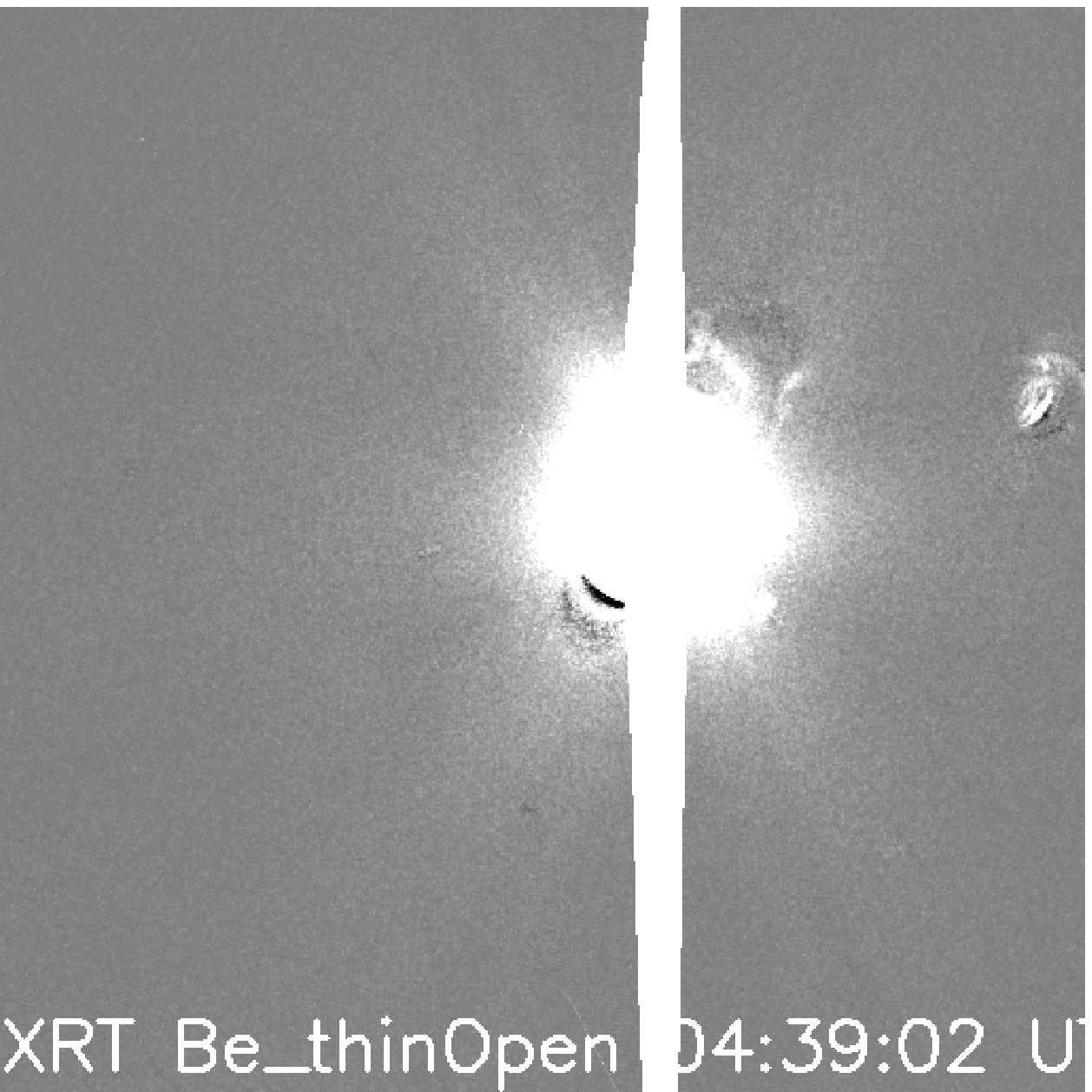}
\includegraphics[height=4.cm]{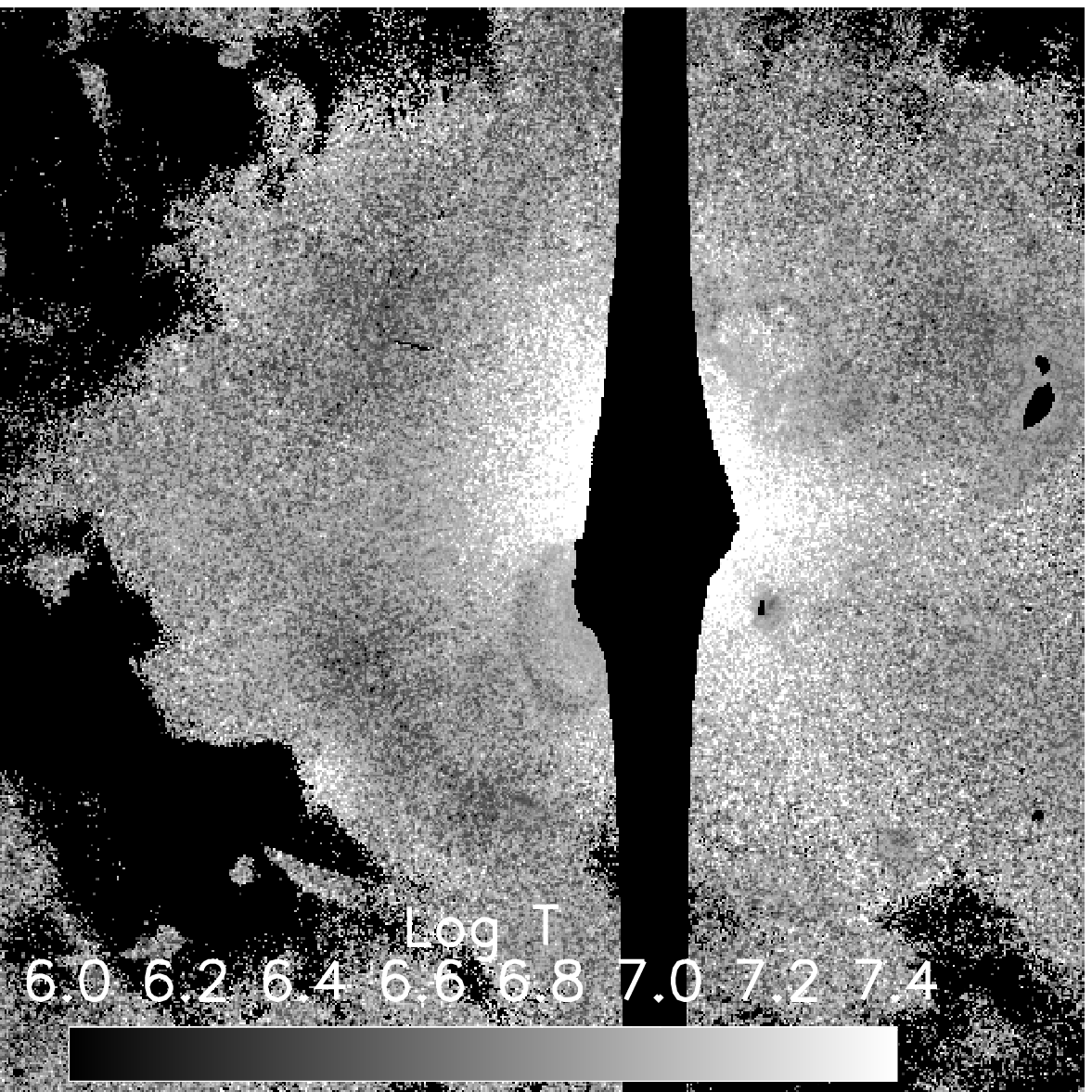}}
\caption{Images obtained through XRT filters C\_poly (first column) and Be\_thin (second column). The first image (through the softest filter) clearly shows the wave front but not the following one. Arrows show the outer edge of the wave front. Third column: temperature map derived from XRT/Hinode observations around 04:40 UT processed by CIFR method. The gray scale is in logarithm temperature. Locations where the temperature is not computed are marked in black. Field of is view is $526 \times 526$ arcsec$^{2}$.}
 \label{temperature2}
 \end{figure}

Observations obtained with XRT/Hinode having quite different temperature sensitivity are given in Figure \ref{temperature2} (two first panels). The observations shown in this Figure, have the small XRT field of view of $526\times526$ arcsec$^{2}$, centered on the flaring active region. The observations are processed as described in the previous Section. The wave front is clearly visible in the C-poly open filter, the thinner, cooler, filter, but not in the Beryllium thin filter, the hardest one. These two filters are broad-band filters having their temperature response peacking at about 8MK and 10 MK resp. A correct estimate of the temperature has to be done using several filters.

Therefore, we use the Combined Improved Filter Ratio (CIFR, Reale et al. 2007) that we applied on XRT observations through all the four filters listed in Section 2 (these observations are available but not presented here) to compute a temperature map of these observations. The XRT calibration from Narukage et al. (2011) is taken into account. The blooming pixels are discarded. The computed temperature map, given in Figure \ref{temperature2}, third column, show an average temperature of 5 $\pm$ 1.75 MK inside the wave front which is higher than the temperature guessed from the EUVI/SECCHI/STEREO observations. This analysis shows that at the wave front location the plasma is much hotter (white) than outside (dark in the temperature map).

Putting together the observations obtained with EUVI and XRT, the structure is certainly multithermal, possibly with two lobes, one around 1.5 MK and the other around 5 MK.

\section{Altitude of the observed wave front}

Despite that we showed in the previous Section that the observed wave front morphology is sensitive to the point of view and the bandpass used to obtain its observation, we here try to derive the altitude of the wave front, assuming that it is a solid and optically thick structure, i.e. we only observe its outer envelope, taking advantage of both points of view provided by the STEREO spacecraft. We used three methods: the procedure scc\_measure.pro of the SolarSoft library written in the Interactive Data Language (sswidl) and the reverse projection method developed for prominences (Artzner et al. 2010) that we further developed by differencing the reverse projected images or by correlating them.

\subsection{scc\_measure method}

\begin{figure}
\includegraphics[width=12.2cm]{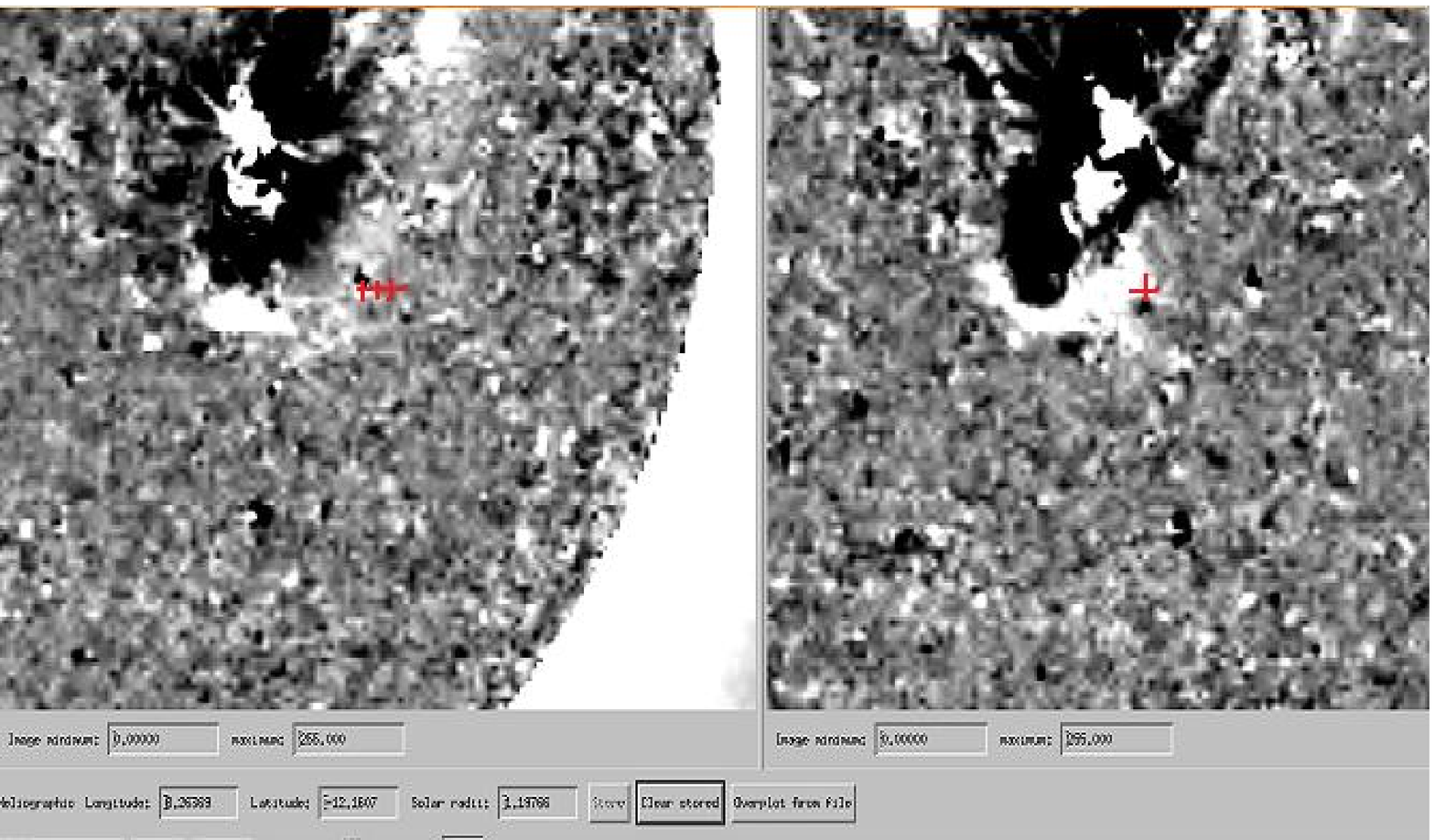} \\ 
\includegraphics[width=12.2cm]{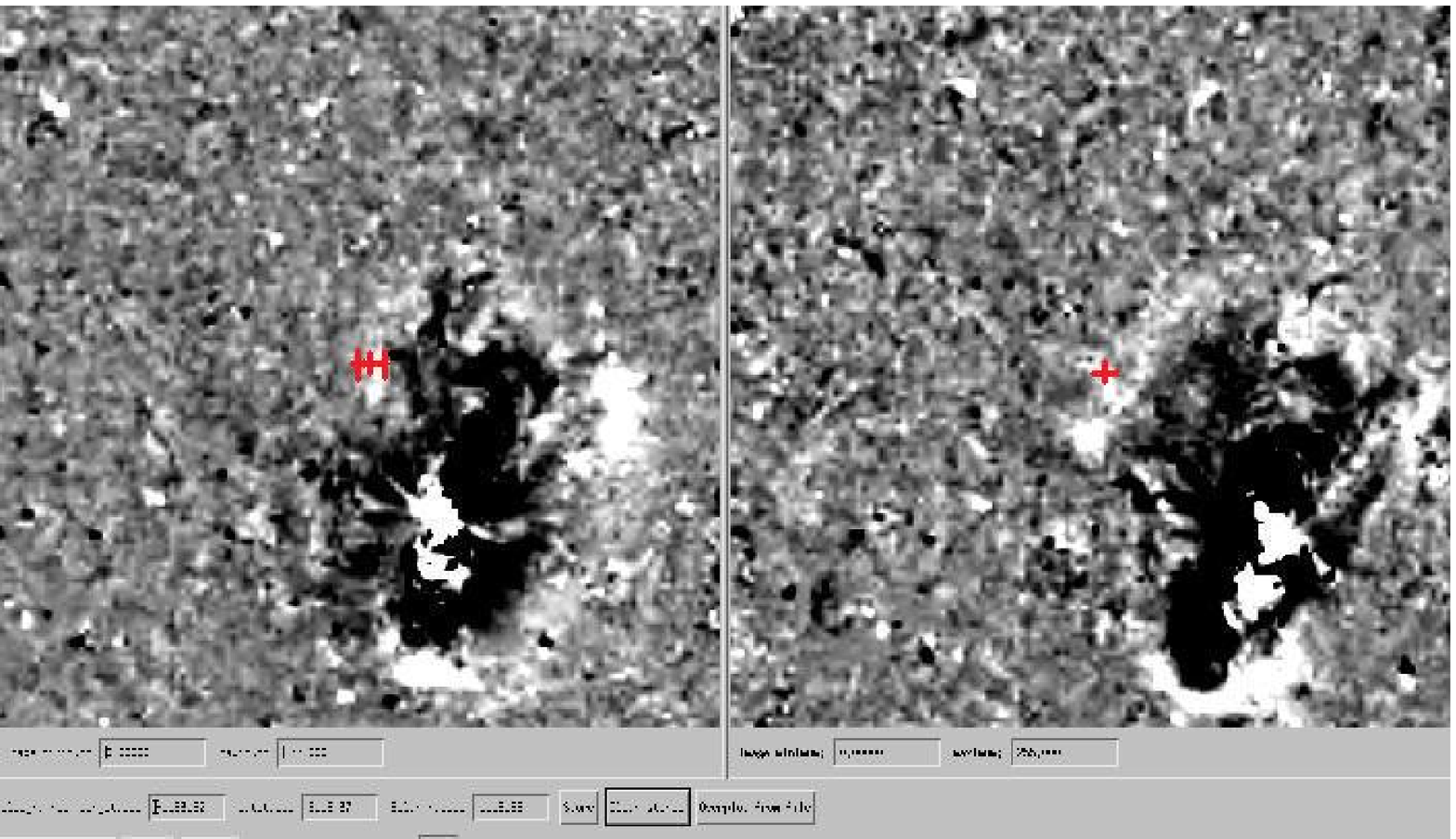}
\caption{Two snapshots of the widget launch by scc\_measure. Each widget has two difference images obtained using STEREO A (right) and STEREO B (left) at 04:38 UT through the filter centered at 171 \AA. The red crosses mark the chosen location of the western (eastern) edge, first (second, resp.) row of the wave front to compute its altitude.}
\label{sccmeasure}
\end{figure}

This procedure is applied to the pair of observations obtained with the 171 \AA~filter at 04:38 UT only. The scc\_measure routine calls images and their header containing the coordinates of the spacecraft and several parameters of the Sun. As these parameters are crucial, and have to be kept intact, instead of correcting the solar rotation of the image showing the wave front, we correct from the solar rotation the pre-event images at 04:31 UT before differencing the images at 04:38 UT with the corrected ones at 04:31 UT. Then we run scc\_measure with the original headers of the images at 04:38 UT and the difference images pair. The scc\_measure routine creates a widget with the two difference images. We choose a particular point in one image obtained with one STEREO spacecraft. This point is inside the slice drawn on the images of Figure \ref{superposition}, used for a first estimate of the altitude of the wave front. A line is drawn on the image of the other STEREO spacecraft to locate the line of sight from the first image. Finally, we select a point on the second image at the intersection of the wave front and that line. Since this procedure seems reasonable for distinct structures such as filaments and loops (Liewer et al., 2009; Bemporad, A., 2011), it is very difficult to find a precise point in the diffuse wave front. Therefore, we study the error made in trying to locate the wave front, and the induced error in the computed altitude. We chose a point located on the image obtained with the STEREO A spacecraft, which coordinates are set inside the file scc\_measure.pro. Then, three possible positions of the same edge of the wave front observed with STEREO B point of view are chosen. The derived altitude at 04:38 UT is 123$\pm$13 Mm at the western edge of the wave front (Figure \ref{sccmeasure} first snapshot) and 107$\pm$18 Mm at the eastern edge of the wave front (Figure \ref{sccmeasure} second snapshot).

These altitudes are quite higher (by 33 Mm and 17 Mm) than the one in Patsourakos et al. (2009). The error is quite larger also (2 to 3 times) than in the same reference. The structure seems to present different altitudes, but as the error leads to two domains of altitudes that intersect themselves, the difference in altitude is not conclusive. For all of these uncertainties, we want to use another method to derive the altitude of the wave front.

\subsection{Reverse projection}
\subsubsection{Method}
\label{method}
This method is fully automated and is used for prominence eruption (Artzner et al. 2010, Gosain et al. 2012). We present all the processes done on the original data in the text below. 

Each observed flat image from STEREO A or STEREO B is a projection in two dimensions of physical structures in three dimensions. In order to possibly reconstruct these structures within the optically thick hypothesis, we firstly project both observed images on a set of multiple closely spaced reference surfaces, and secondly we inspect where these reversed projections reasonably match together. This allows a piece by piece reconstruction, as further indicated into details.
The classical Carrington coordinates at the surface of the Sun are well suited to follow solar sunspots. In order to define generalized Carrington coordinates for physical points located at higher altitudes within the solar corona, as opposed to image points in the observations, we consider a set of reference spheres $S_R$, concentric to the Sun, larger than the solar globe. Typical values for the radius $R$ of these spheres range, by 5 Mm steps, from 695 Mm to 850 Mm, the radius of the upper envelope for the considered event. We also consider, for each point $m_R$ of a given reference sphere, the intersection point $m$ of the sphere at $S_{700Mm}$ with the line segment between $m_R$ and the center of the solar globe. The point $m$ is located by its classical Carrington longitude and latitude $[L ,b]$. We attribute to the point $m_R$ on the reference sphere a set of generalized Carrington coordinates $[R, L ,b]$.

Each observed image consists of an array of image points $m_{x,y}$. These points correspond to different lines of sight from the observer to the solar plasma. The lines of sight for image points located within the apparent solar disk are blocked by the solar globe. There is only one intersection of these lines of sight with any reference sphere. The line of sight for an image point located outside the apparent solar disk is not blocked by the solar globe. There are both intersections, in front and behind the observer's plane of the sky, of this line of sight with any reference sphere. For a given image point, inside or outside of the solar disk, for a given value of the radius of the reference sphere, we compute the three Cartesian space coordinates of the physical point(s) at the intersection(s) of the considered reference sphere with the line of sight from the observer to the plasma corresponding to the image point $m_{x,y}$. In our case the image points of interest are within the solar disk, the hypothetical physical points corresponding to image points are located in front of the solar globe. These Cartesian space coordinates are converted to generalized Carrington coordinates. We consider the generalized Carrington map corresponding to the value $R$ of the radius of the reference sphere. The measured intensity $I_{x,y}$ at image point $m_{x,y}$ is attributed to the corresponding bin in the Carrington map. We choose the bin size in the Carrington map as large as to insure that each bin receive the intensities from several image points. 

First, we compute the reverse projections, as explained above, for images obtained on both STEREO spacecraft in the timeline of the wave front.

Second, we make a ratio difference of one reverse projected image with a similarly reverse projected image obtained prior to the emission of the wave front (at each pixel we apply $(I-I_{before})/(I+I_{before})$) for each couple of observations obtained at the same date with the two STEREO spacecraft.

We choose the maximum radius for the reference sphere at 850 Mm. At larger radius, the overlay of the wave front has not any sense as its western edge observed by STEREO B may be superposed to its eastern edge observed by STEREO A at 04:33 UT.

From this point, we use two different methods to derive the altitude of the coronal front.

\subsubsection{Differencing}
\label{difference}

\begin{figure}
\centerline{\includegraphics[width=4cm]{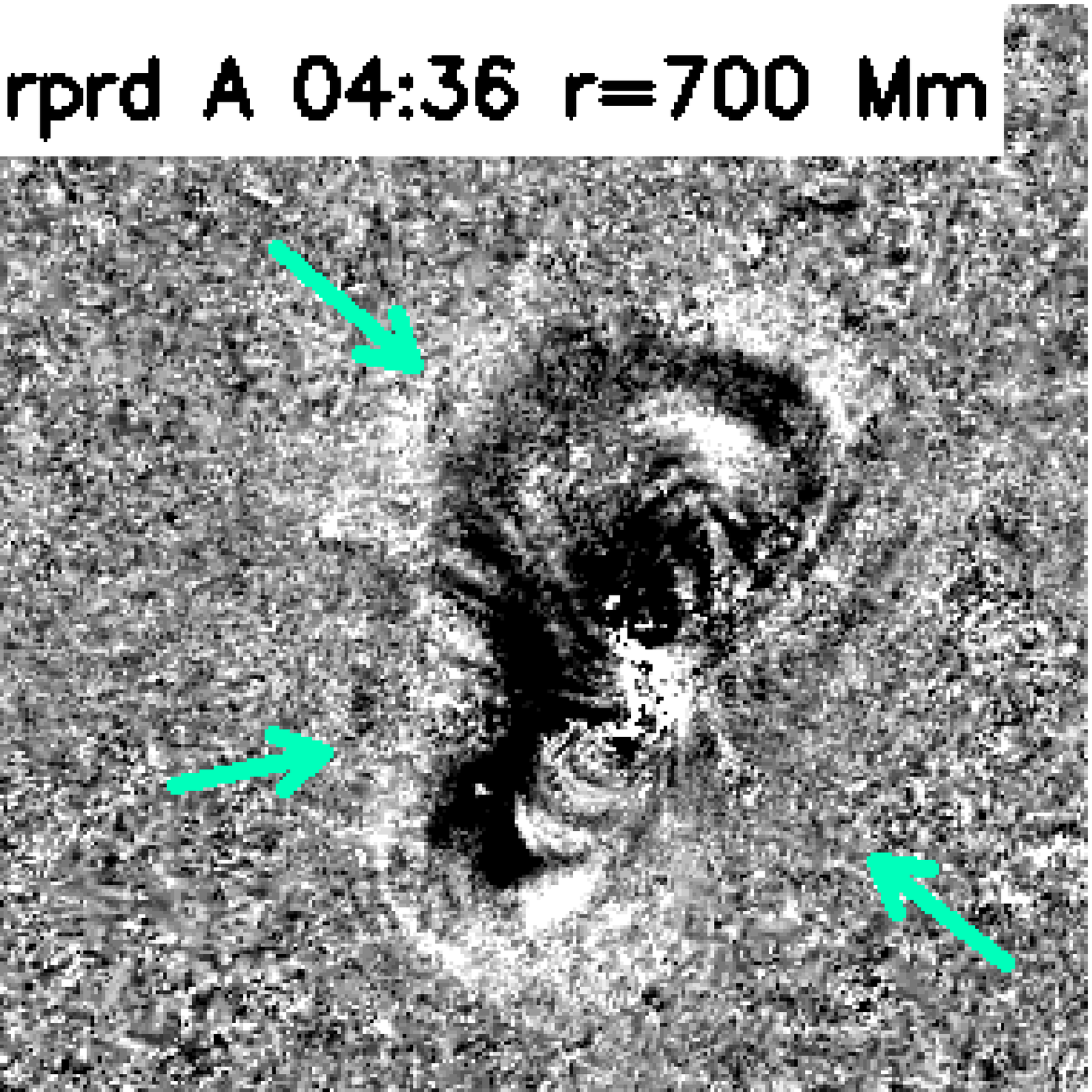}\hspace{0.1cm}
\includegraphics[width=4cm]{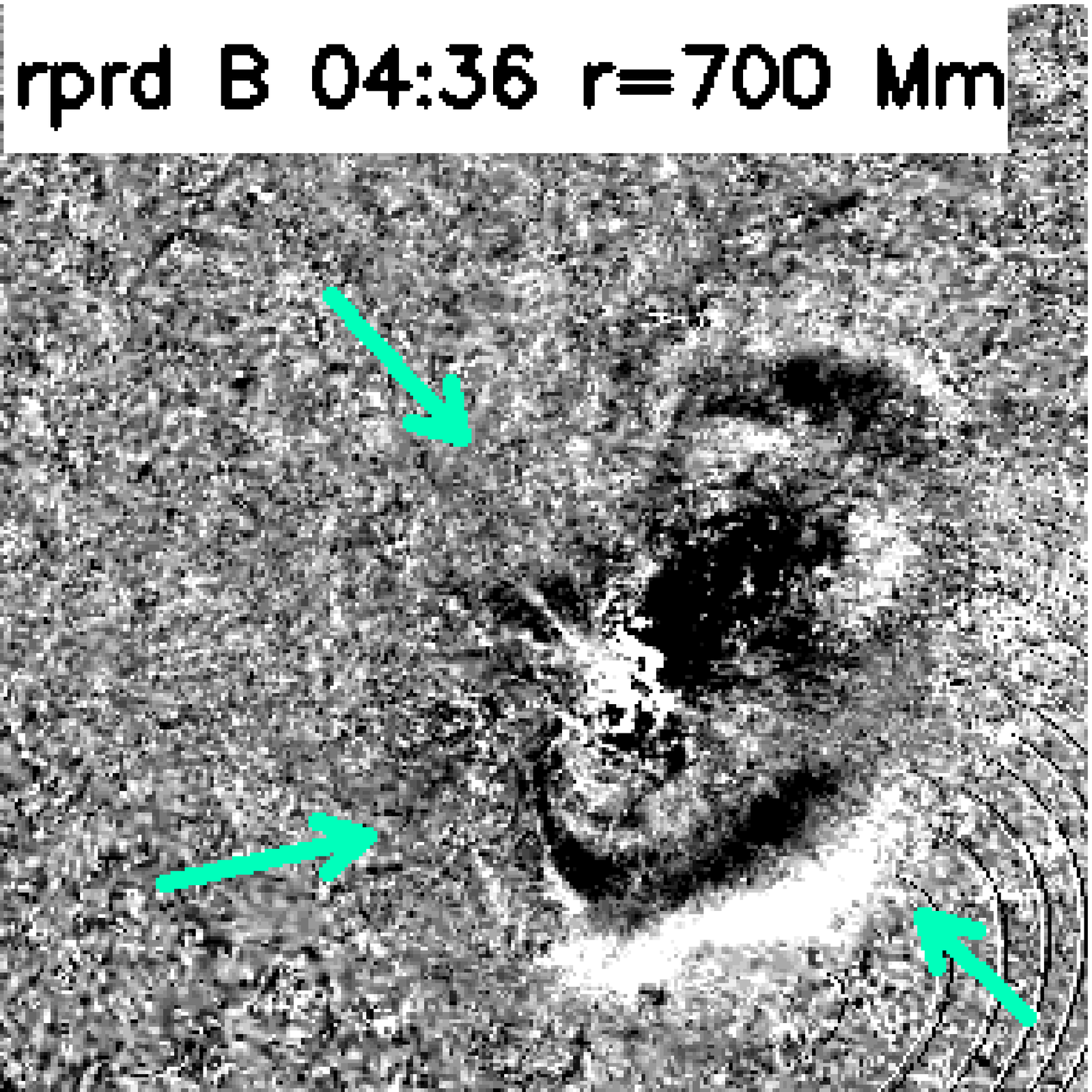}
\includegraphics[width=4cm]{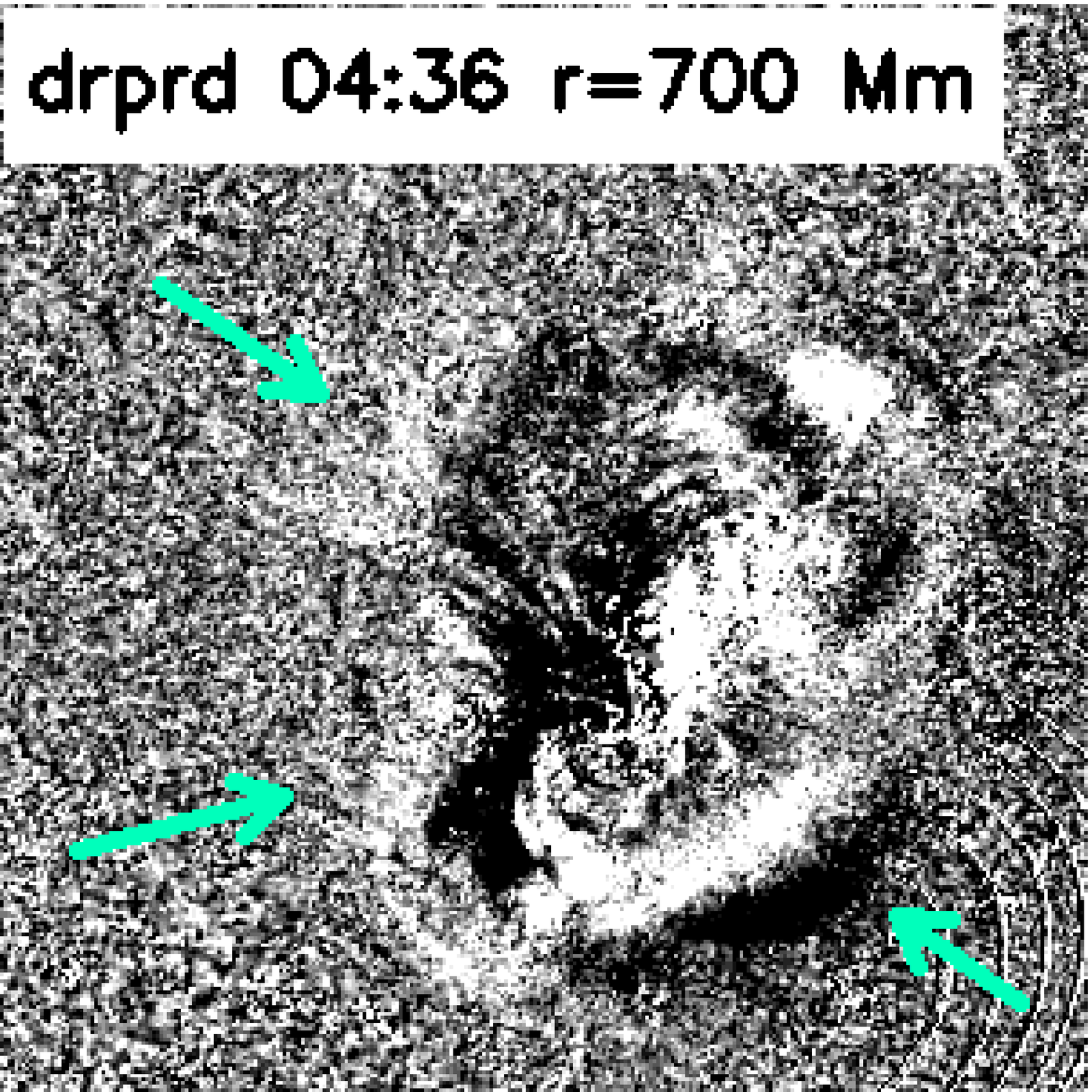}}
\caption{ 171 \AA~STEREO A (first) and B (second) images projected as if all the structures in the image belong to a sphere centered at the Sun center with radius of 700 Mm observed from the earth, the intensity values are ratio difference with an image obtained at 04:31 UT prior to the wave front emission (labelled rprd). Third image is the subtraction of the two previous ones, named in the main manuscript difference reverse projected ratio difference image (drprd). The black and white arcs at the right edge of the second and third images are due to projection leading to pixels with missing values or sum of several original pixels. Arrows indicate the wave front outer edge.}
\label{altitude}
\end{figure}
We verify that at a radius of 700 Mm (solar surface) the reverse projected ratio difference image of STEREO B is very similar to the one of STEREO A, except for the flare region, by differencing their intensities, producing a new image named drprd (difference reverse projected ratio difference) image (Figure \ref{altitude}). In this Figure, structures which are at this altitude appear in gray as the pixels show null intensity. The structures that are at different altitudes appear in black or white, the white parts corresponding to the emission of the wave front observed in the STEREO A field of view and the black parts corresponding to the one observed in the STEREO B field of view. The location of the wave front appears as an elliptic white and black structure that surrounds the flaring active region in the drprd images, meaning that the wave front is above the solar surface. 



\begin{figure}
\centerline{\includegraphics[width=12cm]{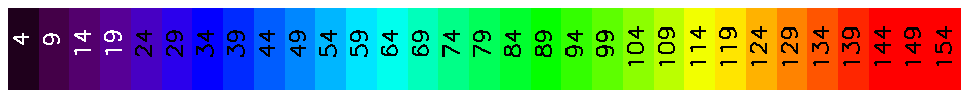}}
\centerline{\includegraphics[width=4cm]{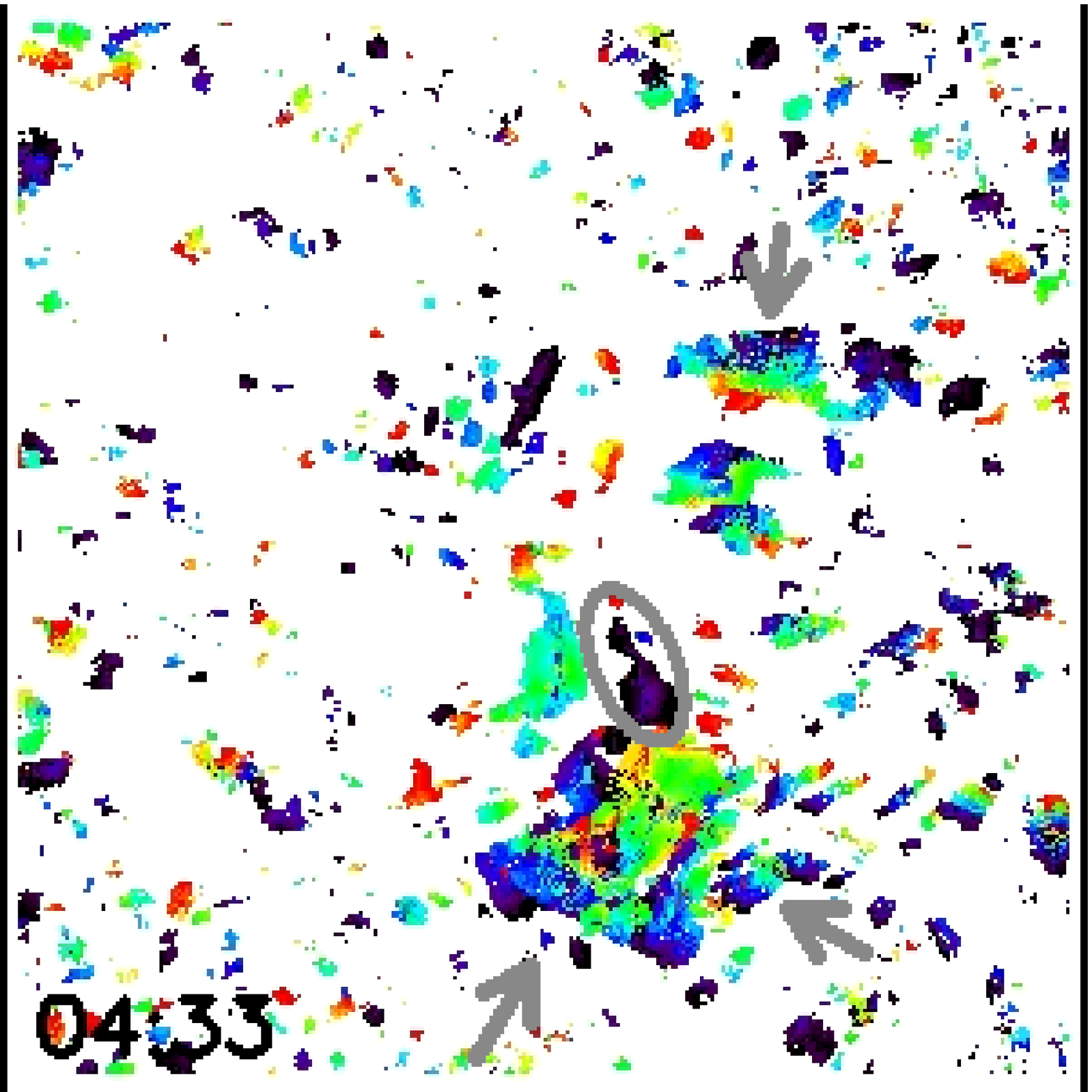}
\includegraphics[width=4cm]{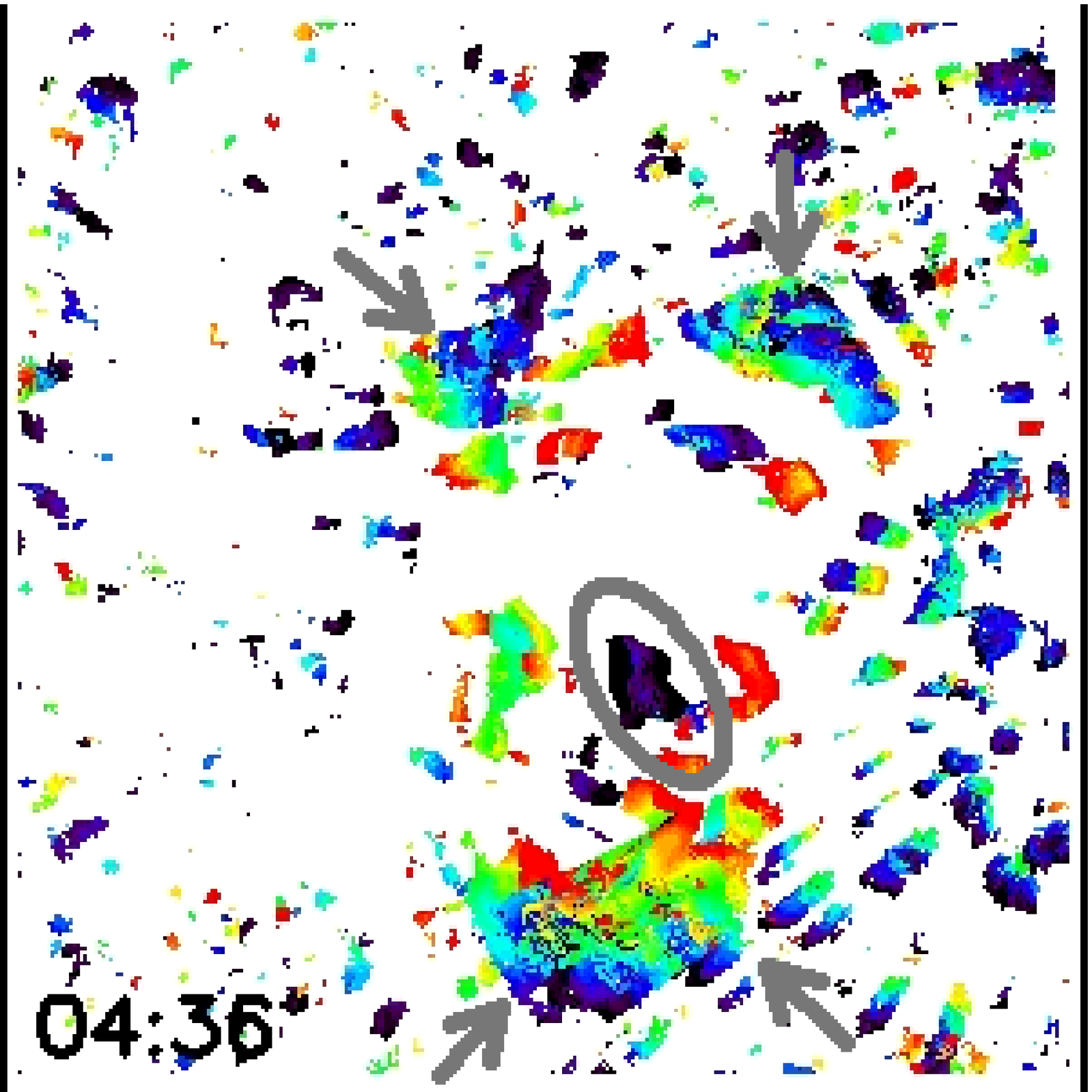}
\includegraphics[width=4cm]{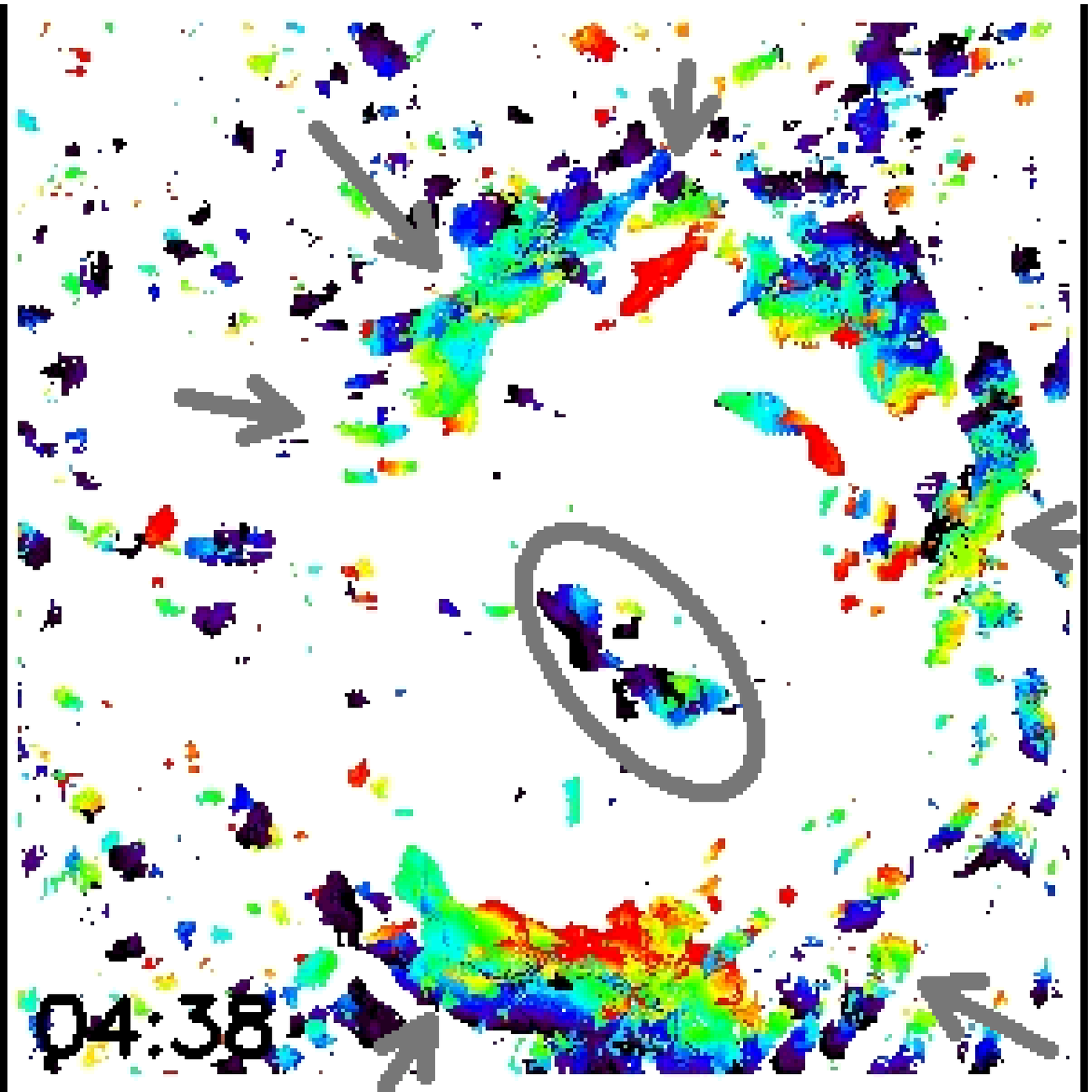}}
\centerline{\includegraphics[width=4cm]{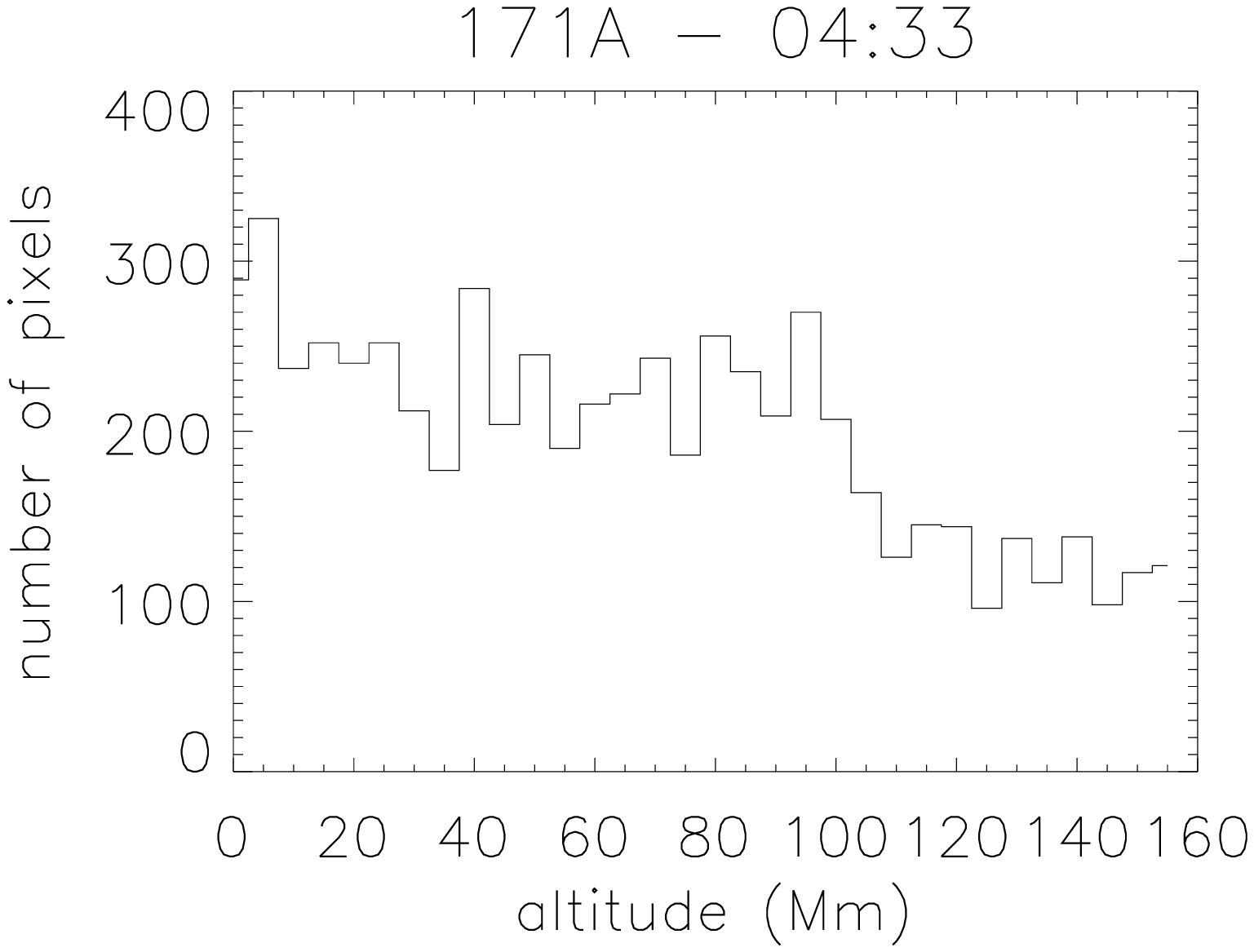}
\includegraphics[width=4cm]{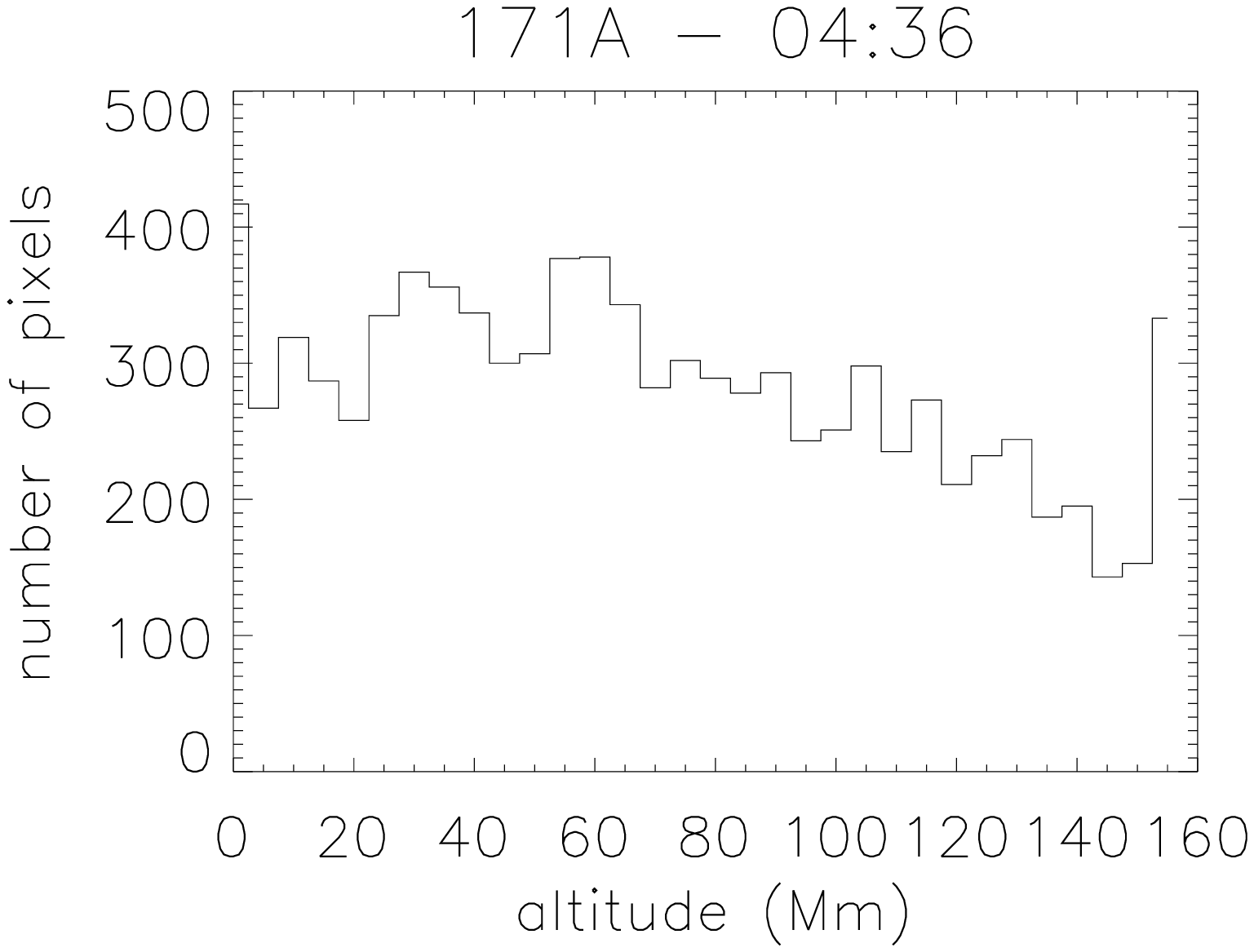}
\includegraphics[width=4cm]{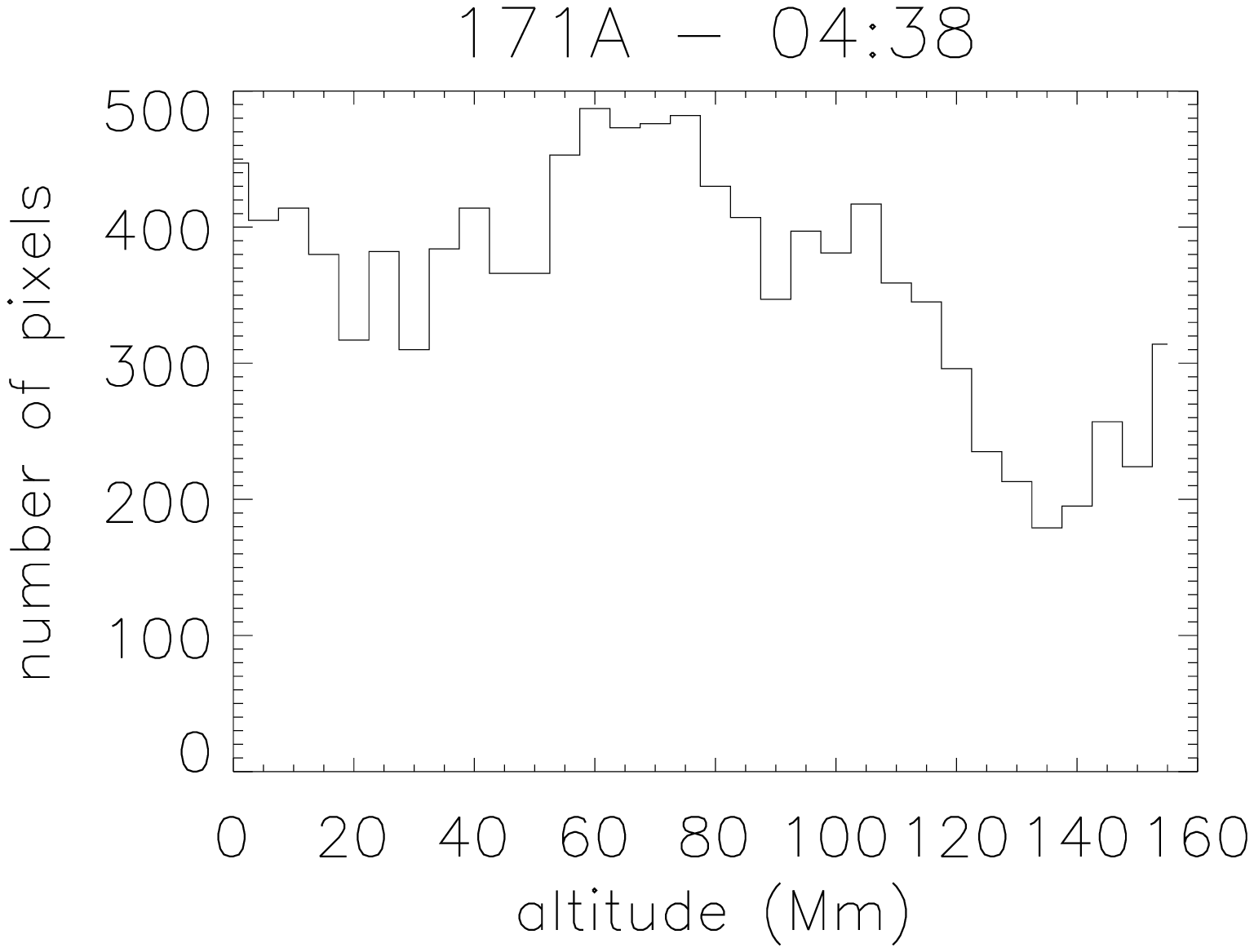}}
\centerline{\includegraphics[width=4cm]{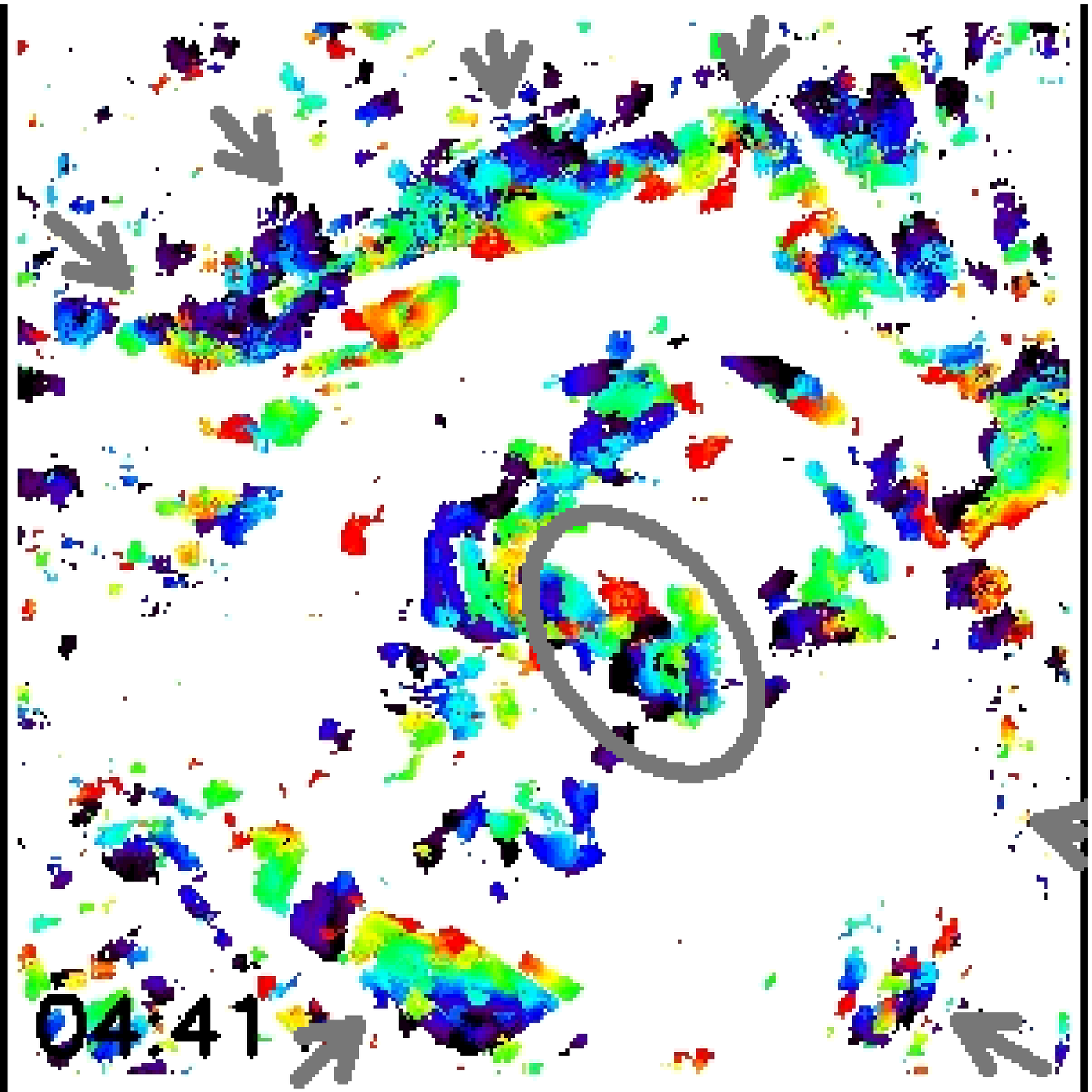}
\includegraphics[width=4cm]{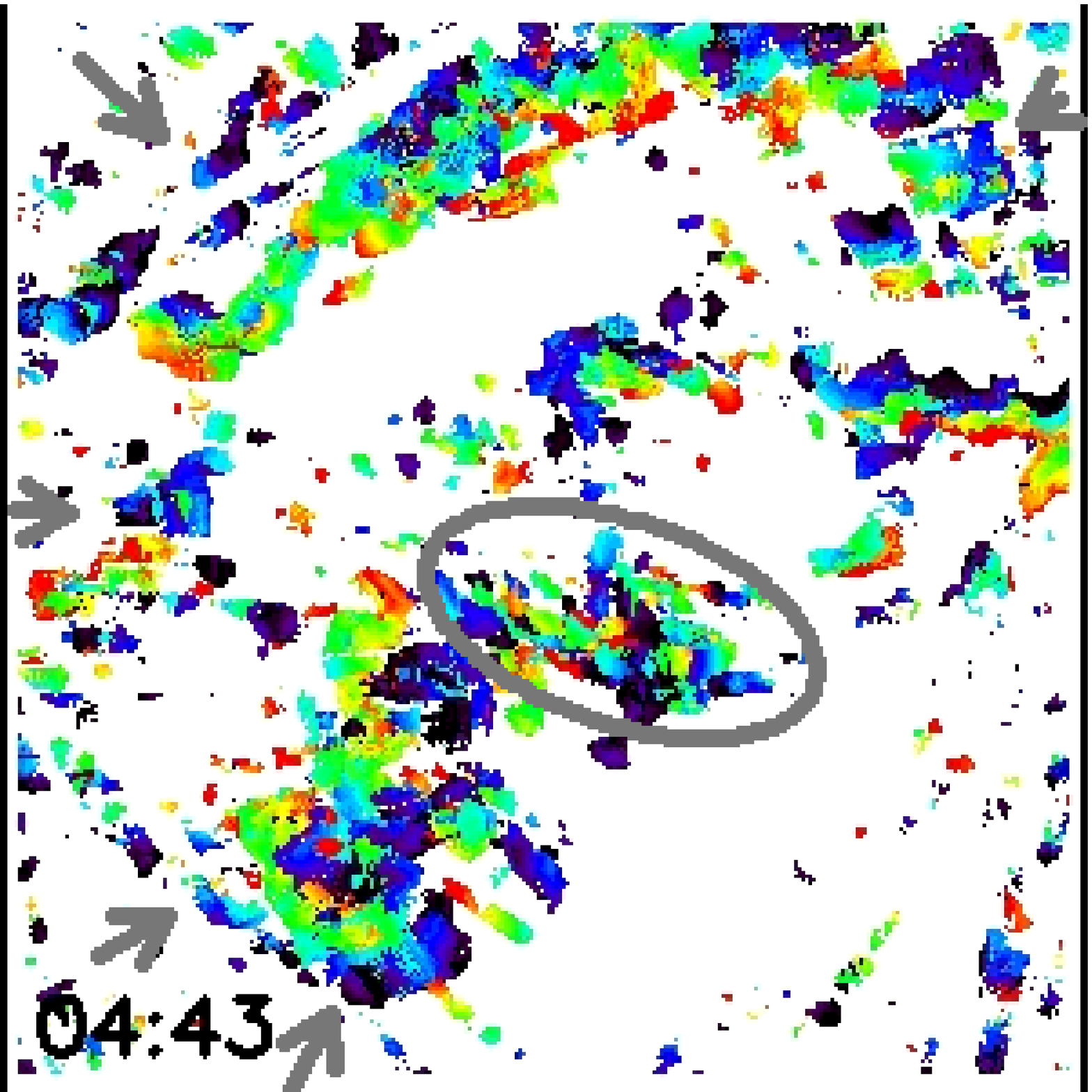}
\includegraphics[width=4cm]{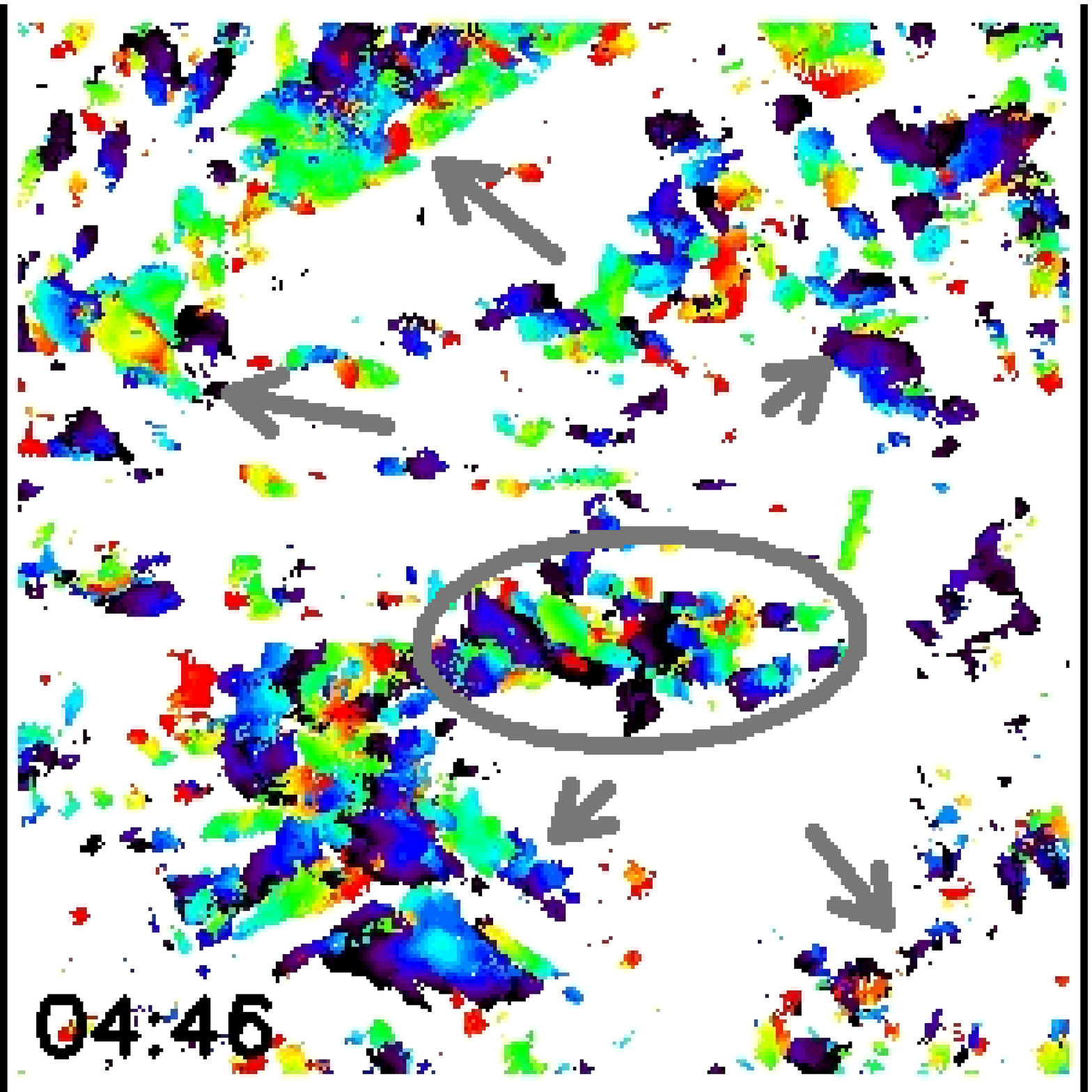}}
\centerline{\includegraphics[width=4cm]{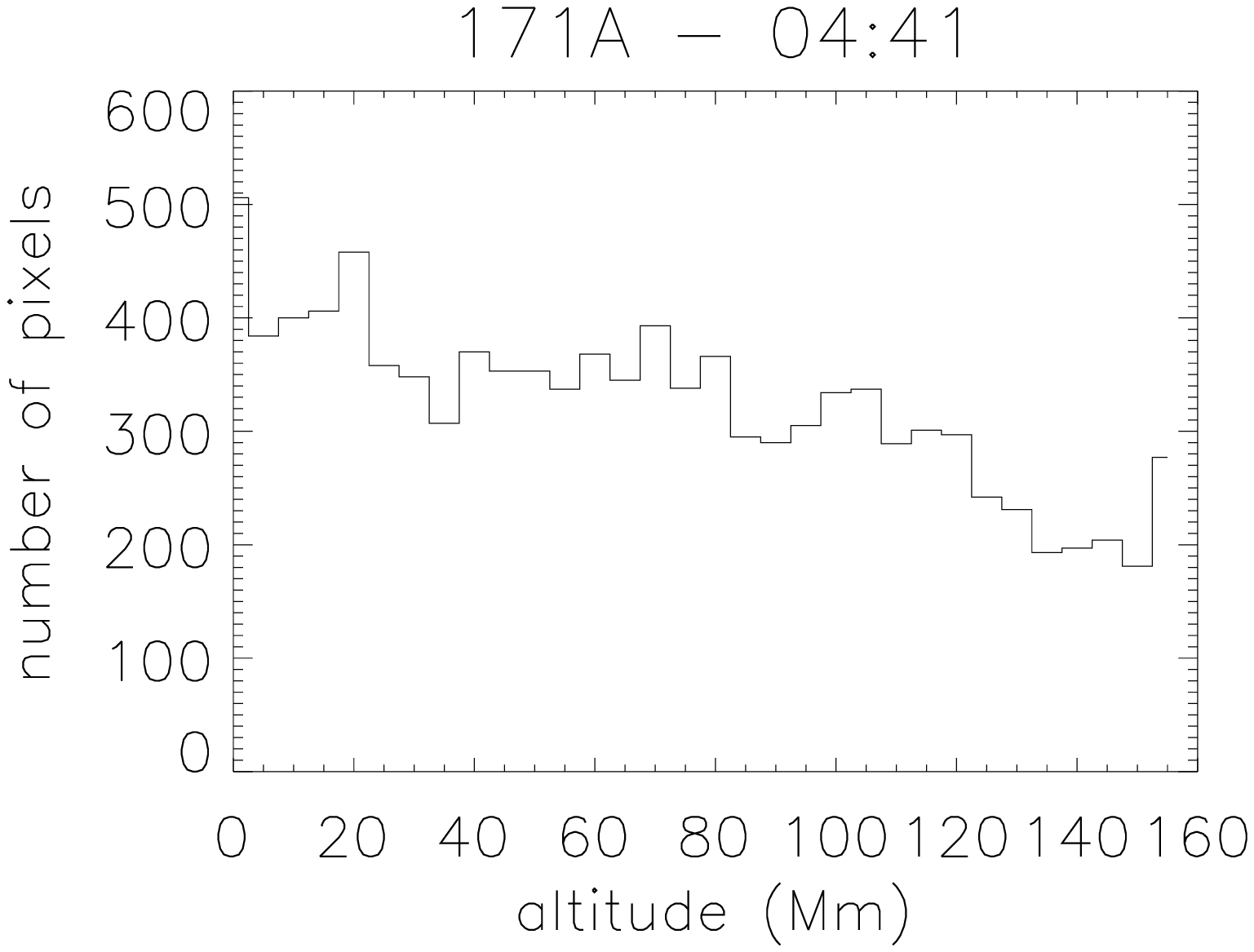}
\includegraphics[width=4cm]{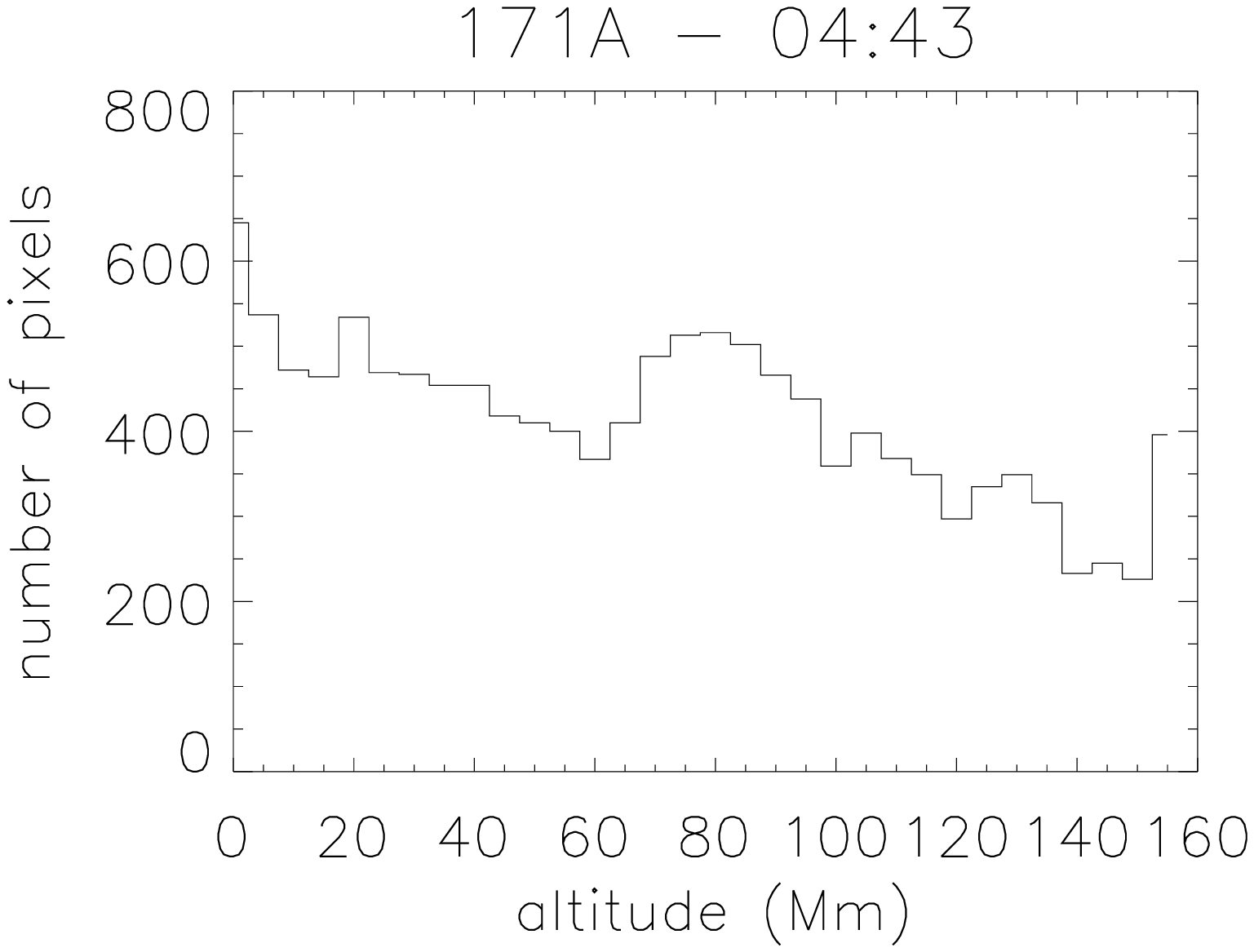}
\includegraphics[width=4cm]{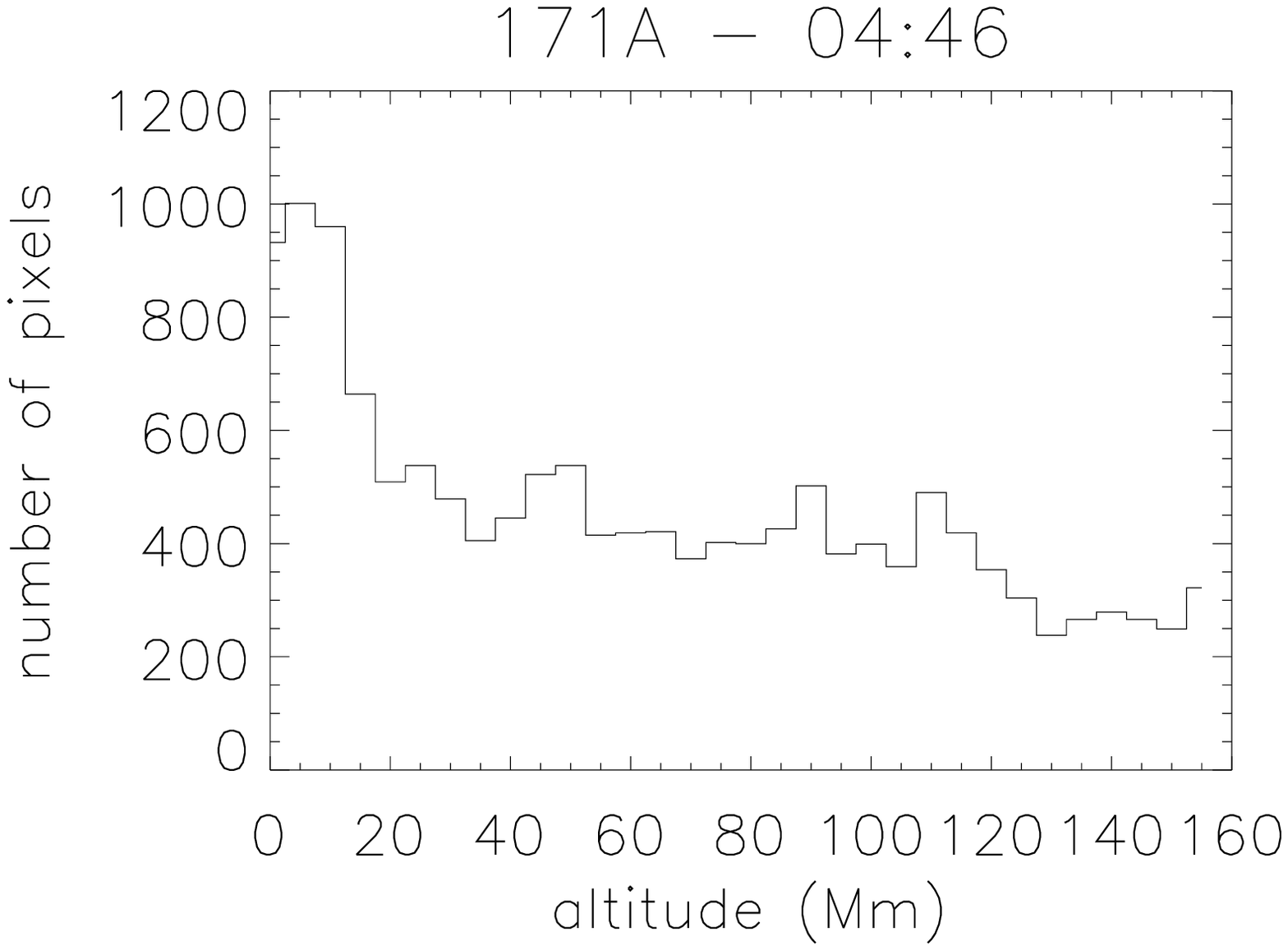}}
\caption{Time series of altitude map obtained using observations at 171 \AA. White corresponds to locations where the altitude is not computed due to the absence of wave front. The altitudes are color coding as in the scale at the top of the Figure, the values are in Mm. The gray circles show the location of the post-flare loops. The gray arrows show the wave front. The plots are histograms of the amount of pixels presenting a particular altitude versus the altitude. Each histogram is obtained using the map in the above panel, except the points outside the wave front.}
\label{cartealtitude171}
\end{figure}

\begin{figure}
\centerline{\includegraphics[width=4cm]{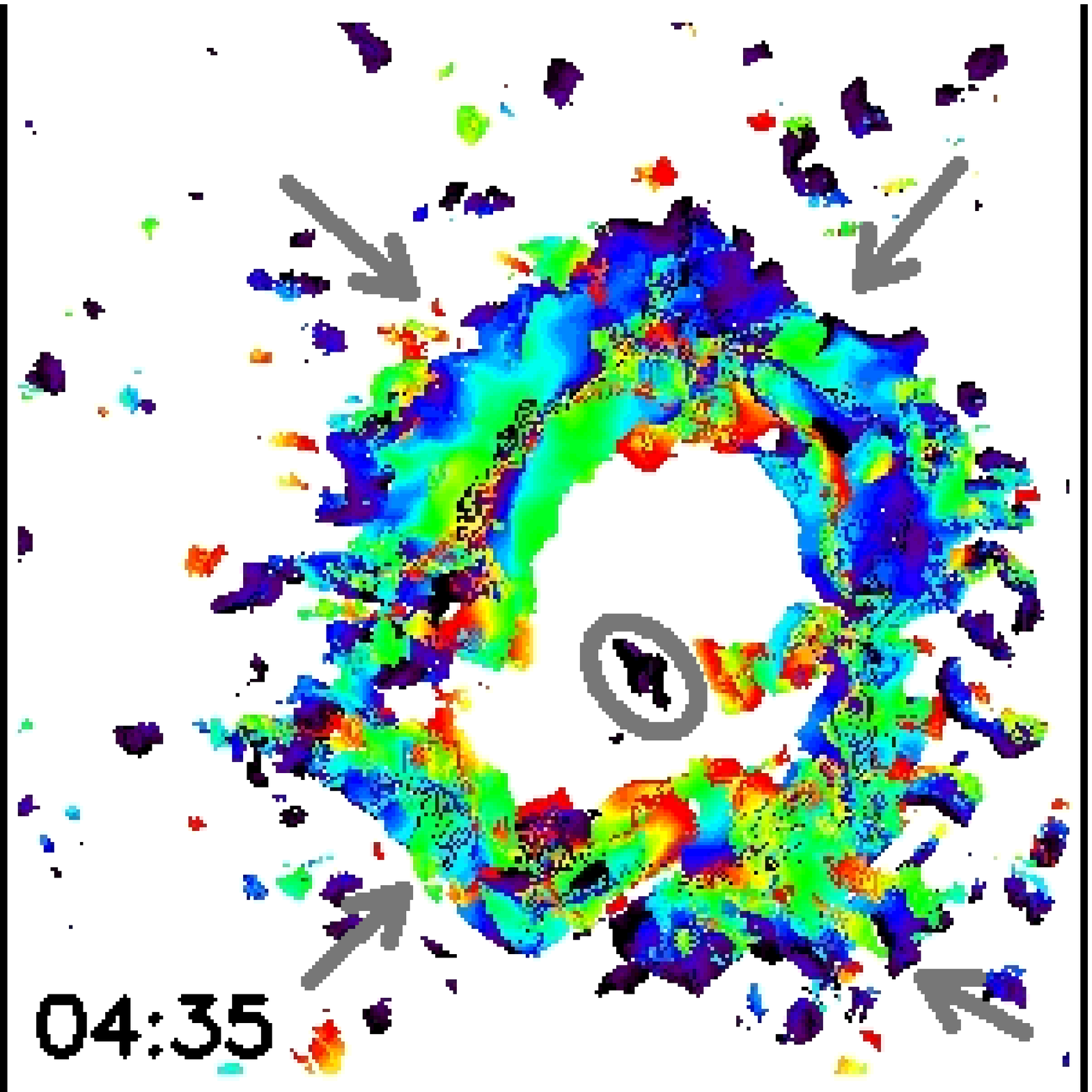}
\includegraphics[width=4cm]{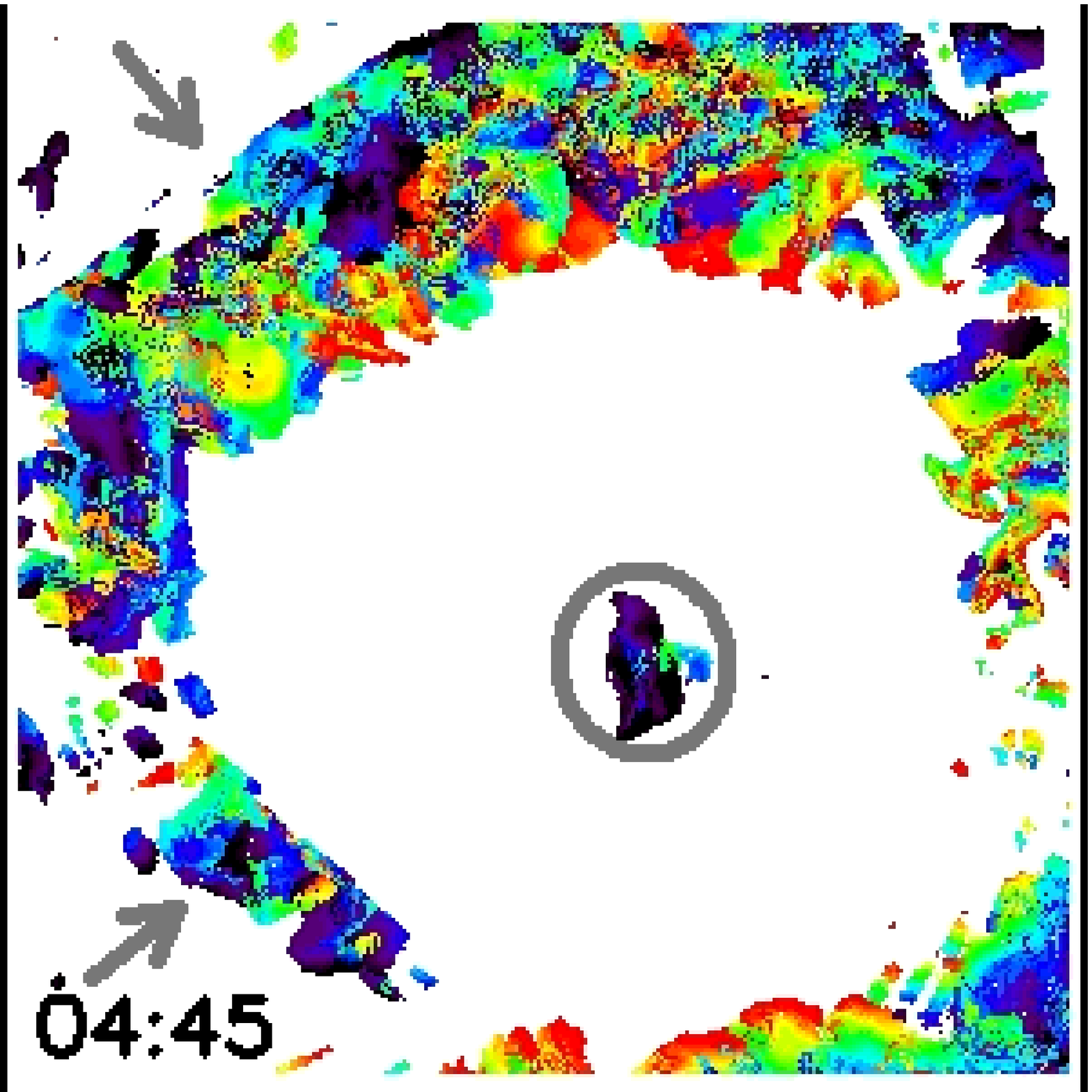}
\includegraphics[width=4cm]{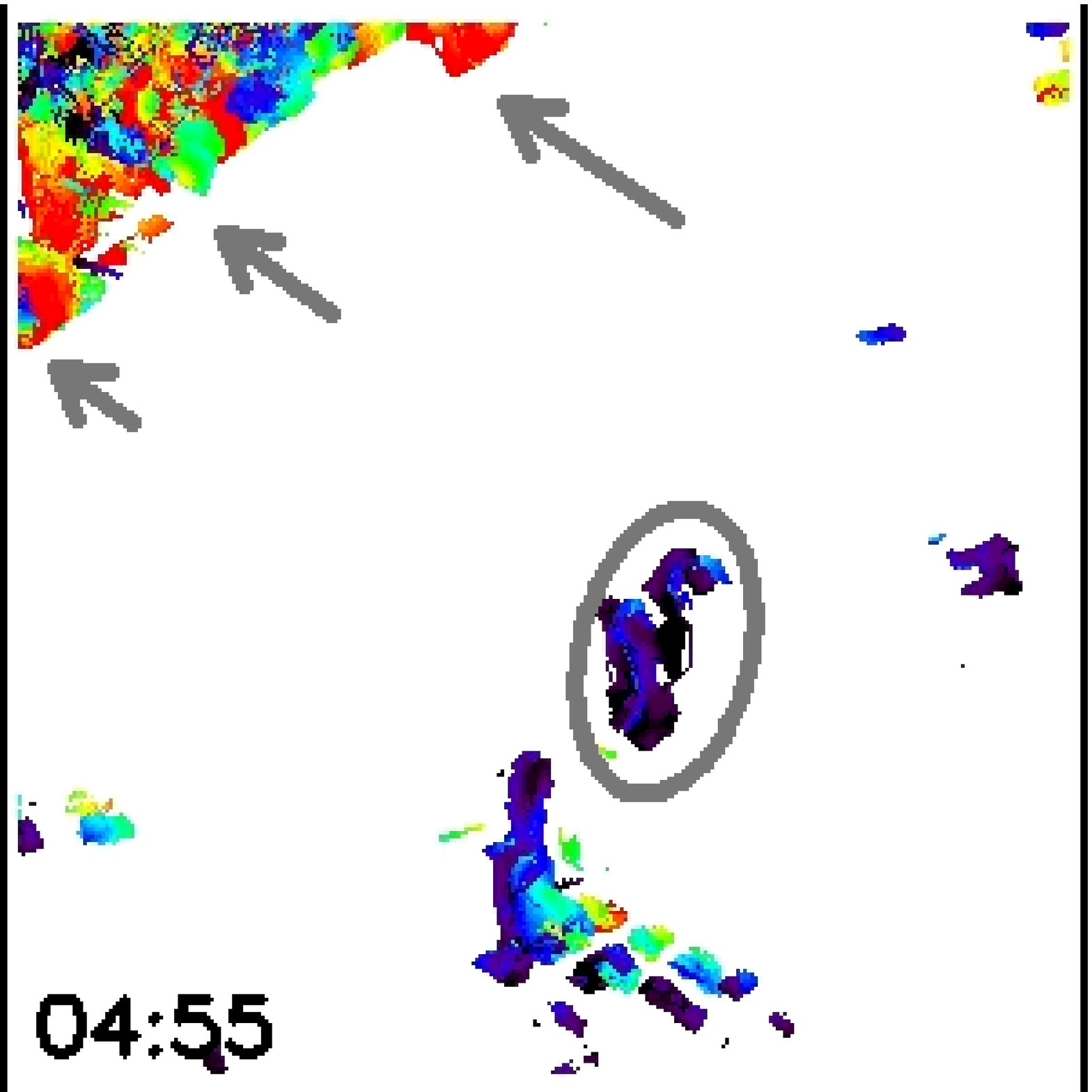}}
\centerline{\includegraphics[width=4cm]{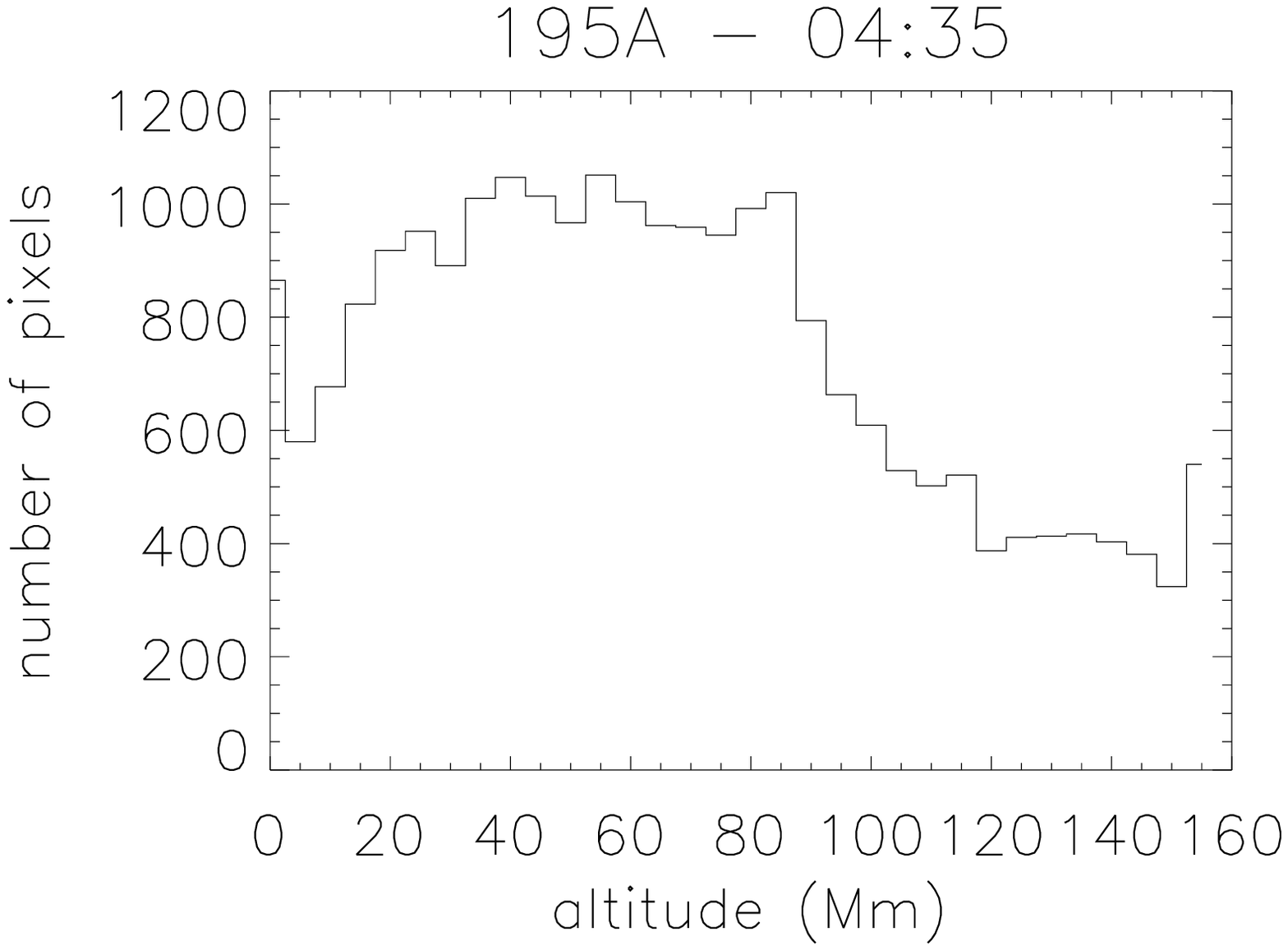}
\includegraphics[width=4cm]{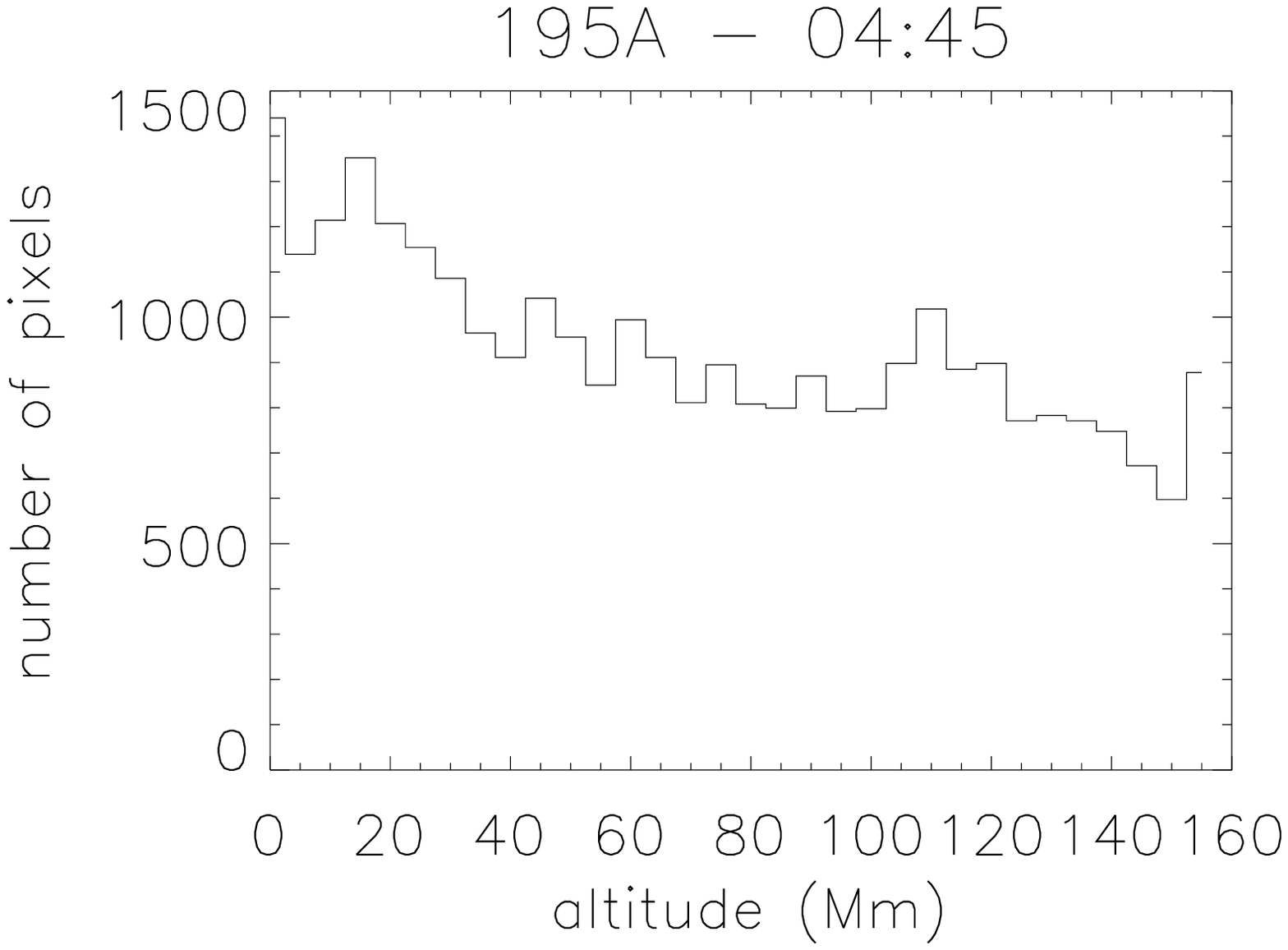}
\includegraphics[width=4cm]{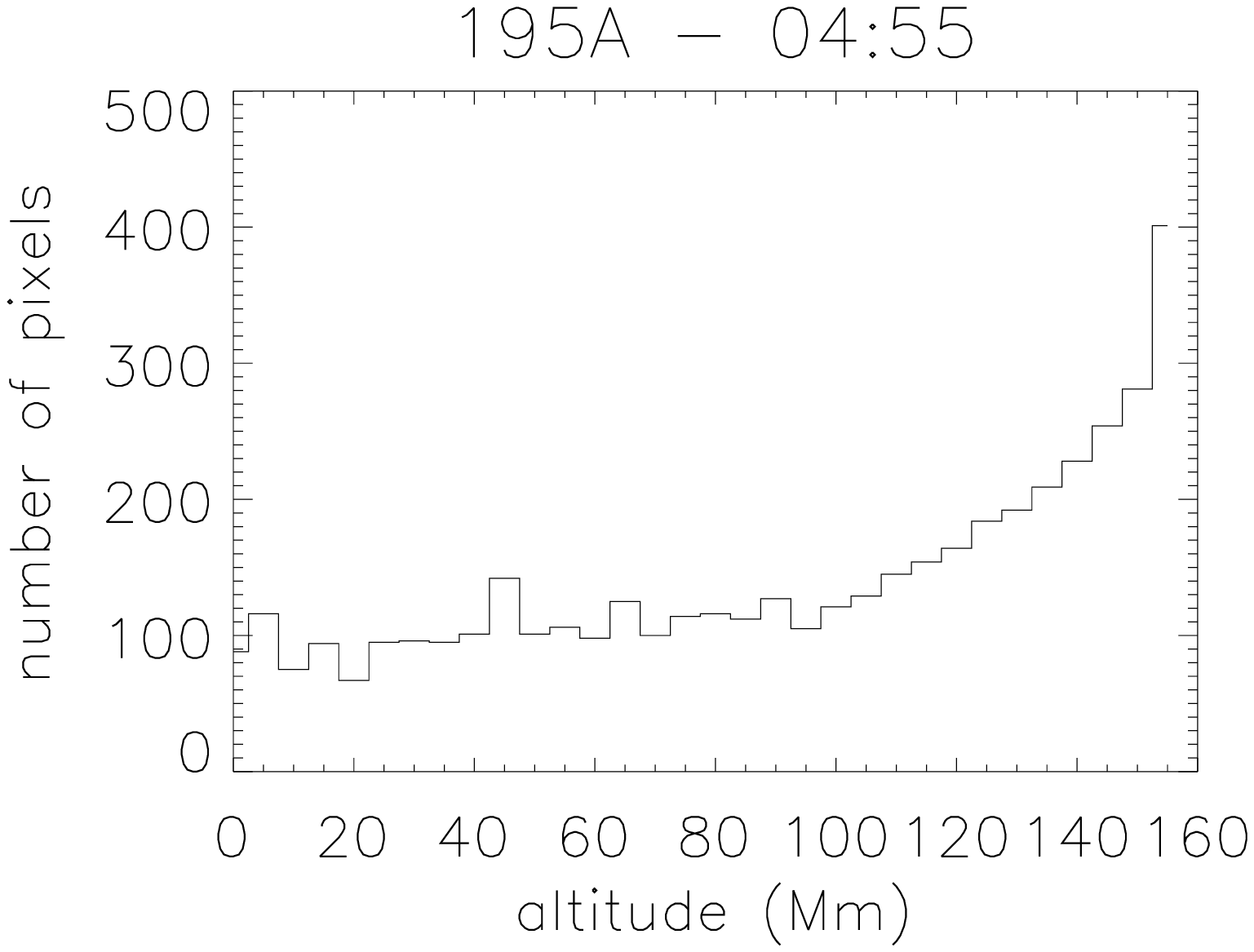}}
\caption{Time series of altitude map obtained using observations at 195 \AA. White corresponds to locations where the altitude is not computed due to the absence of wave front. The altitude color coding is given in the scale at the top of Figure \ref{cartealtitude171}, which values are in Mm. The gray circles show the location of the post-flare loops. The gray arrows show the wave front. The plots are histograms of the amount of pixels presenting a particular altitude versus the altitude. Each histogram is obtained using the map in the above panel, except the points outside the wave front.}
\label{cartealtitude195}
\end{figure}

\begin{figure}
\centerline{\includegraphics[width=4cm]{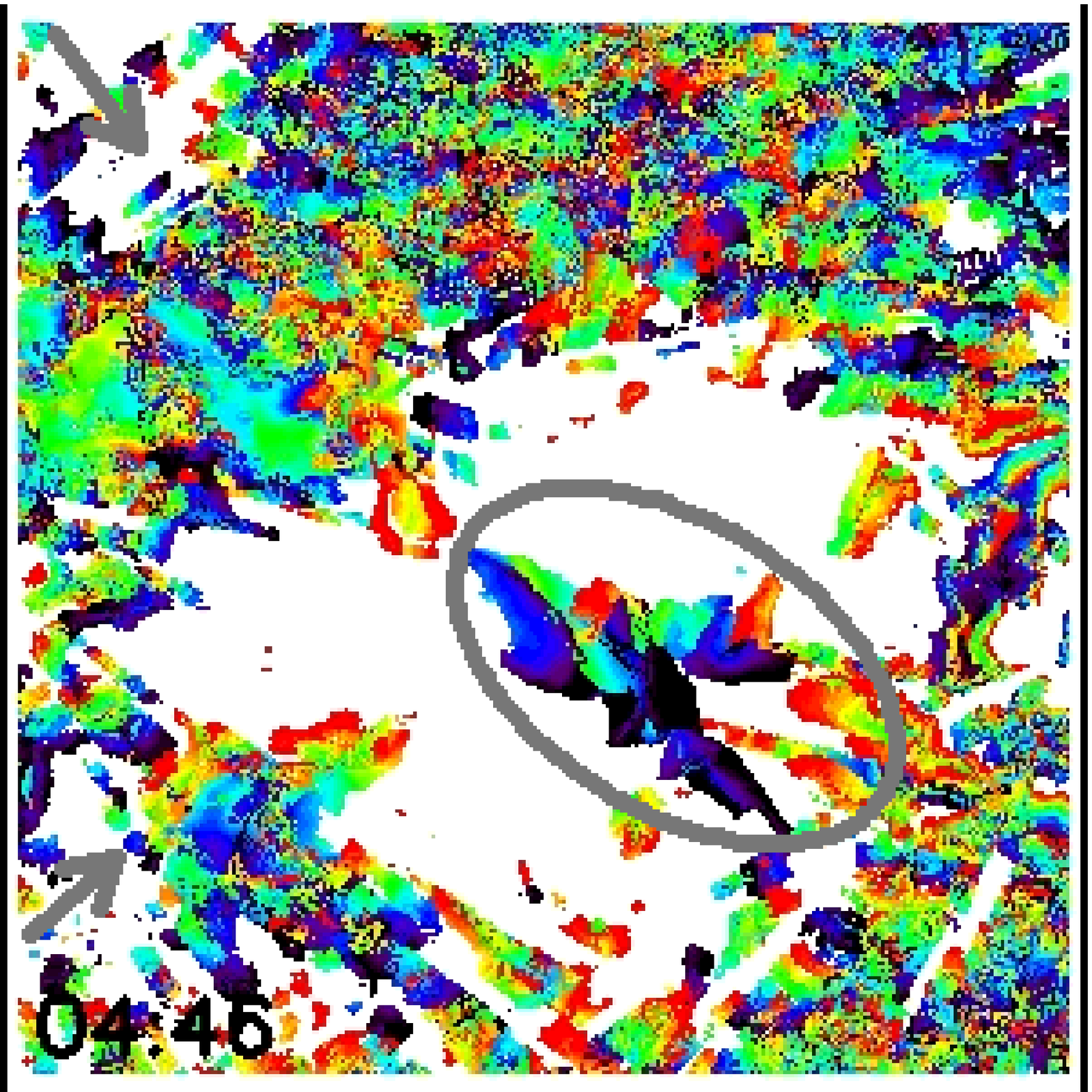}
\includegraphics[width=4cm]{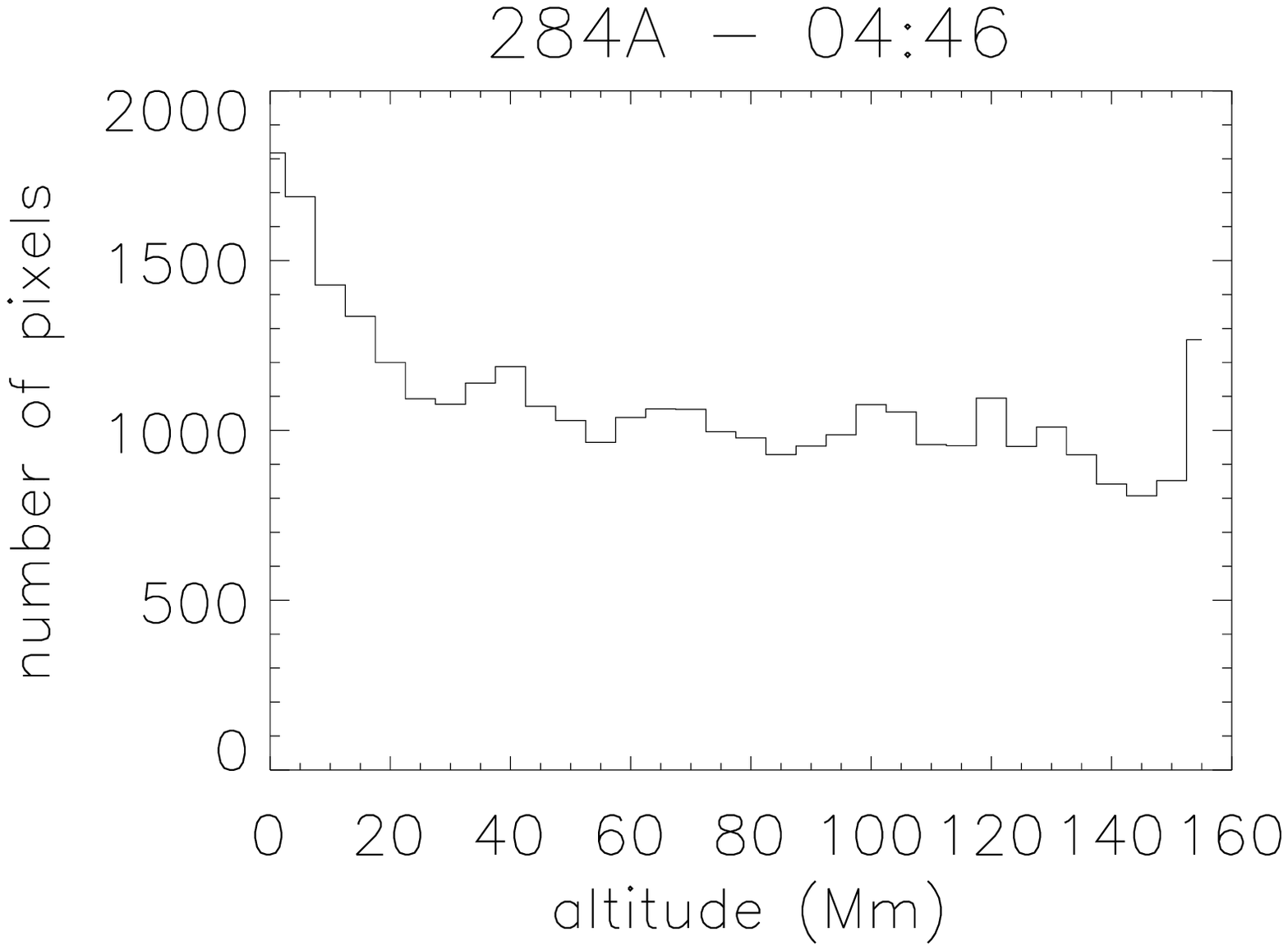}}
\caption{Altitude map obtained using observation at 284 \AA. White corresponds to locations where the altitude is not computed due to the absence of wave front. The altitude color coding is given in the scale at the top of Figure \ref{cartealtitude171}, which values are in Mm. The gray circle shows the location of the post-flare loops. The gray arrows show the wave front. The plot is histogram of the number of pixels presenting a particular altitude versus the altitude obtained using the map in the panel on the left, except the points outside the wave front.}
\label{cartealtitude284}
\end{figure}

Adding this last differencing process as a third step to the two first steps described in the previous Section, we repeat all the processing varying the radius of the sphere of reference from 695 Mm to 850 Mm and for each radius and trying to find the nulls in the drprd image. However, no nulls are found in the processed data due to the noise of the light detection, the motion of the structures, the scattering of the light inside the telescope and the integration along the line of sight of the plasma emission. Therefore, instead of trying to find the nulls in the drprd images, we first average the results of the drprd images over squares of 5$\times$5 pixels to smooth the small spatial variations and to reveal the wave front location.
At each given longitude and latitude position on the Sun, we identified the radius at which the drprd process gives the minimum value, leading to the identification of the altitude of the structure inside the pixel. We present the computed maps of altitude in Figures \ref{cartealtitude171}, \ref{cartealtitude195}, \ref{cartealtitude284}. It is noteworthy that the process always gives only one minimum value. We compute altitudes only in the wave front by excluding from the study every pixel that have not a minimum value found in the images obtained before the last subtraction (in 171 \AA~the minimum value is 0.016, in 195 \AA~is 0.035 and in 284 \AA~is 0.019). This leads to large white area, where the altitude is not computed, in the maps of altitude. The wave front is the colored elliptic structure.

The altitude maps show that the wave front is a structure evolving in altitude. All along the wave front, we find altitudes in the range of 4 Mm to 154 Mm, our limits of computation. The histograms displayed in Figures \ref{cartealtitude171}, \ref{cartealtitude195}, \ref{cartealtitude284}, are very flat. Therefore, computing the average altitude of wave front has very little meaning. However, we can see by eye that the inner border of the wave front is higher (154 Mm is more frequently found) than the outer border. 
These results seem to agree with the sketches given in Figure \ref{sketchdomecone}, the inner border of the wave front would be closer to the top of the structure and its outer border be closer to the base of the corona. 

Nevertheless, there are two limits to this approch. Firstly, the wave front is a broad and faint structure superposed over small scale features resulting in a salt and pepper ratio difference images. Therefore, in a given pixel, the minimum in the drprd process can be attributed either to the broad and faint wave front or to one of the salt and pepper features. Secondly, the line of sight integration effects, described in Section \ref{morphology}, can change the estimation of the altitude. Indeed, Figure \ref{sketchdomecone} explains why the brighter and wider section of the wave front as observed from one of both STEREO spacecraft is fainter and thinner than the same section being viewed from the other STEREO. Thus the reverse projection process aims at overlaying two features that do not have the same shape and brightness, even though they do correspond to the same section of the wave front.
During the reverse projection differencing process with increasing radii for the sphere of reference, the part of the wave front appearing thinner as view from one of the STEREO scans the same edge of the wave front that appears wider from the other STEREO, creating a systematic gradient of altitude that we could not accurately estimate.

It seems that the post flare loops (inside the gray ellipse in the altitude maps in the Figures \ref{cartealtitude171}, \ref{cartealtitude195} and \ref{cartealtitude284}) show increasing altitudes versus time, from 4 to 64 Mm, in 171 \AA. At the same instant, they are higher in 284 \AA~than in 195 \AA~and in 171 \AA~(in order). The wave front is well above the post flare loops. More detailed study of the post-flare loop should be done in a separate article.

From the map of altitude, we compute the average altitude of the wave front at the location comprised in the same slices of Figure \ref{superposition}. The western front seems at lower altitude (27 $\pm$ 9 Mm) than the eastern front (67 $\pm$ 11 Mm) at 04:38 UT in 171 \AA, corresponding to the morphologies of wave fronts described in Section \ref{morphology}. However, these values are very different from the ones found using the scc\_measure method (123 $\pm$ 13 and 107 $\pm$ 18 Mm, resp.). The reverse projection differencing technique seems to fail to properly find the western edge of the wave front certainly due to its very faint intensity.

The morphology of the maps of altitude in the three bandpasses are different: the wave front seems wider in 284 \AA~than in 195 \AA~than in 171 \AA~ and closer to the flare site in 284 \AA~than in 171 \AA~than in 195 \AA. However, we cannot be certain that this effect is real or due to an artifact to be determined.

\subsubsection{Correlation}
To try to find the mean altitude of the wave front, we use the reverse projected ratio difference (rprd) images obtained through the process described in Section \ref{method}. We add the following new processes to the first and second processes in Section \ref{method}:
third, the flare site and the dimming region is removed by replacing the corresponding pixel by a random intensity.
Fourth, we compute the correlation coefficient of the rprd images obtained using the two spacecraft observations implying a given radius of sphere of reference.

\begin{figure}
\centerline{\hfill 171 \AA~\hfill 195 \AA~\hfill 284 \AA~\hfill}
\centerline{\includegraphics[width=4cm]{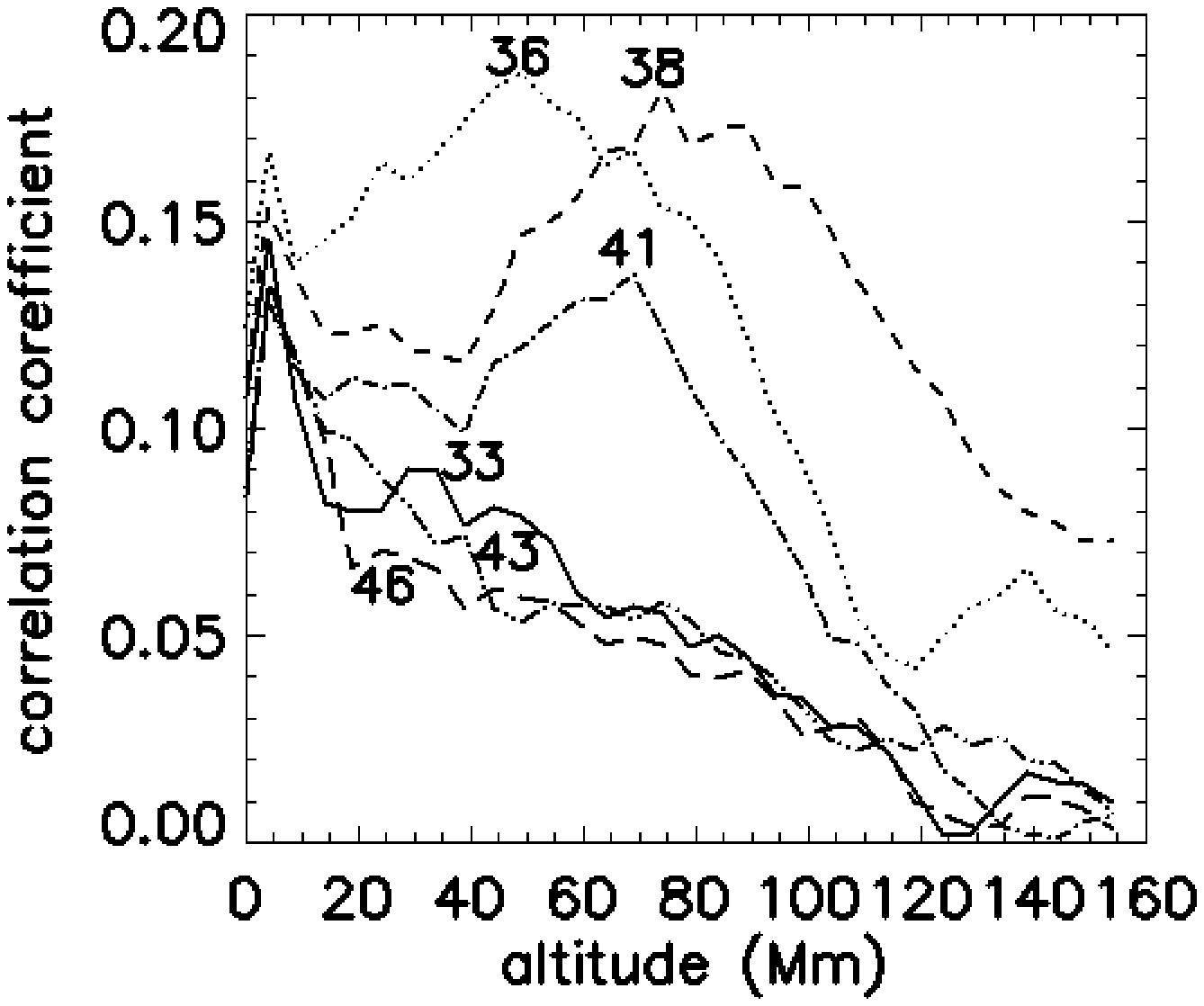}
\includegraphics[width=4cm]{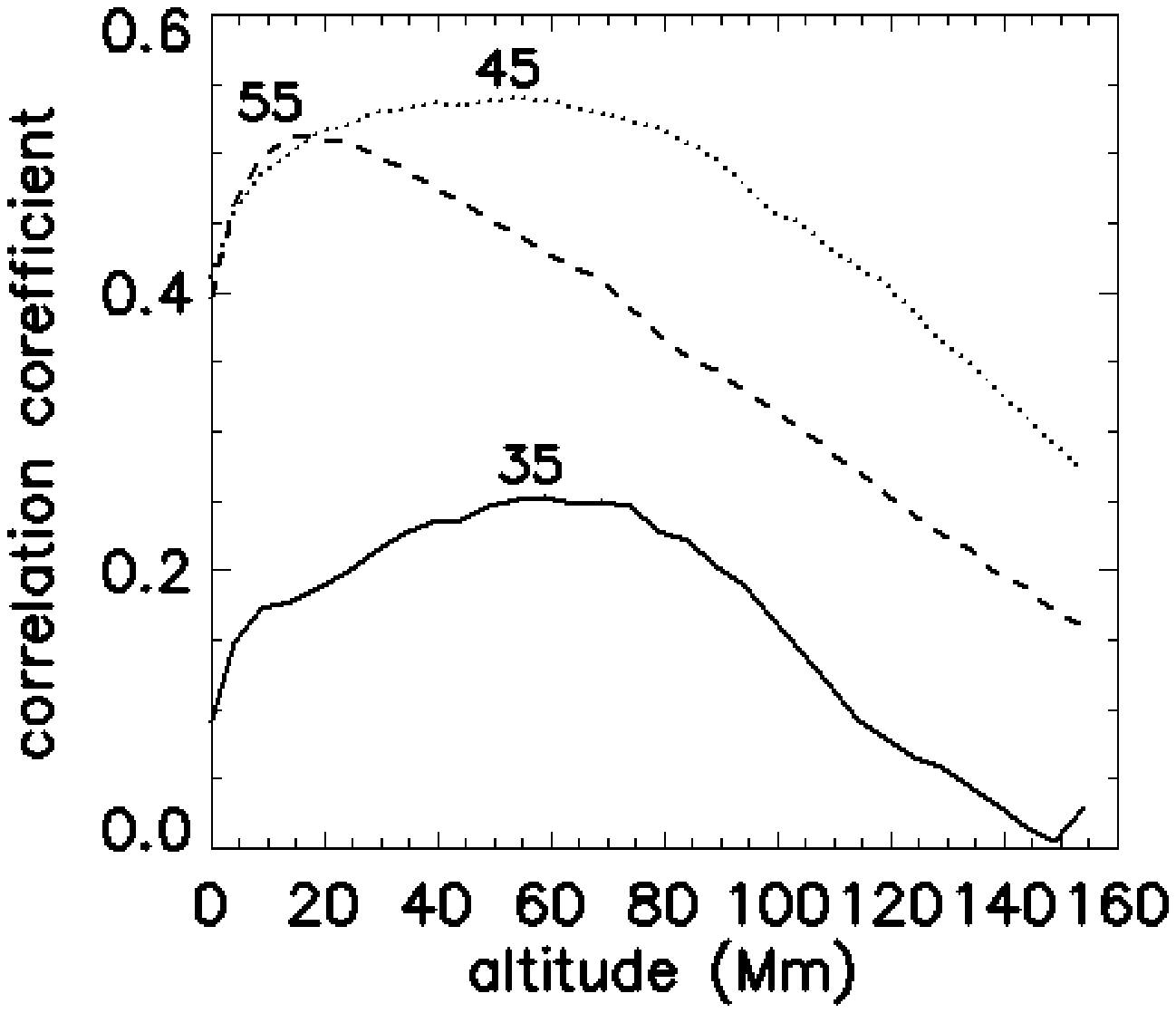}
\includegraphics[width=4cm]{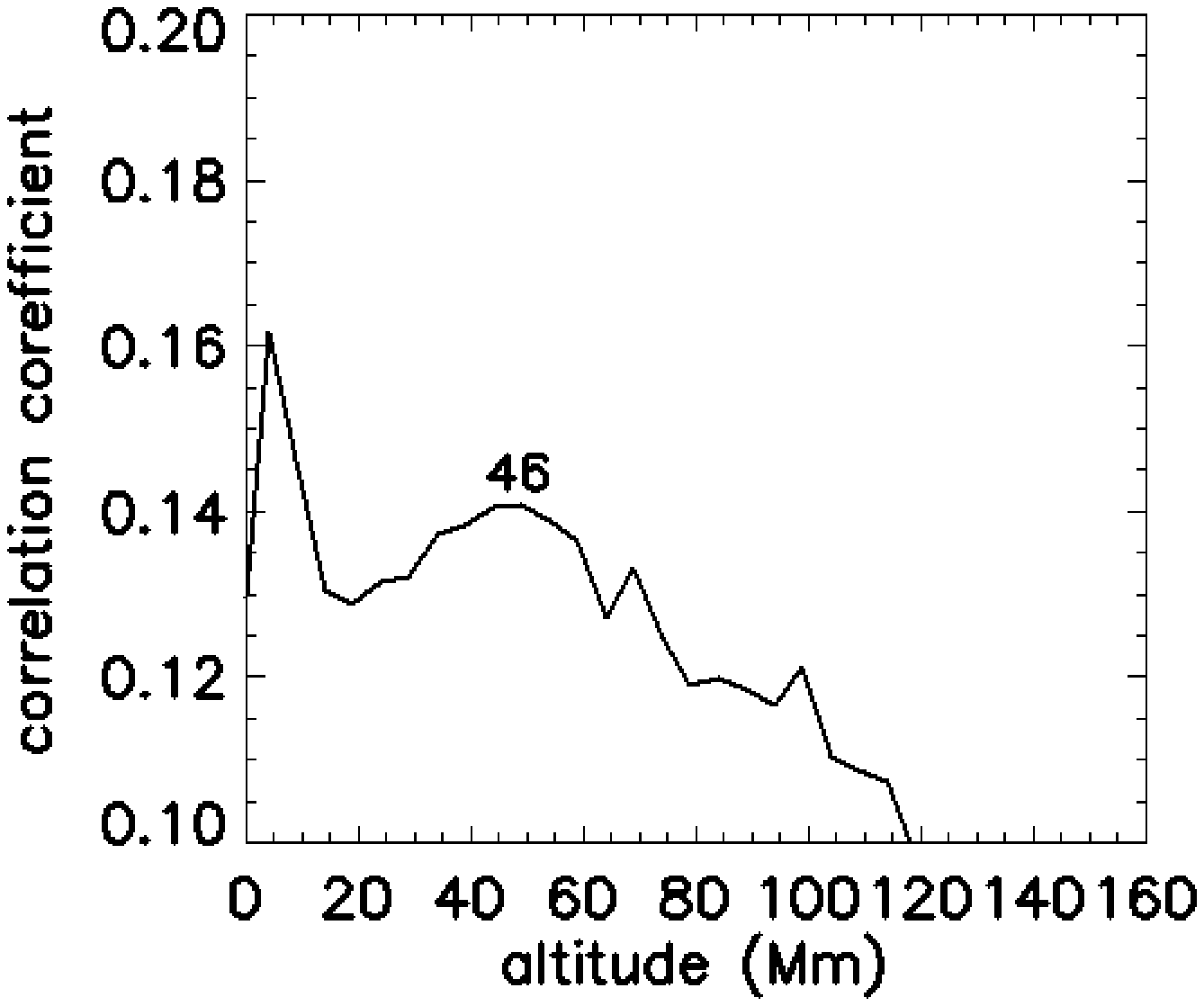}}
\caption{Plot of the correlation coefficient between two ratio difference STEREO A and B images projected on reference sphere having a defined radius versus the altitude of the reference sphere above the solar photosphere. The first (second and third) panel is for 171 \AA~(195 \AA~and 284 \AA, resp.) observations. The numbers above each line show the location of the highest correlation coefficient and the minutes of the time of two studied images. The first peak at 4 Mm in 171 \AA~and 284 \AA~corresponds to the presence of many fine structures, assimilated to noise. The altitude of the wave front increases from 04:33 UT to 04:38 UT in the 171 \AA~panel.}
\label{correlationcoef}
\end{figure}

\begin{figure}
\centerline{171 \AA~      195 \AA~}
\centerline{\includegraphics[width=4cm]{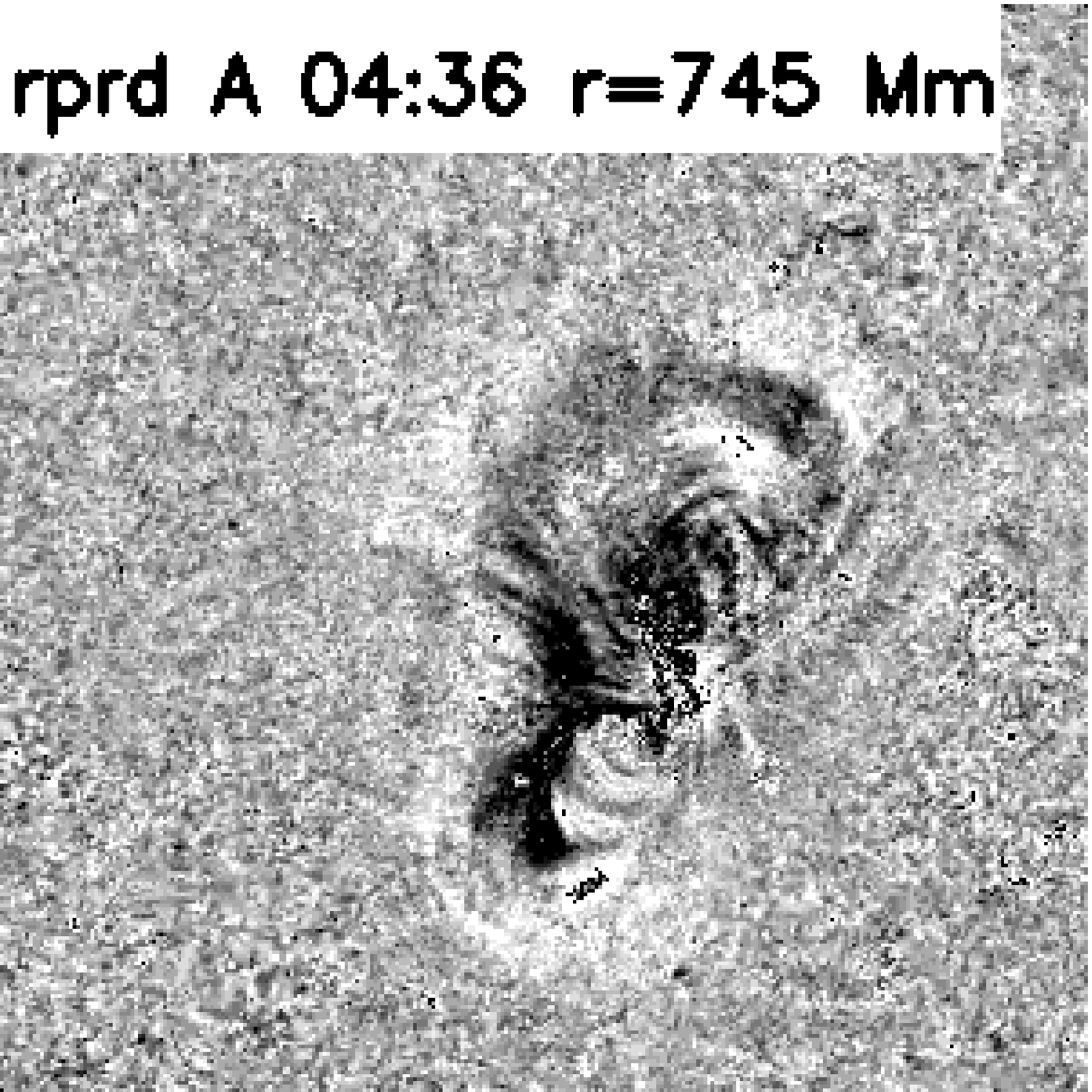}
\includegraphics[width=4cm]{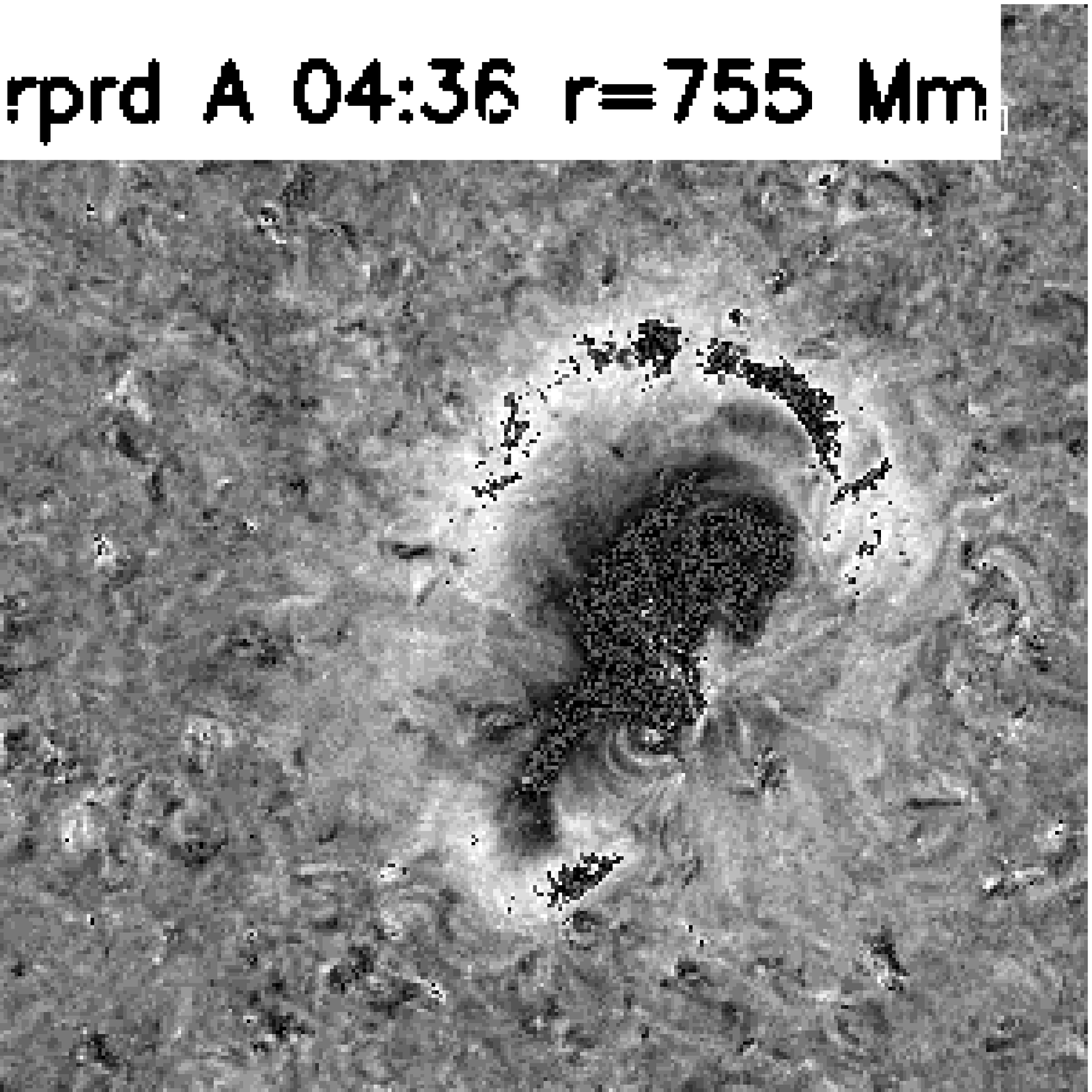}}
\centerline{\includegraphics[width=4cm]{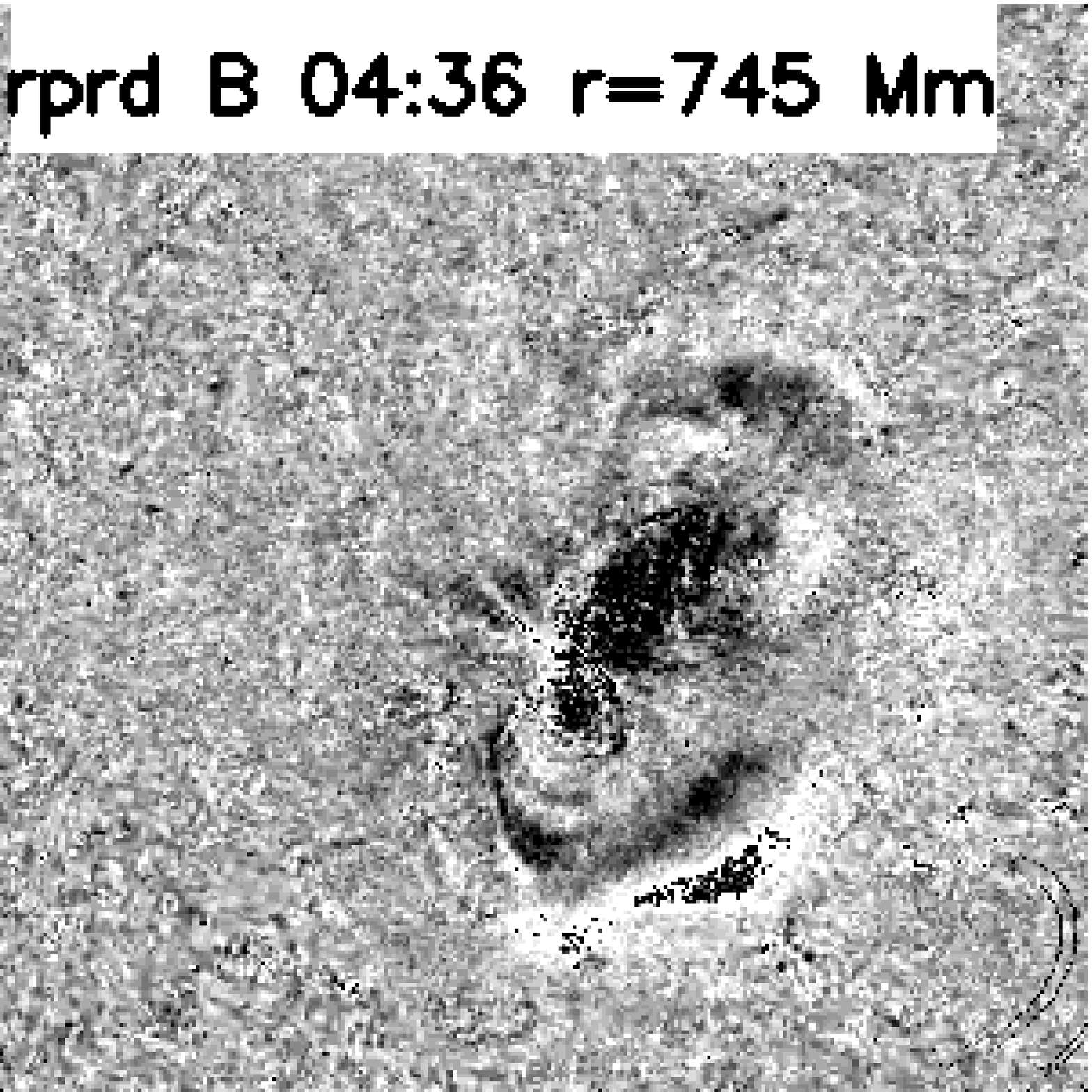}
\includegraphics[width=4cm]{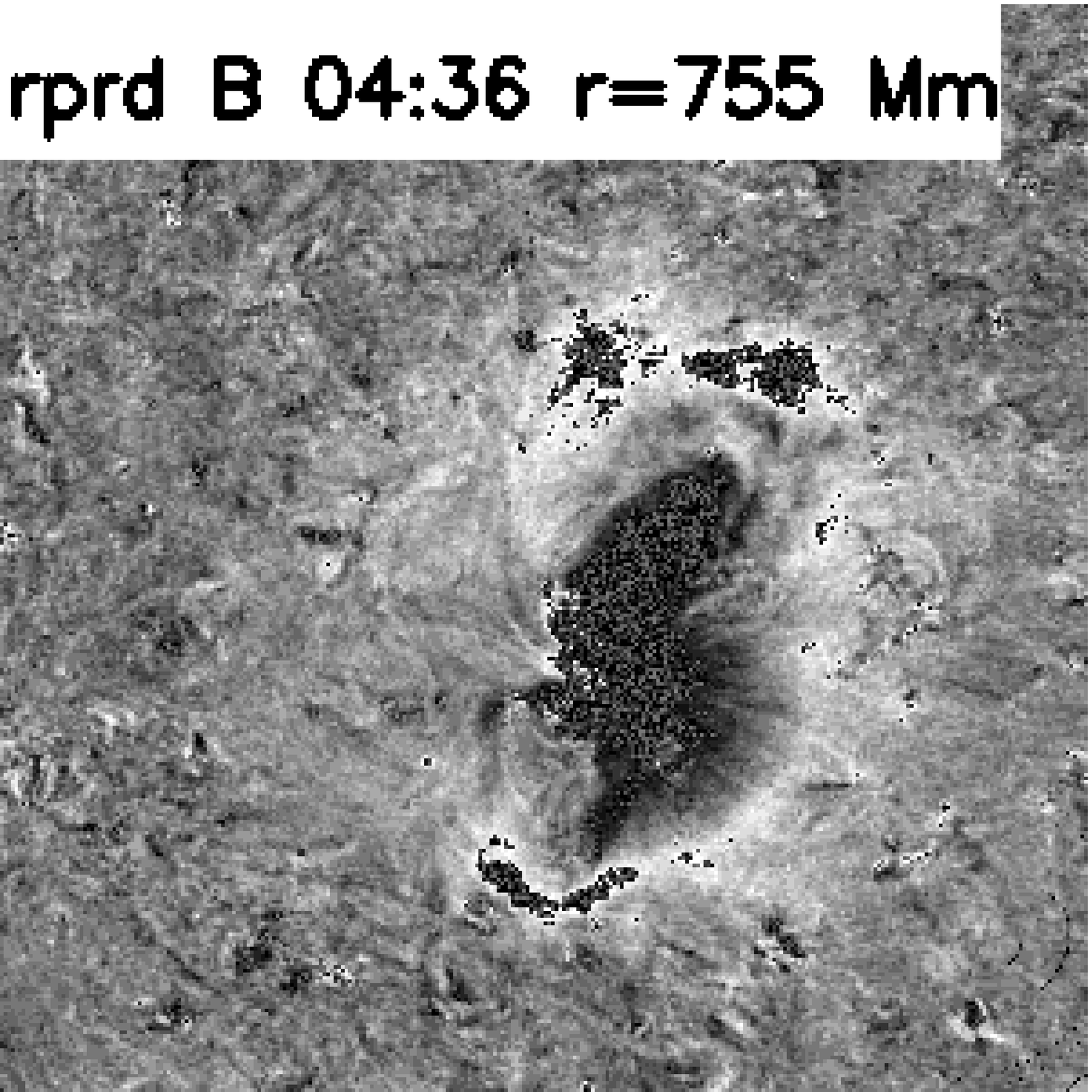}}
\caption{Columns are couples of reverse projected ratio difference (rprd) images obtained with the two STEREO A and B at the same time, at the radius of reference sphere for reverse projection that corresponds to the highest correlation coefficient.}
\label{correlation}
\end{figure}

The highest correlation coefficient gives the radius of reverse projection that corresponds to the average altitude of the wave front (plots of the correlation coefficient versus the altitude is given in Figure \ref{correlationcoef}). However, in the two last images of the wave front obtained using the 171 \AA~filter, the correlation process failed to be sensitive to the wave front emission as it becomes very faint. So, we made the correlation of these observations by eye.
Examples of reverse projected ratio difference images having the best correlation are shown in Figure \ref{correlation}. The derived altitudes are shown in Table 3.

\begin{table}
\begin{tabular}{|c|c|c|c|c|c|}
\cline{1-6}
\multicolumn{2}{c|}{171 \AA}&\multicolumn{2}{c|}{195 \AA}&\multicolumn{2}{c|}{284 \AA}\\
\cline{1-6}
 time & height & time & height & time & height \\
 (UT) & (Mm)  & (UT) & (Mm)  & (UT) & (Mm)  \\
\cline{1-6}
4:33& 34 &      &    &      &    \\
4:36& 49 & 4:35 & 59 &      &    \\
4:38& 74 &      &    &      &    \\
4:41& 69 &      &    &      &    \\
4:43& 64* &      &    &      &    \\
4:46& 64* & 4:45 & 54 & 4:46 & 49 \\
    &    & 4:55 & 19 &      &    \\
\cline{1-6}
\end{tabular}
\label{table:correlation}
\caption{Table presenting the average altitude of the wave front computed in the three available bandpass using the reverse projection correlation method. The stars indicate the estimated altitude from the best correlation done by eye.}
\end{table}

\begin{figure}
\centerline{\hfill eastern \hfill western \hfill}
\centerline{\includegraphics[width=4cm]{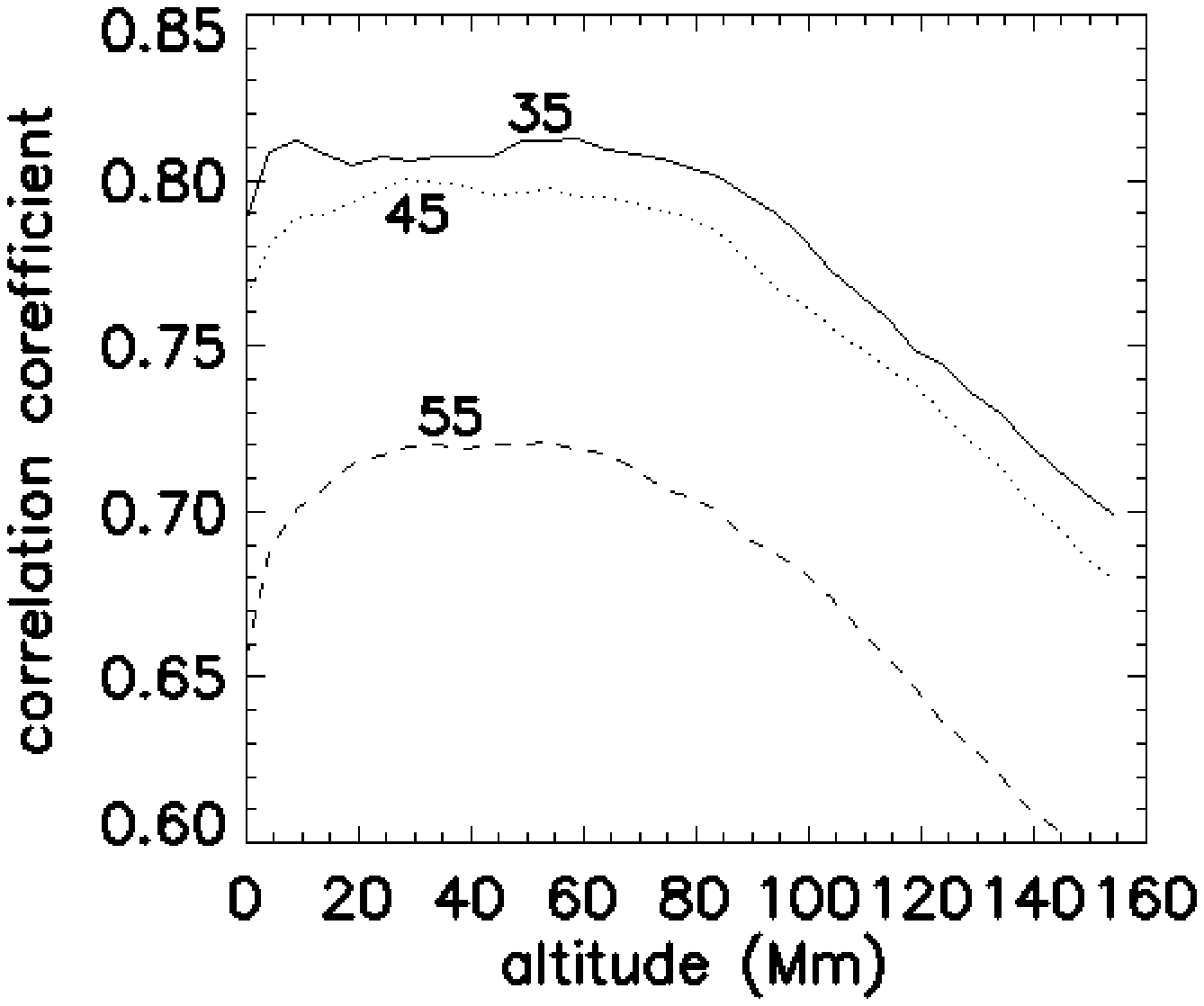}
\includegraphics[width=4cm]{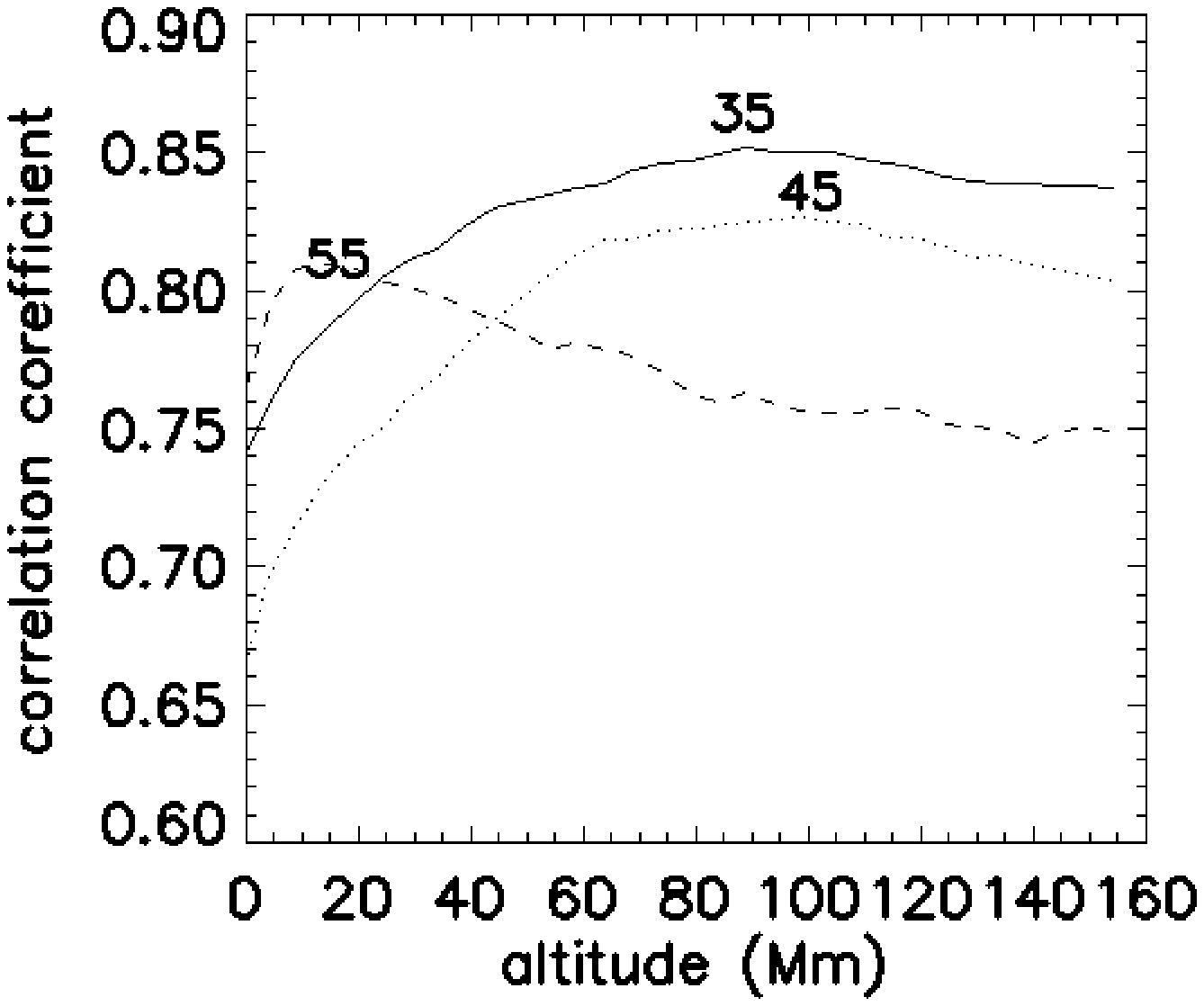}}
\caption{Plot of the correlation coefficient between two ratio difference STEREO A and B images, focused on the eastern (western) part of the wave front on left (right, resp.), obtained using the 195 \AA~filter, projected on reference sphere having a defined radius versus the altitude of the reference sphere above the solar photosphere. The numbers above each line show the location of the highest correlation coefficient and the minutes of the time of two studied images. The altitude of the western part of the wave front increases from 04:35 UT to 04:45 UT while its eastern part has decreasing altitude. The eastern part of the wave front is at lower altitude than the western one.}
\label{correlationcoefeastwest}
\end{figure}

As the step of the radius of the sphere of reference is 5 Mm, the derived altitudes have a minimum error of 5 Mm. The correlation coefficient is not sensitive to the various parameters we put into the different steps of processing, but very sensitive to the relative intensity of the wave front to the surrounding structures (flare, loops, bright points, dimming ...). Even if we tried to remove those structures by finding their typical emission and the typical emission of the wave front, and replacing the intensity of the pixels of the surrounding structures by a random intensity having the mean intensity of the wave front, not all the structures are removed. As the intensities of the wave front are the closest to the surrounding structures, the highest correlation coefficient corresponds to the location of the numerous surrounding structures. This effect is very sensitive in the last image obtained in each filter.

The following results are derived from the study of the plots given in Figure \ref{correlationcoef}. The altitude of the wave front increases rapidly during the first 5 minutes (04:33 - 04:38 UT) in 171 \AA~from 34 Mm to 74 Mm. The wave front lies at about the same altitude in 195 \AA~at 04:35 UT and in 171 \AA~at 04:36 UT. After 04:38 UT, it goes back to about 64 Mm during the 5 following minutes. The last available image at 195 \AA~shows a wave front very low in altitude: 19 Mm, corresponding to the altitude of the low coronal structures.

The same study performed on the eastern and western parts of the wave front is shown in Figure \ref{correlationcoefeastwest}. As the structure is very diffuse in 171 \AA~and in 284 \AA, the 195 \AA~observations analysis is conclusive only. The western wave front has increasing altitude from 04:35 UT to 04:45 UT while the eastern wave front, more diffuse, does not show this increase during the same interval of time. The western wave front is higher than the eastern one (89 and 59 Mm respectively at 04:35 UT). These altitudes are significantly different than the ones given by the scc\_measure and difference methods possibly because of the larger portion of the wave front taken to derive the altitude in this last correlation method but also due to the difficulty of each method to find the signal of the wave front itself.

\section{Discussion}

We present a study of the altitude of a coronal wave using STEREO mission observations. This mission consists of two spacecraft moving on the Earth orbit, one ahead and the other one trailing behind. The separation angle of the two spacecraft, at the time the observation reported here, permits to derive the altitude of observed structures using triangulation techniques. On December, 7th, 2007, a coronal wave is observed. It shows different morphologies under the two points of view due to integration along the line of sight of the emission of the thin plasma in its 3D morphology. The wave front is observed through different filters. The interpretation of these data is not straightforward.

A study of the emission of the wave front in the available coronal bandpasses shows that the plasma temperature would be better observed using a bandpass having emission line at high temperature (1.5-5 MK). The coronal wave front is certainly multithermal.

We try to use 3 different techniques to estimate the altitude of the coronal wave. Each of them present artifacts. (1) The scc\_measure.pro procedure in the SolarSoft is dependent on the process of visualization of the wave front (i.e. the contrast of visualization might highlight or hide parts of the wave front border) and on what is estimated as being the same point in one image and in the other one. (2) The reverse projection differencing method is also dependent on the line of sight integration of the light emission but also on random results. The effect of random results should be quantified but was not possible to define at this time. However, this process highlights certainly the complexity of the wave front which is composed of both underlying structures that illuminate on the passage of a fainter structure and of this fainter structure itself. (3) The reverse projection correlation technique permits to obtain an average altitude of the wave front if the fine structure of the wave front is correctly removed from this estimation which is not straighforward to do. The difference in the results coming from these technics are due to the artifacts of the technics themselves. Therefore, none of these technics can give reliable results if taken alone. However mixing them, it is possible to derive the tendency of the results. The results given hereafter come from the mixed analysis of all the results given by the scc\_measure, reverse projection differencing method, and reverse projection correlation techniques. All these results are in agreement with the first estimates of the 3D morphology given in Section \ref{morphology}.

It is found that the altitude of the observed wave front is higher than the post flare loops but remain rather low (34-154 Mm), putting together extrema given by the three techniques. This altitude rapidly increases during the first 5 minutes (from 34 to 74 Mm) then goes back to the low corona as shown by the reverse projection correlation technique.
This can be explained by the dilution of the emitting plasma in the coronal wave front during the first instants. During these first instants, the observations are sensitive to the faint coronal wave front that expands. Then, as this structure expands in all 3D directions, its plasma density decreases, its light emission fades also, but the lower structures that are perturbed by its passage, continue to be illuminated. The highest substructure in the wave front reaches 154 Mm, as shown by the reverse projection differencing technique. No conclusion can be given about the 3D anisotropy of the wave front due to the difficulty of each method to be sensitive to the wave front, difficulty increased when studying separately the western and eastern part of the wave front.


The altitude of the wave front seems low (34-154 Mm), even lower than in Patsourakos and Vourlidas (2009) who find that the wave front lies at 90 Mm. Patsourakos and Vourlidas (2009) conclude that this low altitude is in contradiction with the current shell model given in Delann\'ee et al. (2008). In this later article, the expansion of magnetic field line and the current shell, produced by this expansion simulated to reproduce some characteristics of a CME, is compared to two wave fronts. The simulation for the current shell model is dimensionalized to the size of the flaring active region and to its magnetic field strength. The derived altitude is about 280-407 Mm. To be correctly compared to the measured altitude of the wave front in the present study, the current shell simulation has to be dimensionalized to the active region that produced the wave front on December 7th, 2007. A brief analysis of the magnetogram gives a distance of 28 Mm between the two polarities of this active region. The superposition of the current shell model to the observed wave front gives a dimensionless altitude of 6 at the time step 76, i.e. 10 alfv\`en time after the beginning of the eruption. The maximum magnetic field strength is about 200 G. Using the relations found in Delann\'ee et al. (2008 pages 141 and 142), those numbers correspond to a wave front at 84 Mm in altitude 6 minutes after the beginning of the eruption. The two estimates of the altitude of this wave front, the one coming from the wave front observation (34-154 Mm) and the one from the dimensionalyzed analysis of the simulation of the current shell model (84 Mm), are surprisingly in good agreement taking into account the many parameters (line of sight, emission line temperature, etc.) that influence the results. 

In conclusion, the wave front definitely rises for the first 5 minutes which corresponds to any model implying a rising border, a dome of a magnetosonic wave (e.g. Afanasyev and Uralov 2011, Schmidt and Offman 2010, Uchida 1974) or an ice cone of a CME (Delann\'ee et al. 2008, Chen et al. 2002).

\begin{acks}
Susanna Parenti acknowledges the support from the Belgian Federal Science Policy Office through the ESAPRODEX programme.

This CME catalog is generated and maintained at the CDAW Data Center by NASA and The Catholic University of America in cooperation with the Naval Research Laboratory. SOHO is a project of international cooperation between ESA and NASA.

The STEREO/SECCHI data used here are produced by an international consortium of the Naval Research Laboratory (USA), Lockheed Martin Solar and Astrophysics Lab (USA), NASA Goddard Space Flight Center (USA), Rutherford Appleton Laboratory (UK), University of Birmingham (UK), Max-Planck-Institut f\"ur Sonnensystemforschung (Germany), Centre Spatiale de Liège (Belgium), Institut d'Optique Th\'eorique et Appliqu\'ee (France), and Institut d'Astrophysique Spatiale (France).

Hinode is a Japanese mission developed and launched by ISAS/JAXA, with NAOJ as domestic partner and NASA and STFC (UK) as international partners. It is operated by these agencies in co-operation with ESA and NSC (Norway).
\end{acks}
\end{article} 
\end{document}